\documentclass[Journal]{IEEEtran}

\usepackage[cmex10]{amsmath}
\usepackage{amssymb}
\usepackage{graphicx}
\usepackage{verbatim}
\usepackage{mathdots}
\usepackage{mathtools}

\usepackage{psfrag}

\usepackage{cite}
\usepackage{hyperref}
\usepackage[caption=false,font=footnotesize]{subfig}
\usepackage{stfloats}
\fnbelowfloat
\usepackage{mathrsfs}
\usepackage[utf8]{inputenc}
\usepackage[T1]{fontenc}

\usepackage{enumitem}
\usepackage{dsfont}
\newcommand{\overbar}[1]{\mkern 1.5mu\overline{\mkern-1.5mu#1\mkern-1.5mu}\mkern 1.5mu}

\newcounter{muni}
\newenvironment{remunerate}
               {\begin{list}{{\upshape 
               \arabic{muni}.}}{\usecounter{muni}
                \setlength{\leftmargin}{0pt}
                \setlength{\itemindent}{15pt}}}{\end{list}}

\makeatletter
\newcommand{\labitem}[2]{%
\def\@itemlabel{#1}
\item
\def\@currentlabel{#1}\label{#2}}
\makeatother

\newenvironment{smallarray}[1]
 {\null\,\vcenter\bgroup\scriptsize
  \arraycolsep=.2em
  \hbox\bgroup$\array{@{}#1@{}}}
 {\endarray$\egroup\egroup\,\null}

\def\smallint{\begingroup\textstyle \int\endgroup}

\begin{document}

\title{On the optimal control of passive or non-expansive systems}
\author{Timothy H. Hughes \thanks{Timothy H.\ Hughes is with the College of Engineering, Mathematics and Physical Sciences, University of Exeter, Penryn Campus, Penryn, Cornwall, TR10 9EZ, UK, \texttt{t.h.hughes@exeter.ac.uk}}\thanks{\copyright \hspace{0.1cm} 2017 IEEE.}}


\newtheorem{theorem}{Theorem}
\newtheorem{lemma}[theorem]{Lemma}
\newtheorem{proposition}[theorem]{Proposition}
\newtheorem{corollary}[theorem]{Corollary}
\newtheorem{definition}[theorem]{Definition}
\newtheorem{remark}[theorem]{Remark}
\providecommand{\abs}[1]{\left\lvert#1\right\rvert}

\newtheorem{thmapp}{Theorem}[section]
\newtheorem{lemapp}[thmapp]{Lemma}
\newtheorem{defapp}[thmapp]{Definition}

\maketitle

\begin{abstract}
The positive-real and bounded-real lemmas solve two important linear-quadratic optimal control problems for passive and non-expansive systems, respectively. The lemmas assume controllability, yet a passive or non-expansive system can be uncontrollable. In this paper, we solve these optimal control problems without making any assumptions. In particular, we show how to extract the greatest possible amount of energy from a passive but not necessarily controllable system (e.g., a passive electric circuit) using state feedback. A complete characterisation of the set of solutions to the linear matrix inequalities in the positive-real and bounded-real lemmas is also obtained. In addition, we obtain necessary and sufficient conditions for a system to be non-expansive that augment the bounded-real condition with new conditions relevant to uncontrollable systems.
\end{abstract}


\section{Introduction}
The positive-real and bounded-real lemmas are recognised as two of the most fundamental results in systems and control. They relate to two important optimal control problems, for passive and non-expansive systems, respectively \cite{JWDSP1, JWDSP2, AndVong, MOLLQOC}. In the positive-real lemma, the solution to the optimal control problem gives the least upper bound on the energy that can be extracted from a passive system. The lemmas also provide results on the solutions of important classes of Linear Matrix Inequalities (LMIs) and Algebraic Riccati Equations (AREs), the theory of spectral factorization, and the concepts of positive-real and bounded-real functions. But the classical versions of these lemmas consider only controllable systems.

In \cite{THTPLSNA}, it was emphasised that a passive system (e.g., a passive electric circuit) can be uncontrollable, and a theory of passive linear systems was developed that does not assume controllability. In contrast to other papers on this subject (see \cite[Section 3.3]{DSAC} and the discussion following Theorem \ref{thm:pbtnsc} in the present paper), \cite{THTPLSNA} did not introduce any alternative assumptions. But it did not consider the related optimal control problem, nor did it consider non-expansive systems. It is the purpose of this paper to solve the optimal control problems considered in the positive-real and bounded-real lemmas in the absence of any assumptions. In so doing, we characterise the set of solutions to the LMIs in these two lemmas, and we show how to use state feedback to extract the greatest amount of energy from a passive (not necessarily controllable) linear system. Also, in contrast with the case of controllable systems, we show that for there to exist a solution to the LMI in the bounded-real lemma, and for a system to be non-expansive, it is necessary \emph{but not sufficient} for the $\mathcal{H}_{\infty}$ norm of the system's transfer function to be bounded above by one. We also provide a necessary and sufficient condition, by introducing two new conditions relevant to uncontrollable systems.

The paper is structured as follows. Section \ref{sec:np} introduces the notation, and contains preliminary system theoretic results that are formalised using the behavioral approach \cite{JWIMTSC}. In Section \ref{sec:prl}, we review the classical positive-real lemma and the associated optimal control problem. We then state the main results concerning this optimal control problem in Theorems \ref{thm:pbtsc}, \ref{thm:pbtsced} and \ref{thm:pbtnsc}, which are proved in Sections \ref{sec:pbaae} and \ref{sec:cae}. The theorems explicitly characterise the solution to the optimal control problem in terms of an ARE (relevant when the transfer function $H$ satisfies $\lim_{\xi \rightarrow \infty}(H(\xi) + H(-\xi)^{T}) > 0$), and a spectral factorization of $H(\xi) + H(-\xi)^{T}$ (relevant in the general case). Section \ref{sec:brl} contains analogous results relevant to non-expansive systems (Theorems \ref{thm:nebtsc}, \ref{thm:nebtnsc} and \ref{thm:nebtsc2}, which are proved in Sections \ref{sec:nebas} and \ref{sec:compas}). In particular, we define the new concept of a \emph{bounded-real pair} of polynomial matrices, which appears in our new necessary and sufficient condition for a system to be non-expansive. Finally, some intermediate results are provided in Appendices \ref{sec:abas}--\ref{sec:ecaesl}.

\section{Notation and preliminaries}
\label{sec:np}

The notation in the paper is as follows. $\mathbb{R}$ ($\mathbb{C}$)  denotes the real (complex) numbers; $\mathbb{C}_{+}$ ($\overbar{\mathbb{C}}_{+}$) denotes the open (closed) right-half plane; $\mathbb{C}_{-}$ ($\overbar{\mathbb{C}}_{-}$) denotes the open (closed) left-half plane. If $\lambda \in \mathbb{C}$, then $\Re{(\lambda)}$ ($\Im{(\lambda)}$) denotes its real (imaginary) part, and $\bar{\lambda}$ its complex conjugate. $\mathbb{R}[\xi]$ ($\mathbb{R}(\xi)$) denotes the polynomials (rational functions) in the indeterminate $\xi$ with real coefficients. $\mathbb{R}^{m \times n}$ (resp., $\mathbb{C}^{m \times n}$, $\mathbb{R}^{m \times n}[\xi]$, $\mathbb{R}^{m \times n}(\xi)$) denotes the $m \times n$ matrices with entries from $\mathbb{R}$ (resp., $\mathbb{C}$, $\mathbb{R}[\xi]$, $\mathbb{R}(\xi)$). If $H \in \mathbb{R}^{m \times n}$, $\mathbb{C}^{m \times n}$, $\mathbb{R}^{m \times n}[\xi]$, or $\mathbb{R}^{m \times n}(\xi)$, then $H^{T}$ denotes its transpose, and if $H$ is nonsingular (i.e., $\det(H) \neq 0$) then $H^{-1}$ denotes its inverse. $\mathbb{R}^{n \times n}_{s}$ denotes the real $n \times n$ symmetric matrices. The block column (block diagonal) matrix with entries $H_{1}, \ldots , H_{n}$ is denoted $\text{col}(H_{1} \hspace{0.15cm} \cdots \hspace{0.15cm} H_{n})$ ($\text{diag}(H_{1} \hspace{0.15cm} \cdots \hspace{0.15cm} H_{n})$). We will use horizontal and vertical lines to indicate the partition in block matrix equations (e.g., see (\ref{eq:sfl2})). If $M \in \mathbb{C}^{m \times m}$, then $M > 0$ ($M \geq 0$) indicates that $M$ is Hermitian positive (non-negative) definite, and $\text{spec}(M) \coloneqq \lbrace \lambda \in \mathbb{C} \mid \text{det}(\lambda I {-} M) = 0\rbrace$. 

A $V \in \mathbb{R}^{n \times n}[\xi]$ is called \emph{unimodular} if its determinant is a non-zero constant (equivalently, $V^{-1} \in \mathbb{R}^{n \times n}[\xi]$). The matrices $P \in \mathbb{R}^{m \times n}[\xi]$ and $Q \in \mathbb{R}^{m \times q}[\xi]$ are called \emph{left co-prime} if $\text{rank}(\begin{bmatrix}P& -Q\end{bmatrix}(\lambda))$ is the same for all $\lambda \in \mathbb{C}$.  

If $H \in \mathbb{R}^{m \times n}(\xi)$, then (i) $H^{\star}(\xi) \coloneqq H(-\xi)^{T}$; (ii) $\text{normalrank}(H) \coloneqq \max_{\lambda \in \mathbb{C}}(\text{rank}(H(\lambda)))$; and (iii) $H$ is called \emph{proper} if $\lim_{\xi \rightarrow \infty}(H(\xi))$ exists, and \emph{strictly proper} if $\lim_{\xi \rightarrow \infty}(H(\xi)) = 0$. If $Z \in \mathbb{R}^{r \times n}(\xi)$ and $H \coloneqq Z^{\star}Z$, then $Z$ is called a \emph{spectral factor} of $H$ if (i) $Z$ is analytic in $\mathbb{C}_{+}$; and (ii) $Z(\lambda)$ has full row rank for all $\lambda \in \mathbb{C}_{+}$. If $H \in \mathbb{R}^{m \times n}(\xi)$, then $\|H\|_{\infty}$ denotes its $\mathscr{H}_{\infty}$ norm, and it is called bounded-real if $\|H\|_{\infty} \leq 1$ (i.e., $H$ satisfies $I - H(\bar{\lambda})^{T}H(\lambda) \geq 0$ for all $\lambda \in \overbar{\mathbb{C}}_{+}$). If $m=n$, then $H$ is called positive-real if (i) $H$ is analytic in $\mathbb{C}_{+}$; and (ii) $H(\bar{\lambda})^{T} + H(\lambda) \geq 0$ for all $\lambda \in \mathbb{C}_{+}$. 

We let $\mathcal{L}_{2}^{\text{loc}}\left(\mathbb{R}, \mathbb{R}^{k}\right)$ denote the ($k$-vector-valued) locally square integrable functions, and if $\mathbf{w} \in \mathcal{L}_{2}^{\text{loc}}\left(\mathbb{R}, \mathbb{R}^{k}\right)$ then $\mathbf{w}^{T}$ denotes the function satisfying $\mathbf{w}^{T}(t) = \mathbf{w}(t)^{T}$ for all $t \in \mathbb{R}$.

We will consider state-space systems of the form
\begin{align}
&\hspace*{-0.1cm} \mathcal{B}_{s} = \lbrace (\mathbf{u}, \mathbf{y}, \mathbf{x}) \in \mathcal{L}_{2}^{\text{loc}}\left(\mathbb{R}, \mathbb{R}^{n}\right) {\times} \mathcal{L}_{2}^{\text{loc}}\left(\mathbb{R}, \mathbb{R}^{m}\right) {\times} \mathcal{L}_{2}^{\text{loc}}\left(\mathbb{R}, \mathbb{R}^{d}\right) \nonumber\\
& \hspace{1.0cm} \text{such that } \tfrac{d\mathbf{x}}{dt} = A\mathbf{x} + B\mathbf{u} \text{ and } \mathbf{y} = C\mathbf{x} + D\mathbf{u}\rbrace, \nonumber \\
&\hspace*{-0.1cm}\text{with } A \in \mathbb{R}^{d \times d}, B \in \mathbb{R}^{d \times n}, C \in \mathbb{R}^{m \times d} \text{ and } D \in \mathbb{R}^{m \times n}, \label{eq:sssg}
\end{align}
and we interpret differentiation in a weak sense (see \cite[Section 2.3.2]{JWIMTSC}). In particular, for any given $\mathbf{u} \in \mathcal{L}_{2}^{\text{loc}}\left(\mathbb{R}, \mathbb{R}^{n}\right)$ and $\mathbf{x}_{0} \in \mathbb{R}^{d}$, there exists $(\mathbf{u}, \mathbf{y}, \mathbf{x}) \in \mathcal{B}_{s}$ such that
\begin{align}
\mathbf{x}(t) &= e^{A(t-t_{0})}\mathbf{x}_{0} + \smallint_{t_{0}}^{t}e^{A(t-\tau)}B\mathbf{u}(\tau) d\tau, \text{ and} \label{eq:xvocf}\\
\mathbf{y}(t) &= Ce^{A(t-t_{0})}\mathbf{x}_{0} + D\mathbf{u}(t) + \smallint_{t_{0}}^{t}Ce^{A(t-\tau)}B\mathbf{u}(\tau) d\tau , \label{eq:yvocf}
\end{align}
for all $t \geq t_{0}$. Moreover, if $(\mathbf{u}, \mathbf{y}, \mathbf{x}) \in \mathcal{B}_{s}$, then there exists $\mathbf{x}_{0} \in \mathbb{R}^{d}$ such that $\mathbf{x}$ satisfies (\ref{eq:xvocf}) for (almost) all $t \geq t_{0}$, so $\mathbf{x}(t)$ is determined by (\ref{eq:xvocf}) in this interval (and $\mathbf{x}(t_{0}) = \mathbf{x}_{0}$).

The external behavior of (\ref{eq:sssg}) is denoted by
\begin{equation}
\mathcal{B}_{s}^{(\mathbf{u},\mathbf{y})} \coloneqq \lbrace (\mathbf{u}, \mathbf{y}) \mid \exists \mathbf{x} \text{ such that } (\mathbf{u},\mathbf{y},\mathbf{x}) \in \mathcal{B}_{s}\rbrace, \label{eq:ebbs}
\end{equation}
and has the properties outlined in the following two lemmas, which are easily shown from results in \cite{THBRSF}. 

\begin{lemma}
\label{lem:ssr}
Let $\mathcal{B}_{s}$ be as in (\ref{eq:sssg}) and $\mathcal{A}(\xi) {\coloneqq} \xi I {-} A$. There exist polynomial matrices $P,Q,M,N,U,V,E,F, G$ such that
\begin{enumerate}[label=\arabic*., ref=\arabic*, leftmargin=0.5cm]
\setlength{\itemsep}{3pt}
\item $\begin{bmatrix}M& N \\ U& V\end{bmatrix}\begin{bmatrix}-D& I& -C\\ -B& 0& \mathcal{A}\end{bmatrix} = \begin{bmatrix}-P& Q& 0\\ -E& -F& G\end{bmatrix}$; \label{nl:utsse1}
\item $\begin{bmatrix}M& N \\ U& V\end{bmatrix}$ is unimodular; and \label{nl:utsse2}
\item $G$ is nonsingular. \label{nl:utsse3}
\end{enumerate}

Furthermore, whenever conditions \ref{nl:utsse1}--\ref{nl:utsse3} are satisfied, then the external behavior $\mathcal{B}_{s}^{(\mathbf{u},\mathbf{y})}$ in (\ref{eq:ebbs}) satisfies $\mathcal{B}_{s}^{(\mathbf{u},\mathbf{y})} = \mathcal{B}$, where
\begin{equation}
\hspace*{-0.4cm} \mathcal{B} {=} \lbrace (\mathbf{u}, \mathbf{y}) {\in} \mathcal{L}_{2}^{\text{loc}}(\mathbb{R}, \mathbb{R}^{n}) {\times} \mathcal{L}_{2}^{\text{loc}}(\mathbb{R}, \mathbb{R}^{m}) \mid P(\tfrac{d}{dt})\mathbf{u} {=} Q(\tfrac{d}{dt})\mathbf{y} \rbrace\label{eq:bsio}
\end{equation}
and we call $(A,B,C,D)$ a \emph{realization} of $(P,Q)$.
\end{lemma}

\begin{lemma}
\label{lem:ssr2}
Let $\mathcal{B}$ be as in (\ref{eq:bsio}) with $P \in \mathbb{R}^{m \times n}[\xi]$ and $Q \in \mathbb{R}^{m \times m}[\xi]$ where $Q$ is nonsingular and $Q^{-1}P$ is proper. Then there exists $\mathcal{B}_{s}$ as in (\ref{eq:sssg}) such that $\mathcal{B} = \mathcal{B}_{s}^{(\mathbf{u}, \mathbf{y})}$. Furthermore, whenever $\mathcal{B}_{s}$ in (\ref{eq:sssg}) satisfies $\mathcal{B} = \mathcal{B}_{s}^{(\mathbf{u}, \mathbf{y})}$, then there exist polynomial matrices $M,N,U,V,E,F$ and $G$ satisfying conditions \ref{nl:utsse1}--\ref{nl:utsse3} of Lemma \ref{lem:ssr}.
\end{lemma}

\begin{remark}
If $\mathcal{B}_{s}$ in (\ref{eq:sssg}) and $\mathcal{B}$ in (\ref{eq:bsio}) satisfy $\mathcal{B}_{s}^{(\mathbf{u}, \mathbf{y})} = \mathcal{B}$, then $H(\xi) \coloneqq D + C(\xi I - A)^{-1}B$ satisfies $Q^{-1}P = H$. However, the condition $Q^{-1}P = H$ only guarantees that $\mathcal{B}_{s}^{(\mathbf{u}, \mathbf{y})}$ takes the form of (\ref{eq:bsio}) if $P$ and $Q$ are left coprime.\hfill$\triangle$
\end{remark}

A system $\mathcal{B}$ is called \emph{controllable} if, for any two trajectories $\mathbf{w}_{1}, \mathbf{w}_{2} \in \mathcal{B}$ and $t_{0} \in \mathbb{R}$, there exists $\mathbf{w} \in \mathcal{B}$ and $t_{1} \geq t_{0}$ such that $\mathbf{w}(t) = \mathbf{w}_{1}(t)$ for all $t \leq t_{0}$ and $\mathbf{w}(t) = \mathbf{w}_{2}(t)$ for all $t \geq t_{1}$ \cite[Definition 5.2.2]{JWIMTSC}; and \emph{stabilizable} if for any $\mathbf{w}_{1} \in \mathcal{B}$ there exists $\mathbf{w} \in \mathcal{B}$ such that $\mathbf{w}(t) = \mathbf{w}_{1}(t)$ for all $t \leq t_{0}$ and $\lim_{t \rightarrow \infty}\mathbf{w}(t) = 0$ \cite[Definition 5.2.29]{JWIMTSC}. The behavior $\mathcal{B}$ in (\ref{eq:bsio}) is controllable (resp., stabilizable) if and only if $P$ and $Q$ are left coprime (resp., $\text{rank}(\begin{bmatrix}P& {-}Q\end{bmatrix}(\lambda))$ is the same for all $\lambda \in \overbar{\mathbb{C}}_{+}$) \cite[Theorems 5.2.10 and 5.2.30]{JWIMTSC}. We call the pair $(A,B)$ controllable (resp., stabilizable) if $\mathcal{B}_{s}$ is controllable (resp., stabilizable), which holds if and only if $\text{rank}(\begin{bmatrix}\lambda I - A& B\end{bmatrix}) = d$ for all $\lambda \in \mathbb{C}$ (resp., $\lambda \in \overbar{\mathbb{C}}_{+}$). 

Finally, if $\mathcal{B}_{s}$ takes the form of (\ref{eq:sssg}), then we call the pair $(C,A)$ \emph{observable} if $(\mathbf{u}, \mathbf{y}, \mathbf{x}) \in \mathcal{B}_{s}$ and $(\mathbf{u}, \mathbf{y}, \hat{\mathbf{x}}) \in \mathcal{B}_{s}$ imply $\mathbf{x} = \hat{\mathbf{x}}$ \cite[Definition 5.3.2]{JWIMTSC}. If, in addition, $\mathcal{B}_{s}^{(\mathbf{u}, \mathbf{y})}$ takes the form of (\ref{eq:bsio}), then we call $(A,B,C,D)$ an \emph{observable realization} for $(P,Q)$. With the notation
\begin{equation}
V_{o} \coloneqq \text{col}\begin{pmatrix}C& CA& \cdots & CA^{d-1}\end{pmatrix}, \label{eq:vo}
\end{equation}
then $(C,A)$ is observable if and only if $\text{rank}(V_{o}) = d$ \cite[Theorem 5.3.9]{JWIMTSC}.
\begin{remark}
\label{rem:bsieco}
It is easily shown that if $\mathcal{B}_{s}$ is controllable (resp., stabilizable) then so too is $\mathcal{B}_{s}^{(\mathbf{u},\mathbf{y})}$. Furthermore, if $(C,A)$ is observable and $\mathcal{B}_{s}^{(\mathbf{u},\mathbf{y})}$ is controllable (resp., stabilizable), then $\mathcal{B}_{s}$ is controllable (resp., stabilizable).\hfill$\triangle$
\end{remark}

\section{Passive systems}
\label{sec:prl}
The positive-real lemma considers the optimal control problem concerning the \emph{available energy} for a \emph{passive} system:
\begin{definition}[Available energy, Passive system]
\label{def:saprl}
Let $\mathcal{B}_{s}$ be as in (\ref{eq:sssg}) with $m = n$. For any given $\mathbf{x}_{0} \in \mathbb{R}^{d}$, let
\begin{multline*}
\mathcal{E}_{+}^{\sigma_{p}}(\mathbf{x}_{0}) = \lbrace \smallint_{t_{0}}^{t_{1}}{-(\mathbf{u}^{T}\mathbf{y})(t) dt} \mid t_{1} \geq t_{0}, (\mathbf{u}, \mathbf{y}, \mathbf{x}) \in \mathcal{B}_{s}, \\
\text{and } \mathbf{x}(t_{0}) =  \mathbf{x}_{0}\rbrace.
\end{multline*}
Then the \emph{available energy} $S_{a}^{\sigma_{p}}$ satisfies (i) $S_{a}^{\sigma_{p}}(\mathbf{x}_{0}) = \sup (\mathcal{E}_{+}^{\sigma_{p}}(\mathbf{x}_{0}))$ if $\mathcal{E}_{+}^{\sigma_{p}}(\mathbf{x}_{0})$ is bounded above; and (ii) $S_{a}^{\sigma_{p}}(\mathbf{x}_{0}) = \infty$ otherwise. If $S_{a}^{\sigma_{p}}(\mathbf{x}_{0}) < \infty$ for all $\mathbf{x}_{0} \in \mathbb{R}^{d}$, then $\mathcal{B}_{s}^{(\mathbf{u}, \mathbf{y})}$ is called \emph{passive}.
\end{definition}

In words, the available energy is the least upper bound on the energy that can be extracted from $t_{0}$ onwards.  

The positive-real lemma provides the solution (if it exists) to the optimal control problem in Definition \ref{def:saprl}, and several necessary and sufficient conditions for passivity. These relate: 
(a) the existence of real matrices $X \geq 0$ such that
\begin{equation}
\label{eq:lmipb}
\Omega(X) \coloneqq \begin{bmatrix}-A^{T}X - XA& C^{T}-XB\\ C-B^{T}X& D + D^{T}\end{bmatrix}
\end{equation}
satisfies $\Omega(X) \geq 0$; (b) whether the transfer function
\begin{equation}
H(\xi) \coloneqq D + C(\xi I - A)^{-1}B \label{eq:gd}
\end{equation}
is positive-real; and (c) a second optimal control problem concerning the \emph{required energy}, defined as follows
\begin{definition}[Required energy]
\label{def:srprl}
Let $\mathcal{B}_{s}$ be as in (\ref{eq:sssg}) with $m = n$. For any given $\mathbf{x}_{0} \in \mathbb{R}^{d}$, let
\begin{multline*}
\mathcal{E}_{-}^{\sigma_{p}}(\mathbf{x}_{0}) = \lbrace \smallint_{t_{1}}^{t_{0}}{(\mathbf{u}^{T}\mathbf{y})(t) dt} \mid t_{1} \leq t_{0}, (\mathbf{u}, \mathbf{y}, \mathbf{x}) \in \mathcal{B}_{s}, \\
\mathbf{x}(t_{1}) = 0 \text{ and } \mathbf{x}(t_{0}) =  \mathbf{x}_{0}\rbrace.
\end{multline*}
Then the \emph{required energy} $S_{r}^{\sigma_{p}}$ satisfies (i) $S_{r}^{\sigma_{p}}(\mathbf{x}_{0}) = \sup (\mathcal{E}_{-}^{\sigma_{p}}(\mathbf{x}_{0}))$ if $\mathcal{E}_{-}^{\sigma_{p}}(\mathbf{x}_{0})$ is bounded above; and (ii) $S_{r}^{\sigma_{p}}(\mathbf{x}_{0}) = \infty$ otherwise.
\end{definition}

Also, if $D + D^{T}>0$, then, with the notation
\begin{align}
\Gamma(X) \coloneqq -A^{T}X-XA \hspace*{4.2cm} \nonumber \\ - (C^{T}-XB)(D+D^{T})^{-1}(C-B^{T}X), \label{eq:gamxd} \\
\text{and } A_{\Gamma}(X) \coloneqq A-B(D+D^{T})^{-1}(C-B^{T}X), \label{eq:aprsm}
\end{align}
the conditions (a)--(c) also relate to the spectral properties of $A_{\Gamma}(X)$ for solutions $X$ to the ARE $\Gamma(X) = 0$. Critically to this paper, it is typically assumed that $(A,B)$ is controllable and $(C,A)$ is observable.
\begin{lemma}[Positive-real lemma]
\label{thm:prlsc}
Let $\mathcal{B}_{s}$ be as in (\ref{eq:sssg}) with $m=n$, $(A,B)$ controllable and $(C,A)$ observable; let $S_{a}^{\sigma_{p}}$ and $S_{r}^{\sigma_{p}}$ be as in Definitions \ref{def:saprl} and \ref{def:srprl}, and let $\Omega$ and $H$ be as in (\ref{eq:lmipb})--(\ref{eq:gd}). The following are equivalent:
\begin{enumerate}[label=\arabic*., ref=\arabic*, leftmargin=0.5cm]
\item $S_{a}^{\sigma_{p}}(\mathbf{x}_{0}) < \infty$ for all $\mathbf{x}_{0} \in \mathbb{R}^{d}$ (i.e., $\mathcal{B}_{s}^{(\mathbf{u}, \mathbf{y})}$ is passive). \label{nl:prlc1}
\item $S_{r}^{\sigma_{p}}(\mathbf{x}_{0})< \infty$  for all $\mathbf{x}_{0} \in \mathbb{R}^{d}$. \label{nl:prlc2}
\item $H$ is positive-real. \label{nl:prlc3}
\item There exists $X \in \mathbb{R}^{d \times d}_{s}$ such that $X \geq 0$ and $\Omega(X) \geq 0$.\label{nl:prlc4} 
\item $S_{a}^{\sigma_{p}}(\mathbf{x}_{0}) = \tfrac{1}{2}\mathbf{x}_{0}^{T}X_{-}\mathbf{x}_{0}$ and $S_{r}^{\sigma_{p}}(\mathbf{x}_{0}) = \tfrac{1}{2}\mathbf{x}_{0}^{T}X_{+}\mathbf{x}_{0}$, where $X_{-}, X_{+} \in \mathbb{R}_{s}^{d \times d}$ satisfy (i) $\Omega(X_{-}) \geq 0$ and $\Omega(X_{+}) \geq 0$; and (ii) if $X \in \mathbb{R}_{s}^{d \times d}$ satisfies $\Omega(X) \geq 0$, then $0<X_{-} \leq X \leq X_{+}$. \label{nl:prlc5}
\end{enumerate}

If, in addition, $D{+}D^{T} {>} 0$ and $\Gamma(X), A_{\Gamma}(X)$ are as in (\ref{eq:gamxd})--(\ref{eq:aprsm}), then \ref{nl:prlc1}--\ref{nl:prlc5} are equivalent to each of the following:
\begin{enumerate}[label=\arabic*., ref=\arabic*, leftmargin=0.5cm]
\setcounter{enumi}{5}
\item There exists a unique $X_{-} \in \mathbb{R}^{d \times d}_{s}$ satisfying (i) $X_{-} \geq 0$; (ii) $\Gamma(X_{-}) = 0$; and (iii) $\text{spec}(A_{\Gamma}(X_{-})) \in \overbar{\mathbb{C}}_{-}$.\label{nl:brlnsc3}
\item There exists a unique $X_{+} \in \mathbb{R}^{d \times d}_{s}$ satisfying (i) $X_{+} \geq 0$; (ii) $\Gamma(X_{+}) = 0$; and (iii) $\text{spec}(A_{\Gamma}(X_{+})) \in \overbar{\mathbb{C}}_{+}$.\label{nl:brlnsc4}
\end{enumerate}
Moreover, if conditions \ref{nl:brlnsc3} and \ref{nl:brlnsc4} hold, then $S_{a}^{\sigma_{p}}(\mathbf{x}_{0}) = \tfrac{1}{2}\mathbf{x}_{0}^{T}X_{-}\mathbf{x}_{0}$ and $S_{r}^{\sigma_{p}}(\mathbf{x}_{0}) = \tfrac{1}{2}\mathbf{x}_{0}^{T}X_{+}\mathbf{x}_{0}$.
\end{lemma}

\begin{IEEEproof}
See \cite[Sections 3--5]{JWDSP2}.
\end{IEEEproof}

It was shown in \cite{THTPLSNA} that, if controllability and observability are not assumed, then the positive-real condition is necessary \emph{but not sufficient} for there to exist a solution to the LMI in the positive-real lemma (condition \ref{nl:prlc4} in Lemma \ref{thm:prlsc}). A new condition was provided in terms of the polynomial matrices $P, Q$ describing the external behavior (see Lemma \ref{lem:ssr}). Specifically, it was shown that there exists a solution to the LMI if and only if $(P,Q)$ are a positive-real pair, defined as follows. 
\begin{definition}[Positive-real pair]
\label{def:prp}
Let $P, Q \in \mathbb{R}^{n \times n}[\xi]$. 
We call $(P, Q)$ a \emph{positive-real pair} if the following hold:
\begin{enumerate}[label=(\alph*)]
\item $P(\lambda)Q(\bar{\lambda})^{T} + Q(\lambda)P(\bar{\lambda})^{T} \geq 0$ for all $\lambda \in \mathbb{C}_{+}$. \label{nl:prpc1}
\item $\text{rank}(\begin{bmatrix}P& {-}Q\end{bmatrix}(\lambda)) = n$ for all $\lambda \in \overbar{\mathbb{C}}_{+}$. \label{nl:prpc2} 
\item If $\mathbf{p} \in \mathbb{R}^{n}[\xi]$ and $\lambda \in \mathbb{C}$ satisfy $\mathbf{p}^{T}(PQ^{\star} + QP^{\star}) = 0$ and $\mathbf{p}(\lambda)^{T}\begin{bmatrix}P& {-}Q\end{bmatrix}(\lambda) = 0$, then $\mathbf{p}(\lambda) = 0$. \label{nl:prpc3}
\end{enumerate}
\end{definition}
\begin{remark}
\label{rem:pmbs}
If $\mathcal{B}_{s}$ is as in (\ref{eq:sssg}), then $\mathcal{B}_{s}^{(\mathbf{u}, \mathbf{y})}$ takes the form indicated in Lemma \ref{lem:ssr}. With $P, Q$ as defined in Lemma \ref{lem:ssr}, $Q$ is nonsingular, and $H \coloneqq Q^{-1}P$ satisfies (\ref{eq:gd}). Then, condition \ref{nl:prpc1} of Definition \ref{def:prp} is equivalent to $H$ being positive-real \cite[Sections 4--5]{THTPLSNA}. Also, condition \ref{nl:prpc2} is equivalent to the stabilizability of $\mathcal{B}_{s}^{(\mathbf{u}, \mathbf{y})}$. Finally, a physical interpretation of condition \ref{nl:prpc3} is given in \cite[Sections 4--5]{THTPLSNA}. This condition relates to the fact that, if (i) $(\mathbf{u}, \mathbf{y}, \mathbf{x}) \in \mathcal{B}_{s}$ and $t_{1} > t_{0}$ satisfy $\mathbf{x}(t_{0}) = \mathbf{x}(t_{1}) = 0$ and $\smallint_{t_{0}}^{t_{1}}(\mathbf{u}^{T}\mathbf{y})(t)dt = 0$; (ii) $(0, \hat{\mathbf{y}}, \hat{\mathbf{x}}) \in \mathcal{B}_{s}$; and (iii) $\alpha \in \mathbb{R}$, then $(\alpha \mathbf{u}, \alpha \mathbf{y} {+} \hat{\mathbf{y}}, \alpha \mathbf{x} {+} \hat{\mathbf{x}}) \in \mathcal{B}_{s}$. It can then be shown that, if $\mathcal{B}_{s}^{(\mathbf{u}, \mathbf{y})}$ is passive, then $\smallint_{t_{0}}^{t_{1}}(\mathbf{u}^{T}\hat{\mathbf{y}})(t)dt = 0$.\hfill$\triangle$ 
\end{remark}

In this paper, we develop the results in \cite{THTPLSNA} to solve the optimal control problem of extracting the greatest possible amount of energy from a passive system, and to characterise the set of solutions to the LMI considered in the positive-real lemma, in the absence of any controllability or observability assumptions. The main results in this section are in the next three theorems, which will be proved in Sections \ref{sec:pbaae} and \ref{sec:cae}.

\begin{theorem}
\label{thm:pbtsc}
Let $\mathcal{B}_{s}$ and $\mathcal{B}_{s}^{(\mathbf{u},\mathbf{y})}$ be as in (\ref{eq:sssg}) and (\ref{eq:ebbs}) with $m = n$; let $S_{a}^{\sigma_{p}}$ be as in Definition \ref{def:saprl}; and let $V_{o}$ and $\Omega$ be as in (\ref{eq:vo}) and (\ref{eq:lmipb}). The following are equivalent:
\begin{enumerate}[label=\arabic*., ref=\arabic*, leftmargin=0.5cm]
\item $S_{a}^{\sigma_{p}}(\mathbf{x}_{0}) < \infty$ for all $\mathbf{x}_{0} \in \mathbb{R}^{d}$ (i.e., $\mathcal{B}_{s}^{(\mathbf{u}, \mathbf{y})}$ is passive). \label{nl:ptc1}
\item The external behavior $\mathcal{B}_{s}^{(\mathbf{u}, \mathbf{y})}$ takes the form of (\ref{eq:bsio}), where $(P, Q)$ is a positive-real pair. \label{nl:ptc3}
\item There exists $X \in \mathbb{R}^{d \times d}_{s}$ such that $X \geq 0$ and $\Omega(X) \geq 0$.\label{nl:ptc4}
\item $S_{a}^{\sigma_{p}}(\mathbf{x}_{0}) = \tfrac{1}{2}\mathbf{x}_{0}^{T}X_{-}\mathbf{x}_{0}$, where $X_{-} \in \mathbb{R}^{d \times d}_{s}$ satisfies (i) $X_{-} \geq 0$; (ii) $\Omega(X_{-}) \geq 0$; (iii) if $\mathbf{z} \in \mathbb{R}^{d}$, then $V_{o}\mathbf{z} = 0 \iff X_{-}\mathbf{z} = 0$; and (iv) if $X \in \mathbb{R}^{d \times d}_{s}$ satisfies $X \geq 0$ and $\Omega(X) \geq 0$, then $X_{-} \leq X$.\label{nl:ptc5}
\end{enumerate}
Moreover, if $(C,A)$ is observable and the above conditions hold, then (i) $\text{spec}(A) \in \overbar{\mathbb{C}}_{-}$; and (ii) if $X \in \mathbb{R}_{s}^{d \times d}$ satisfies $\Omega(X) \geq 0$, then $X_{-} \leq X$.
\end{theorem}

\begin{remark}
We note from Theorem \ref{thm:pbtsc} that, for a passive system, $\mathbf{z} \in \mathbb{R}^{d}$ satisfies $V_{o}\mathbf{z} = 0$ if and only if $S_{a}^{\sigma_{p}}(\mathbf{z}) = 0$. In words, the available energy of the state $\mathbf{z}$ is zero if and only if $\mathbf{z}$ is an unobservable mode (we call $\mathbf{z}$ an unobservable mode if $(0,0,\mathbf{x}) \in \mathcal{B}_{s}$ where $\mathbf{x}(t) = e^{At}\mathbf{z}$ for all $t \in \mathbb{R}$).\hfill$\triangle$
\end{remark}

The next theorem provides an explicit expression for the available energy for the case with $D+D^{T} > 0$.
\begin{theorem}
\label{thm:pbtsced}
Let $\mathcal{B}_{s}$ be as in (\ref{eq:sssg}) with $m = n$; let $S_{a}^{\sigma_{p}}$ be as in Definition \ref{def:saprl}; let $V_{o}, \Gamma$ and $A_{\Gamma}$ be as in (\ref{eq:vo}), (\ref{eq:gamxd}) and (\ref{eq:aprsm}), respectively; and let $D{+}D^{T} {>} 0$. The following are equivalent:
\begin{enumerate}[label=\arabic*., ref=\arabic*, leftmargin=0.5cm]
\item $S_{a}^{\sigma_{p}}(\mathbf{x}_{0}) < \infty$ for all $\mathbf{x}_{0} \in \mathbb{R}^{d}$ (i.e., $\mathcal{B}_{s}^{(\mathbf{u}, \mathbf{y})}$ is passive). \label{nl:ptnsc1}
\item There exists $X_{-} \in \mathbb{R}^{d \times d}_{s}$ satisfying (i) $X_{-} \geq 0$; (ii) $\Gamma(X_{-}) = 0$; (iii)  if $\mathbf{z} \in \mathbb{R}^{d}$ satisfies $V_{o}\mathbf{z} = 0$, then $X_{-}\mathbf{z} = 0$; and (iv) if $\lambda \in \mathbb{C}_{+}$ and $\mathbf{z} \in \mathbb{C}^{d}$ satisfy $A_{\Gamma}(X_{-})\mathbf{z} = \lambda \mathbf{z}$, then $V_{o}\mathbf{z} = 0$. \label{nl:ptnsc2}
\end{enumerate}
Moreover, if these conditions hold, then $S_{a}^{\sigma_{p}}(\mathbf{x}_{0}) {=} \tfrac{1}{2}\mathbf{x}_{0}^{T}X_{-}\mathbf{x}_{0}$.
\end{theorem}

The final theorem provides an explicit expression for the available energy in the general case.
\begin{theorem}
\label{thm:pbtnsc}
Let $\mathcal{B}_{s}$ be as in (\ref{eq:sssg}) with $m = n$; let $S_{a}^{\sigma_{p}}$ be as in Definition \ref{def:saprl}; and let $V_{o}$ and $H$ be as in (\ref{eq:vo}) and (\ref{eq:gd}). The following are equivalent:
\begin{enumerate}[label=\arabic*., ref=\arabic*, leftmargin=0.5cm]
\item $S_{a}^{\sigma_{p}}(\mathbf{x}_{0}) < \infty$ for all $\mathbf{x}_{0} \in \mathbb{R}^{d}$ (i.e., $\mathcal{B}_{s}^{(\mathbf{u}, \mathbf{y})}$ is passive). \label{nl:pts2c1}
\item There exists $X_{-} \in \mathbb{R}^{d \times d}_{s}$ satisfying (i) $X_{-} \geq 0$; (ii)  if $\mathbf{z} \in \mathbb{R}^{d}$ satisfies $V_{o}\mathbf{z} = 0$, then $X_{-}\mathbf{z} = 0$; and (iii) there exist real matrices $L$ and $W$ such that
\begin{itemize}[leftmargin=0.8cm]
\item[(iiia)] $-A^{T}X_{-} - X_{-}A = L^{T}L$, $C-B^{T}X_{-} = W^{T}L$, and $D + D^{T} = W^{T}W$; and
\item[(iiib)] $Z(\xi) {\coloneqq} W {+} L(\xi I {-} A)^{-1}B$ is a spectral factor of $H {+} H^{\star}$.
\end{itemize}\label{nl:pts2c2}
\end{enumerate}
Moreover, if these conditions hold, then $S_{a}^{\sigma_{p}}(\mathbf{x}_{0}) {=} \tfrac{1}{2}\mathbf{x}_{0}^{T}X_{-}\mathbf{x}_{0}$.
\end{theorem}

In proving Theorems \ref{thm:pbtsc}, \ref{thm:pbtsced} and \ref{thm:pbtnsc}, we show how to compute the available energy $S_{a}^{\sigma_{p}}$ and obtain a linear state feedback law such that, with $\mathbf{x}_{0} \coloneqq \mathbf{x}(t_{0})$, then $\smallint_{t_{0}}^{\infty}-(\mathbf{u}^{T}\mathbf{y})(t)dt$ is arbitrarily close to $S_{a}^{\sigma_{p}}(\mathbf{x}_{0})$ (see Remark \ref{rem:oc}).

We next present an example to illustrate the distinction between the results in this section and other papers in the literature that deal with similar objectives. 

It has long been recognised that the controllability and observability assumptions in the positive-real lemma are unduly restrictive, and there have been many notable attempts to relax these assumptions. A comprehensive summary is provided in \cite[Section 3.3]{DSAC} (see also \cite{cam_pass} for additional properties of the LMI $\Omega(X) \geq 0$). These results focus on the equivalence of the positive-real condition with the existence of solutions $X \in \mathbb{R}^{d \times d}_{s}$ to an LMI (similar to condition \ref{nl:ptc4} of Theorem \ref{thm:pbtsc}) or an ARE (similar to condition \ref{nl:ptnsc2} of Theorem \ref{thm:pbtsced}) \cite{CLJKYPL, SKSPRCPLTI, POPRL, KPRLNMR}. None of these papers explicitly consider the optimal control problem in Definition \ref{def:saprl}. Also, each of these papers introduce alternative assumptions that are not necessary for guaranteeing a solution to the optimal control problem. These assumptions include: (i) $\text{spec}(A) \in \mathbb{C}_{-}$ \cite{POPRL, KPRLNMR}; (ii) $(A,B)$ is stabilizable \cite{CLJKYPL, SKSPRCPLTI}; (iii) $H + H^{\star}$ is nonsingular \cite{SKSPRCPLTI}; (iv) $H(j\omega) + H(-j \omega)^{T} > 0$ for all $\omega \in \mathbb{R}$ \cite{CLJKYPL, KPRLNMR} (note that this implies (iii)); and (v) $(C,A)$ is observable \cite{CLJKYPL, SKSPRCPLTI}. A key objective of this paper is to avoid such assumptions entirely. 

We also note that several papers have sought to demonstrate the equivalence of the conditions (a) $H(j\omega) + H(-j \omega)^{T} \geq 0$ for all $\omega \in \mathbb{R}$; and (b) there exists $X \in \mathbb{R}^{d \times d}_{s}$ (not necessarily non-negative definite) such that $\Omega(X) \geq 0$ \cite{FPPRLE, FPRLVNA}. The papers \cite{camwb, PBDUS} consider a similar problem using the formalism of the behavioral approach. These papers again introduce additional assumptions. Specifically, \cite{FPPRLE} assumes that $A$ is \emph{unmixed}; and \cite{FPRLVNA} assumes \emph{sign controllability}. Both of these conditions imply the assumption (vi) $\begin{bmatrix}j \omega I - A& B\end{bmatrix}$ has full row rank for all $\omega \in \mathbb{R}$. Also, \cite{PBDUS} assumes conditions (iii), (v) and (vi); and \cite{camwb} considers only single-input single-output systems that satisfy conditions (v) and (vi).

\begin{figure}[t!]
\scriptsize
\begin{center}
\begin{psfrags}
\psfrag{nb}[t][b]{$\displaystyle \begin{array}{l} 
\frac{d\mathbf{x}}{dt} = A\mathbf{x} + Bi, v = C\mathbf{x} + Di, \\
\mathbf{x} {=} \text{col}\begin{pmatrix}i_{1}& i_{2}& v_{3}& v_{4}\end{pmatrix} \\
A {=} \begin{bmatrix}-1& -1& 1& 0\\
-1& -1& 0& 1\\
-1& 0& 0& 0\\
0& -1& 0& 0\end{bmatrix}, B {=} \begin{bmatrix}1\\ 1\\ 1\\ 1\end{bmatrix} \\
C {=} \begin{bmatrix}-1& -1& 1& 1\end{bmatrix}, D {=} 1 \\
\begin{bmatrix}j& -j& 1& -1\end{bmatrix}\!\begin{bmatrix}jI{-}A& B\end{bmatrix} {=} 0 \\
H(\xi) {\coloneqq} D+C(\xi I - A)^{-1}B {=} 1.
\end{array}$}
\psfrag{nc}[t][b]{$\displaystyle \begin{array}{l}
\frac{d\mathbf{x}}{dt} = \hat{A}\mathbf{x} + \hat{B}i, v = \hat{C}\mathbf{x} + \hat{D}i, \\
\mathbf{x} = \text{col}\begin{pmatrix}i_{1}{+}i_{2}& v_{3}{+}v_{4}& i_{2}& v_{4}\end{pmatrix} \\
\hat{A} {=} \begin{bmatrix}0& 1& 0& 0\\
-1& 0& 0& 0\\
0& 0& 0& 1\\
0& 0& -1& 0\end{bmatrix}, \hat{B}{=} \begin{bmatrix}0\\ 2\\ 0\\ 1\end{bmatrix} \\
\hat{C} {=} \begin{bmatrix}0& 1& 0& 0\end{bmatrix}, \hat{D}{=}0 \\
\begin{bmatrix}-1& j& 2& -2j\end{bmatrix}\!\begin{bmatrix}jI{-}\hat{A}& \hat{B}\end{bmatrix} {=} 0 \\
H(\xi) {\coloneqq} \hat{D}+\hat{C}(\xi I - \hat{A})^{-1}\hat{B} {=} \tfrac{2\xi}{\xi^{2}+1}.
\end{array}$}
\psfrag{nab}[c][c]{}
\psfrag{nbb}[c][c]{}
\psfrag{ncb}[c][c]{}
\psfrag{r5}[c][c]{$v_{5} {=} i_{5}$}
\psfrag{l3}[c][c]{$\displaystyle v_{2} {=} \frac{di_{2}}{dt}$}
\psfrag{l4}[c][c]{$\displaystyle v_{1} {=} \frac{di_{1}}{dt}$}
\psfrag{c1}[c][c]{$\displaystyle \frac{dv_{3}}{dt} {=} i_{3}$}
\psfrag{c2}[c][c]{$\displaystyle \frac{dv_{4}}{dt} {=} i_{4}$}
\psfrag{c5}[c][c]{$\displaystyle \frac{dv_{5}}{dt} {=} i_{5}$}
\includegraphics[page=1, width=0.95\hsize]{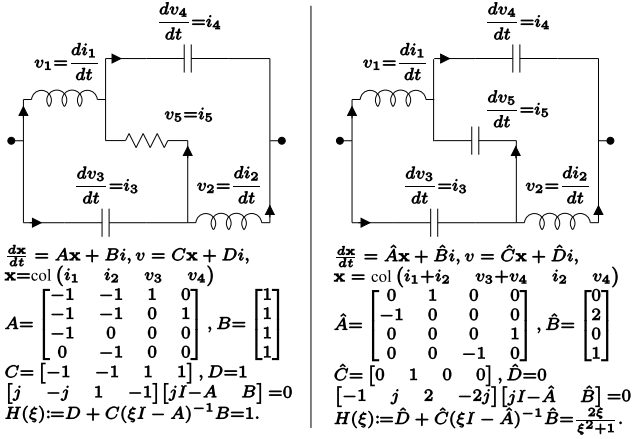}
\end{psfrags}
\end{center}
\caption{Two electric circuits with uncontrollable and unobservable state-space representations.}
\label{fig:nucuor}
\end{figure}

However, there are physical systems that do not satisfy any of the assumptions in these papers. For example, consider the two electric circuits in Fig.\ \ref{fig:nucuor}. Note that, for each of these circuits, the pair $(A,B)$ is not stabilizable. This implies that there is no state feedback law that transfers the internal currents and voltages to zero (however, there is a state feedback law that transfers the external currents and voltages to zero, and so the external behavior $\mathcal{B}_{s}^{(i,v)}$ \emph{is} stabilizable). Also, both circuits violate assumptions (i), (ii) and (vi) in the previous discussion, and the circuit on the right has $H + H^{\star} = 0$ (and so violates assumptions (iii) and (iv)). Now, consider the circuit on the left. Following note \ref{lem:osf}, we let
\begin{equation*}
T \coloneqq \begin{bmatrix}1& 1& -1& -1\\
0& 1& 0& 0\\
0& 0& 1& 0\\
0& 0& 0& 1\end{bmatrix}, \text{ so }T^{-1} = \begin{bmatrix}1& -1& 1& 1\\ 0& 1& 0& 0\\0& 0& 1& 0\\ 0& 0& 0& 1\end{bmatrix},
\end{equation*}
which transform the system into observer staircase form:
\begin{align*}
\hat{A} &\coloneqq TAT^{-1} = \begin{bmatrix}-1& 0& 0& 0\\
-1& 0& -1& 0\\
-1& 1& -1& -1\\
0& -1& 0& 0\end{bmatrix}, \hat{B} \coloneqq TB = \begin{bmatrix}0\\ 1\\ 1\\ 1\end{bmatrix}, \displaybreak[3] \\
\text{and } \hat{C}&\coloneqq CT^{-1} = \begin{bmatrix}-1& 0& 0& 0\end{bmatrix}.
\end{align*}
We note that the final three columns of $V_{o}T^{-1}$ are zero (so this circuit also violates assumption (v)), and it follows from Theorem \ref{thm:pbtsc} that $X_{-} = T^{T}\hat{X}_{-}T$ where $\hat{X}_{-} = \text{diag}\begin{pmatrix}\lambda& 0\end{pmatrix}$ and $\lambda$ is the least real positive number satisfying 
\begin{equation*}
\begin{bmatrix}-\hat{A}^{T}\hat{X}_{-} - \hat{X}_{-}\hat{A}& \hat{C}^{T}-\hat{X}_{-}\hat{B}\\ \hat{C}-\hat{B}^{T}\hat{X}_{-}& D + D^{T}\end{bmatrix} = \begin{bmatrix}2\lambda& 0& -1\\
0& 0_{3{\times}3}& 0\\
-1& 0& 2\end{bmatrix}\geq 0.
\end{equation*}
Thus, $\lambda = \tfrac{1}{4}$, and from Theorem \ref{thm:pbtsc} we conclude that, with $\mathbf{x}_{0} = \mathbf{x}(0)$, then $S_{a}^{\sigma_{p}}(\mathbf{x}_{0}) = \tfrac{1}{8}((i_{1}+i_{2}-v_{3}-v_{4})(0))^{2}$. Note that more energy can be extracted from this system than can be extracted from the system $\mathcal{B} = \lbrace (u, y) \in  \mathcal{L}_{2}^{\text{loc}}\left(\mathbb{R}, \mathbb{R}\right) \times  \mathcal{L}_{2}^{\text{loc}}\left(\mathbb{R}, \mathbb{R}\right) \mid y = u\rbrace$ (for which $\smallint_{t_{0}}^{t_{1}}-u(t)y(t)dt \leq 0$), despite the fact that both systems have the same transfer function.

In Remark \ref{rem:oc}, we will show how to extract the greatest possible amount of energy from this circuit. Following that remark, we let $i = {-}(D+D^{T})^{-1}(C-B^{T}X_{-})\mathbf{x} = \tfrac{1}{2}(i_{1}+i_{2}-v_{3}-v_{4})$. From the variation of the constants formula (\ref{eq:xvocf}),
\begin{equation*}
\left[\!\begin{smallmatrix}i_{1}\\ i_{2}\\ v_{3}\\ v_{4}\end{smallmatrix}\!\right]\!(t) = \tfrac{1}{2}\!\left[\!\begin{smallmatrix}(\cos(t){+}e^{-t})(i_{1}{-}i_{2})(0) {+}\sin(t)(v_{3}{-}v_{4})(0)\\
({-}\cos(t){+}e^{-t})(i_{1}{-}i_{2})(0) {-}\sin(t)(v_{3}{-}v_{4})(0)\\
{-}\sin(t)(i_{1}{-}i_{2})(0) {+} (\cos(t){+}e^{-t})(v_{3}{-}v_{4})(0)\\
\sin(t)(i_{1}{-}i_{2})(0){+}({-}\cos(t){+}e^{-t})(v_{3}{-}v_{4})(0)\end{smallmatrix}\!\right],
\end{equation*}
whereupon $v(t) = -\tfrac{1}{2}e^{-t}((i_{1}+i_{2}-v_{3}-v_{4})(0)) = -i(t)$. It can then be verified that $\smallint_{0}^{\infty}{-i(t)v(t)dt} = \tfrac{1}{8}((i_{1}+i_{2}-v_{3}-v_{4})(0))^{2} = S_{a}^{\sigma_{p}}(\mathbf{x}_{0})$.

Now, consider the circuit on the right of Fig.\ \ref{fig:nucuor}. We let 
\begin{equation*}
A = \begin{bmatrix}0& 1\\ -1& 0\end{bmatrix}, B = \begin{bmatrix}0\\2\end{bmatrix}, C = \begin{bmatrix}0& 1\end{bmatrix}, \text{ and } D = 0.
\end{equation*}
From Theorem \ref{thm:pbtsc}, we find that $S_{a}^{\sigma_{p}}(\mathbf{x}(0)) = \tfrac{1}{4}((i_{1}{+}i_{2})(0)^{2} + (v_{3}{+}v_{4})(0)^{2})$. Again,  Remark \ref{rem:oc} explains how to extract the greatest amount of energy from this circuit. In that remark,
\begin{align*}
&A_{\epsilon} {=} \begin{bmatrix}0& 1\\-1& -2\epsilon\end{bmatrix}\!, B_{\epsilon} {=} \begin{bmatrix}0\\ 2\sqrt{1+\epsilon^{2}}\end{bmatrix}\!, C_{\epsilon} {=} \begin{bmatrix}0& \tfrac{1-\epsilon^{2}}{\sqrt{1+\epsilon^{2}}}\end{bmatrix}\!, D_{\epsilon} {=} \epsilon, \\
&X_{-}^{\epsilon} {=} \tfrac{(1-\epsilon)^{2}}{2(1+\epsilon^{2})}I, u_{\epsilon} {=} -\tfrac{1-\epsilon}{\sqrt{1+\epsilon^{2}}}(v_{3}{+}v_{4}), \text{ and } u {=} {-}(v_{3}{+}v_{4}) {=} {-}y.
\end{align*}
We then let $i = u$ and $v = y$. In this case, $i$ and $v$ are independent of $\epsilon$, and it can be verified that $i(t) = te^{-t}(i_{1}{+}i_{2})(0) + (te^{-t}-e^{-t})(v_{3}+v_{4})(0) = -v(t)$ and $\smallint_{0}^{\infty}-i(t)v(t)dt = \tfrac{1}{4}((i_{1}{+}i_{2})(0)^{2} + (v_{3}{+}v_{4})(0)^{2}) = S_{a}^{\sigma_{p}}(\mathbf{x}_{0})$.

We end this section with a remark about the optimal control problem in Definition \ref{def:srprl} concerning the required energy.
\begin{remark}
The required energy $S_{r}^{\sigma_{p}}(\mathbf{x}_{0})$ is not considered in Theorems \ref{thm:pbtsc}, \ref{thm:pbtsced} and \ref{thm:pbtnsc}. If $\mathcal{B}_{s}$ is as in (\ref{eq:sssg}) and $(A,B)$ is controllable, then $S_{r}^{\sigma_{p}}(\mathbf{x}_{0})$ corresponds to the energy required to transfer the state to $\mathbf{x}_{0}$ from the origin. However, if $\mathcal{B}_{s}$ is not controllable, then there exist $\mathbf{x}_{0} \in \mathbb{R}^{d}$ which cannot be reached from the origin, so the required energy for such states is undefined. Indeed, the controllability of $(A,B)$ is related to the existence of an upper bound to the set of $X \in \mathbb{R}_{s}^{d \times d}$ which satisfy condition \ref{nl:ptc4} of Theorem \ref{thm:pbtsc}. Specifically, if there exist $\mathbf{z} \in \mathbb{C}^{d}$ and $\lambda \in \mathbb{C}$ such that $\mathbf{z}^{T}\begin{bmatrix}\lambda I{-}A& B\end{bmatrix} = 0$, then $\mathbf{z}^{T}\tfrac{d\mathbf{x}}{dt} = \lambda \mathbf{z}^{T}\mathbf{x}$, and so $\mathbf{z}^{T}\mathbf{x}(t) = e^{\lambda(t-t_{0})}\mathbf{z}^{T}\mathbf{x}(t_{0})$ for all $t \in \mathbb{R}$, whence $\mathbf{z}^{T}\mathbf{x}(t_{0}) \neq 0$ implies that $\mathbf{x}(t) \neq 0$ for all $t \in \mathbb{R}$. If, in addition, $\lambda \in \overbar{\mathbb{C}}_{-}$, then there are no trajectories satisfying $\mathbf{z}^{T}\mathbf{x}(t_{0}) \neq 0$ and $\lim_{t \rightarrow -\infty}(\mathbf{x}(t)) = 0$. In fact, for a passive system, the following two conditions are equivalent: (i) the LMI in condition \ref{nl:prlc4} of Lemma \ref{thm:prlsc} has no upper bound; and (ii) there exists $0 \neq \mathbf{z} \in \mathbb{C}^{d}$ and $\lambda \in \overbar{\mathbb{C}}_{-}$ such that $\mathbf{z}^{T}\begin{bmatrix}\lambda I{-}A& B\end{bmatrix} = 0$. To see that (ii) $\Rightarrow$ (i), let $X \in \mathbb{R}_{s}^{d \times d}$ satisfy $\Omega(X) \geq 0$ and $\hat{X} \coloneqq \mathbf{z}\bar{\mathbf{z}}^{T} + \bar{\mathbf{z}}\mathbf{z}^{T}$. Then, for any given $\alpha > 0$, $0 \leq X + \alpha \hat{X} \in \mathbb{R}_{s}^{d \times d}$ satisfies $\Omega(X + \alpha \hat{X}) = \Omega(X) - \alpha(\lambda + \bar{\lambda})\text{diag}\begin{pmatrix}\hat{X}& 0\end{pmatrix} \geq 0$. Conversely, if (ii) does not hold, then there exists $K \in \mathbb{R}^{d \times n}$ such that $\text{spec}(A+BK) \in \mathbb{C}_{+}$. Then, for any given $\mathbf{x}_{0} \in \mathbb{R}^{d}$, there exists a trajectory $(\mathbf{u}, \mathbf{y}, \mathbf{x}) \in \mathcal{B}_{s}$ with $\mathbf{u} = K\mathbf{x}$, $\mathbf{x}(t_{0}) = \mathbf{x}_{0}$ and $\lim_{t \rightarrow -\infty}(\mathbf{x}(t)) = 0$. Finally, for this trajectory, it can be shown that (a) there exists $\hat{X} \in \mathbb{R}_{s}^{d \times d}$ such that $ \smallint_{-\infty}^{t_{0}}(\mathbf{u}^{T}\mathbf{y})(t)dt \leq \tfrac{1}{2}\mathbf{x}_{0}^{T}\hat{X}\mathbf{x}_{0}$; and (b) if $X$ satisfies  condition \ref{nl:prlc4} of Lemma \ref{thm:prlsc}, then $\tfrac{1}{2}\mathbf{x}_{0}^{T}X\mathbf{x}_{0} \leq \smallint_{-\infty}^{t_{0}}(\mathbf{u}^{T}\mathbf{y})(t)dt$. It follows that (i) does not hold.\hfill$\triangle$
\end{remark}

The following two sections provide the proofs of Theorems \ref{thm:pbtsc}, \ref{thm:pbtsced} and \ref{thm:pbtnsc}. Then, in Sections \ref{sec:brl}--\ref{sec:compas}, we state and prove three analogous theorems relevant to \emph{non-expansive systems}.

\section{Passive systems and the available energy}
\label{sec:pbaae}
In this section, we prove Theorem \ref{thm:pbtsc}. The proof uses the concept of \emph{storage functions} and the results in Appendix \ref{sec:sf}.
\begin{IEEEproof}[Proof of Theorem \ref{thm:pbtsc}]
That \ref{nl:ptc3} $\iff$ \ref{nl:ptc4} is shown in \cite{THTPLSNA}. Here, we prove \ref{nl:ptc1} $\Rightarrow$ \ref{nl:ptc5} $\Rightarrow$ \ref{nl:ptc4} $\Rightarrow$ \ref{nl:ptc1}.

{\bf \ref{nl:ptc1} $\boldsymbol{\Rightarrow}$ \ref{nl:ptc5}.} \hspace{0.3cm} First, consider a $k>0$ such that $I+kD$ is nonsingular, and let $\mathbf{z} \in \mathbb{R}^{d}$ be fixed but arbitrary. Then, let
\begin{equation}
\tilde{C}\coloneqq(I+kD)^{-1}C \text{ and } \tilde{A} \coloneqq A - kB(I+kD)^{-1}C;\label{eq:tctad}
\end{equation}
let $\tilde{\mathbf{x}}(t) = e^{\tilde{A}(t-t_{0})}\mathbf{z}$ for all $t \in \mathbb{R}$; and let $\tilde{\mathbf{u}} = -k\tilde{C}\tilde{\mathbf{x}}$ and $\tilde{\mathbf{y}} = \tilde{C}\tilde{\mathbf{x}}$. It can be verified that $(\tilde{\mathbf{u}}, \tilde{\mathbf{y}}, \tilde{\mathbf{x}}) \in \mathcal{B}_{s}$, $\tilde{\mathbf{x}}(t_{0}) = \mathbf{z}$, and $\smallint_{t_{0}}^{t_{1}}{-(\tilde{\mathbf{u}}^{T}\tilde{\mathbf{y}})(t)dt} = k\smallint_{t_{0}}^{t_{1}}{(\tilde{\mathbf{y}}^{T}\tilde{\mathbf{y}})(t)dt} \geq 0$. Thus, $S_{a}^{\sigma_{p}}(\mathbf{x}_{0}) = \tfrac{1}{2}\mathbf{x}_{0}^{T}X_{-}\mathbf{x}_{0}$ for some $X_{-} \in \mathbb{R}^{d \times d}_{s}$ with $X_{-} \geq 0$ by Lemma \ref{lem:sfqao} (it is conventional to include the $\tfrac{1}{2}$). It remains to show that $X_{-}$ satisfies conditions \ref{nl:ptc5}(ii)--(iv).

The proof of  \ref{nl:ptc5}(iii) is inspired by \cite[Proof of Lemma 1]{JWDSP2}. Note initially that, since $X_{-} \geq 0$, then any given $\mathbf{z} \in \mathbb{R}^{d}$ satisfies $X_{-}\mathbf{z} = 0 \iff \mathbf{z}^{T}X_{-}\mathbf{z} = 0 \iff S_{a}^{\sigma_{p}}(\mathbf{z}) = 0$, whence $V_{o}\mathbf{z} = 0 \Rightarrow X_{-}\mathbf{z} = 0$ by Lemma \ref{lem:saop}. Now, let $\tilde{\mathbf{u}}$, $\tilde{\mathbf{y}}$, and $\tilde{\mathbf{x}}$ be as in the previous paragraph, so $\smallint_{t_{0}}^{t_{1}}{-(\tilde{\mathbf{u}}^{T}\tilde{\mathbf{y}})(t)dt} = k \smallint_{t_{0}}^{t_{1}}{(\tilde{\mathbf{y}}^{T}\tilde{\mathbf{y}})(t)dt} \leq S_{a}^{\sigma_{p}}(\mathbf{z})$. If $X_{-}\mathbf{z} = 0$, then $0 = S_{a}^{\sigma_{p}}(\mathbf{z}) \geq  k \smallint_{t_{0}}^{t_{1}}{(\tilde{\mathbf{y}}^{T}\tilde{\mathbf{y}})(t)dt}$, so $\tilde{\mathbf{y}}(t) = \tilde{C}e^{\tilde{A}(t-t_{0})}\mathbf{z} = 0$ for all $t \geq t_{0}$. This implies $\tilde{C}\tilde{A}^{k}\mathbf{z} = 0$ ($k = 0, 1, 2, \ldots$), which implies $CA^{k}\mathbf{z} = 0$ ($k = 0, 1, 2, \ldots$), hence $V_{o}\mathbf{z} = 0$.

To prove \ref{nl:ptc5}(ii), note from Lemma \ref{lem:avstor} that $S_{a}^{\sigma_{p}}$ is a storage function (with respect to $\mathbf{u}^{T}\mathbf{y}$). Thus, $S_{a}^{\sigma_{p}}(\mathbf{x}(t_{1})) \leq \smallint_{t_{0}}^{t_{1}}{(\mathbf{u}^{T}\mathbf{y})(t)dt} + S_{a}^{\sigma_{p}}(\mathbf{x}(t_{0}))$ for all $(\mathbf{u}, \mathbf{y}, \mathbf{x}) \in \mathcal{B}_{s}$ and $t_{1} \geq t_{0}$. From the variation of the constants formula (\ref{eq:xvocf})--(\ref{eq:yvocf}), for any given $\mathbf{x}_{0} \in \mathbb{R}^{d}$ and $\mathbf{u}_{0} \in \mathbb{R}^{n}$, there exists a $(\mathbf{u}, \mathbf{y}, \mathbf{x}) \in \mathcal{B}_{s}$ with $\mathbf{x}$ differentiable, $\mathbf{x}(t_{0}) = \mathbf{x}_{0}$, and $\mathbf{u}(t_{0}) = \mathbf{u}_{0}$. Thus, $(\mathbf{u}^{T}\mathbf{y})(t_{0}) - \tfrac{d}{dt}(S_{a}^{\sigma_{p}}(\mathbf{x}))(t_{0}) = \tfrac{1}{2}\begin{bmatrix}\mathbf{x}_{0}^{T}& \mathbf{u}_{0}^{T}\end{bmatrix}\Omega(X_{-})\text{col}\begin{pmatrix}\mathbf{x}_{0}& \mathbf{u}_{0}\end{pmatrix} \geq 0$, and so $\Omega(X_{-}) \geq 0$.

To prove \ref{nl:ptc5}(iv), note that if $X \geq 0$ and $\Omega(X) \geq 0$, then 
\begin{multline}
2\smallint_{t_{0}}^{t_{1}}{(\mathbf{u}^{T}\mathbf{y})(t)dt} - \left[(\mathbf{x}^{T}X\mathbf{x})(t)\right]_{t_{0}}^{t_{1}} \\
= \smallint_{t_{0}}^{t_{1}}{(\begin{bmatrix}\mathbf{x}^{T}& \mathbf{u}^{T}\end{bmatrix}\Omega(X)\text{col}\begin{pmatrix}\mathbf{x}& \mathbf{u}\end{pmatrix})(t)dt}, \label{eq:ibpsff}
\end{multline}
so $\tfrac{1}{2}\mathbf{x}(t_{1})^{T}X\mathbf{x}(t_{1}) \leq \smallint_{t_{0}}^{t_{1}}{(\mathbf{u}^{T}\mathbf{y})(t)dt} + \tfrac{1}{2}\mathbf{x}(t_{0})^{T}X\mathbf{x}(t_{0})$. With $S(\mathbf{x}_{0}) \coloneqq \tfrac{1}{2}\mathbf{x}_{0}^{T}X\mathbf{x}_{0}$ for all $\mathbf{x}_{0} \in \mathbb{R}^{d}$, it follows that $S$ is a storage function. Thus, $\mathbf{x}_{0}^{T}X\mathbf{x}_{0} \geq \mathbf{x}_{0}^{T}X_{-}\mathbf{x}_{0}$ for all $\mathbf{x}_{0} \in \mathbb{R}^{d}$ by Lemma \ref{lem:avstor}, which implies $X \geq X_{-}$.

{\bf \ref{nl:ptc5} $\boldsymbol{\Rightarrow}$ \ref{nl:ptc4}.} \hspace{0.3cm} Immediate.

{\bf \ref{nl:ptc4} $\boldsymbol{\Rightarrow}$ \ref{nl:ptc1}.} \hspace{0.3cm} Recall from the proof of \ref{nl:ptc1} $\Rightarrow$ \ref{nl:ptc5}(iv) that $S(\mathbf{x}_{0}) = \tfrac{1}{2}\mathbf{x}_{0}^{T}X\mathbf{x}_{0}$ is a storage function (with respect to $\mathbf{u}^{T}\mathbf{y}$). Condition \ref{nl:ptc1} then follows from Lemma \ref{lem:avstor}.

It remains to show that, if $(C,A)$ is observable and conditions \ref{nl:ptc1}--\ref{nl:ptc5} hold, then (i) $\text{spec}(A) \in \overbar{\mathbb{C}}_{-}$; and (ii) if $X \in \mathbb{R}^{d \times d}_{s}$ satisfies $\Omega(X) \geq 0$, then $X_{-} \leq X$. 

Condition (i) follows from \cite[Theorem 3.7.5]{AndVong}, as condition \ref{nl:ptc5} implies that $X_{-} > 0$ and $-A^{T}X_{-} - X_{-}A \geq 0$. 

To see (ii), let $X \in \mathbb{R}^{d \times d}_{s}$ satisfy $\Omega(X) \geq 0$, and note that (\ref{eq:ibpsff}) holds. Then, for any given $\mathbf{x}_{0} \in \mathbb{R}^{d}$ and $\epsilon > 0$, there exists $(\mathbf{u}, \mathbf{y}, \mathbf{x}) \in \mathcal{B}_{s}$ with $\mathbf{x}(t_{0}) = \mathbf{x}_{0}$ and $t_{1} \geq t_{0}$ such that 
\begin{equation*}\smallint_{t_{0}}^{t_{1}}{{-}(\mathbf{u}^{T}\mathbf{y})(t)dt} {=} S_{a}^{\sigma_{p}}(\mathbf{x}_{0}) {-} \epsilon {\leq} \tfrac{1}{2} (\mathbf{x}_{0}^{T}X\mathbf{x}_{0} {-} \mathbf{x}(t_{1})^{T}X\mathbf{x}(t_{1})).
\end{equation*}
We will show that there exists $M \in \mathbb{R}$ (independent of $\epsilon$) such that $\mathbf{x}(t_{1})^{T}X\mathbf{x}(t_{1}) \geq M\epsilon$. This implies that $\mathbf{x}_{0}^{T}X_{-}\mathbf{x}_{0} \leq \mathbf{x}_{0}^{T}X\mathbf{x}_{0} + \epsilon(1-M)$. Since $\epsilon$ can be chosen to be arbitrarily small, we conclude that $X_{-} \leq X$.

To obtain the bound $M$, let $k > 0$ be such that $I+kD$ is nonsingular; let $\tilde{C}$ and $\tilde{A}$ be as in (\ref{eq:tctad}); let $\mathbf{x}(t_{1}) = \mathbf{x}_{1}$; and let $(\mathbf{\tilde{u}}(t), \mathbf{\tilde{y}}(t),\mathbf{\tilde{x}}(t)) = (\mathbf{u}(t), \mathbf{y}(t), \mathbf{x}(t))$ for all $t_{0} \leq t < t_{1}$, and $\mathbf{\tilde{x}}(t) = e^{\tilde{A}(t-t_{1})}\mathbf{x}_{1}$, $\mathbf{\tilde{u}}(t) = -k\tilde{C}\mathbf{x}(t)$, and $\mathbf{\tilde{y}}(t) = \tilde{C}\mathbf{x}(t)$ for all $t \geq t_{1}$. Then $(\mathbf{\tilde{u}}, \mathbf{\tilde{y}}, \mathbf{\tilde{x}}) \in \mathcal{B}_{s}$ with $\mathbf{\tilde{x}}(t_{0}) = \mathbf{x}_{0}$. Next, consider a fixed $T > 0$, and let $\tilde{\mathcal{O}} \coloneqq \smallint_{0}^{T}{e^{\tilde{A}^{T}\tau}\tilde{C}^{T}\tilde{C}e^{\tilde{A}\tau}d\tau}$. From earlier in the proof, $(\tilde{C}, \tilde{A})$ is observable since $(C,A)$ is, and so $\tilde{\mathcal{O}} > 0$. Moreover, 
\begin{equation*}\smallint_{t_{0}}^{t_{1}+T}{{-}(\mathbf{\tilde{u}}^{T}\mathbf{\tilde{y}})(t)dt} = S_{a}^{\sigma_{p}}(\mathbf{x}_{0}) - \epsilon + k \mathbf{x}_{1}^{T}\tilde{\mathcal{O}}\mathbf{x}_{1} \leq S_{a}^{\sigma_{p}}(\mathbf{x}_{0}),
\end{equation*} 
so $\mathbf{x}_{1}^{T}\tilde{\mathcal{O}}\mathbf{x}_{1} \leq \epsilon/k$. Now, let $\lambda > 0$ denote the least eigenvalue of $\tilde{\mathcal{O}}$. Also, if $X \geq 0$ we let $\mu \coloneqq 0$, and otherwise we let $\mu < 0$ be the most negative eigenvalue of $X$. By Rayleigh's quotient, $\mathbf{x}_{1}^{T}X\mathbf{x}_{1} \geq \mu \mathbf{x}_{1}^{T}\mathbf{x}_{1} \geq  \tfrac{\mu}{\lambda}\mathbf{x}_{1}^{T}\tilde{\mathcal{O}}\mathbf{x}_{1} \geq \tfrac{\mu \epsilon}{k \lambda}$, which gives the bound $M \coloneqq \mu /(k \lambda)$.
\end{IEEEproof}

\section{Explicit characterisation of the available energy}
\label{sec:cae}
In this section, we prove Theorems \ref{thm:pbtsced} and \ref{thm:pbtnsc}. We also show how to compute the available energy of a passive system.
\begin{IEEEproof}[Proof of Theorem \ref{thm:pbtsced}]
With the notation
\begin{equation}
\label{eq:Sxd}
S(X) \coloneqq \begin{bmatrix}I & 0\\ (D+D^{T})^{-1}(B^{T}X-C)& I\end{bmatrix},
\end{equation}
then $S(X)$ is nonsingular and
\begin{equation}
S(X)^{T}\Omega(X)S(X) = \begin{bmatrix}\Gamma(X)& 0\\0& D+D^{T}\end{bmatrix}.\label{eq:Stog}
\end{equation}
Thus, $\Omega(X) \geq 0 \iff \Gamma(X) \geq 0$.

{\bf \ref{nl:ptnsc2} $\boldsymbol{\Rightarrow}$ \ref{nl:ptnsc1}.} \hspace{0.3cm} From (\ref{eq:Stog}), $X_{-} \geq 0$ and $\Omega(X_{-}) \geq 0$, so $S_{a}^{\sigma_{p}}(\mathbf{x}_{0}) < \infty$ for all $\mathbf{x}_{0} \in \mathbb{R}^{d}$ by Theorem \ref{thm:pbtsc}.

{\bf \ref{nl:ptnsc1} $\boldsymbol{\Rightarrow}$ \ref{nl:ptnsc2}.} \hspace{0.3cm} Since $S_{a}^{\sigma_{p}}(\mathbf{x}_{0}) < \infty$ for all $\mathbf{x}_{0} \in \mathbb{R}^{d}$, then  $S_{a}^{\sigma_{p}}(\mathbf{x}_{0}) = \tfrac{1}{2}\mathbf{x}_{0}^{T}X_{-}\mathbf{x}_{0}$ for some $X_{-} \in \mathbb{R}^{d \times d}_{s}$ satisfying (i) $X_{-} \geq 0$, (ii) $\Omega(X_{-}) \geq 0$, and (iii) if $\mathbf{z} \in \mathbb{R}^{d}$, then $X_{-}\mathbf{z} = 0 \iff V_{o}\mathbf{z} = 0$, by Theorem \ref{thm:pbtsc}. It remains to show that conditions \ref{nl:ptnsc2}(ii) and \ref{nl:ptnsc2}(iv) are also satisfied.

To show condition \ref{nl:ptnsc2}(ii), we let $\sigma(\mathbf{u}, \mathbf{y}) = \mathbf{u}^{T}\mathbf{y}$. From the proof of Theorem \ref{thm:pbtsc}, $\sigma$ satisfies the conditions of Lemma \ref{lem:sfqao}, so (\ref{eq:saad1}) holds from the proof of that lemma. Also, for any given $t_{2} \geq t_{1} \geq t_{0}$ and $(\mathbf{u}, \mathbf{y}, \mathbf{x}) \in \mathcal{B}_{s}$ with $\mathbf{x}(t_{0}) = \mathbf{x}_{0}$, 
\begin{align*}
&\hspace*{-0.3cm}\smallint_{t_{0}}^{t_{2}}{-\sigma(\mathbf{u},\mathbf{y})(t) dt} \leq \smallint_{t_{0}}^{t_{1}}{-\sigma(\mathbf{u},\mathbf{y})(t) dt}+S_{a}^{\sigma_{p}}(\mathbf{x}(t_{1})) \\
&\hspace*{0.3cm}= S_{a}^{\sigma_{p}}(\mathbf{x}_{0}) - \tfrac{1}{2}\smallint_{t_{0}}^{t_{1}}{(\begin{bmatrix}\mathbf{x}^{T}& \mathbf{u}^{T}\end{bmatrix}\Omega(X_{-})\text{col}\begin{pmatrix}\mathbf{x}& \mathbf{u}\end{pmatrix})(t)dt}.
\end{align*}
By taking the supremum over all $t_{2} \geq t_{1}$ and $\mathbf{u} \in \mathcal{L}_{2}^{\text{loc}}\left(\mathbb{R}, \mathbb{R}^{n}\right)$, and using (\ref{eq:saad1}) from the proof of Lemma \ref{lem:sfqao}, we find that 
\begin{align}
&0 \leq \sup_{\mathbf{u} \in \mathcal{L}_{2}^{\text{loc}}\left(\mathbb{R}, \mathbb{R}^{n}\right)}\smallint_{t_{0}}^{t_{1}}{-(\begin{bmatrix}\mathbf{x}^{T}& \mathbf{u}^{T}\end{bmatrix}\Omega(X_{-})\text{col}\begin{pmatrix}\mathbf{x}& \mathbf{u}\end{pmatrix})(t)dt} \nonumber \\
&\hspace{2.5cm} \text{such that } (\mathbf{u}, \mathbf{y}, \mathbf{x}) \in \mathcal{B}_{s}, \mathbf{x}(t_{0}) = \mathbf{x}_{0}, \label{eq:supc}
\end{align}
for any given $\mathbf{x}_{0} \in \mathbb{R}^{d}$ and $t_{1} \geq t_{0} \in \mathbb{R}$. Since $\Omega(X_{-}) \geq 0$, then the above inequality must be satisfied with equality. We let $\mathbf{v} \coloneqq \mathbf{u} + (D{+}D^{T})^{-1}(C-B^{T}X_{-})\mathbf{x}$, so (\ref{eq:Sxd})--(\ref{eq:supc}) imply
\begin{align*}
0 &= \inf_{\mathbf{v} \in \mathcal{L}_{2}^{\text{loc}}\left(\mathbb{R}, \mathbb{R}^{n}\right)}\smallint_{t_{0}}^{t_{1}}{(\mathbf{x}^{T}\Gamma(X_{-})\mathbf{x} + \mathbf{v}^{T}(D{+}D^{T})\mathbf{v})(t)dt} \\
& \text{such that } \mathbf{x} \in \mathcal{L}_{2}^{\text{loc}}(\mathbb{R}, \mathbb{R}^{d}), \tfrac{d\mathbf{x}}{dt} {=} A_{\Gamma}(X_{-})\mathbf{x} {+} B\mathbf{v}, \mathbf{x}(t_{0}) {=} \mathbf{x}_{0}.
\end{align*} 
From \cite[Section 2.3]{AndMoore}, for any given $t_{1} \geq t_{0}$, the above infimum is equal to $\mathbf{x}_{0}^{T}P(t_{0}{-}t_{1})\mathbf{x}$, where $P$ is an absolutely continuous matrix function that satisfies $P(0) = 0$ and
\begin{multline}
-\tfrac{dP}{dt} = PA_{\Gamma}(X_{-}) + A_{\Gamma}(X_{-})^{T}P \\ - PB(D+D^{T})^{-1}B^{T}P + \Gamma(X_{-}). \label{eq:dre}
\end{multline}
Since $\mathbf{x}_{0} \in \mathbb{R}^{d}$ can be chosen arbitrarily, then $P(t) = \tfrac{dP}{dt}(t) = 0$ for all $t < 0$, and so $\Gamma(X_{-}) = 0$ by (\ref{eq:dre}).

To show condition \ref{nl:ptnsc2}(iv), we consider the cases: (i) $(C,A)$ observable; and (ii) $(C,A)$ not observable.

{\bf Case (i): $(C,A)$ observable.} \hspace{0.3cm} We note that $\begin{bmatrix}\lambda I -A & B\end{bmatrix}$ has full row rank for all $\lambda \in \overbar{\mathbb{C}}_{+}$ (see Remarks \ref{rem:bsieco} and \ref{rem:pmbs}). This implies that $\begin{bmatrix}\lambda I -A_{\Gamma}(X_{-}) & B\end{bmatrix}$ has full row rank for all $\lambda \in \overbar{\mathbb{C}}_{+}$, so $(A_{\Gamma}(X_{-}), B)$ is stabilizable. The proof of this condition is then identical to \cite[Lemma 7]{MOLLQOC}.

{\bf Case (ii): $(C,A)$ not observable.} \hspace{0.3cm} Consider the observer staircase form (see note \ref{lem:osf}), and let $\mathcal{B}_{s}$ and $\hat{S}_{a}^{\sigma}$ be as in Lemma \ref{lem:saop} (with $\sigma(\mathbf{u}, \mathbf{y}) = \mathbf{u}^{T}\mathbf{y}$). It follows from Lemma \ref{lem:saop} that $X_{-} = T^{T}\text{diag}\begin{pmatrix}\hat{X}_{-}& 0\end{pmatrix}T$ where $\hat{X}_{-} \in \mathbb{R}^{\hat{d} \times \hat{d}}_{s}$ with $\tfrac{1}{2}\hat{\mathbf{x}}_{0}^{T}\hat{X}_{-}\hat{\mathbf{x}}_{0} = \hat{S}_{a}^{\sigma_{p}}(\hat{\mathbf{x}}_{0})$ for all $\hat{\mathbf{x}}_{0} \in \mathbb{R}^{\hat{d}}$. With 
\begin{align}
&\hat{\Gamma}(\hat{X}) {\coloneqq} {-}A_{11}^{T}\hat{X}{-}\hat{X}A_{11} {-} (C_{1}^{T}{-}\hat{X}B_{1})(D{+}D^{T})^{{-}1}(C_{1}{-}B_{1}^{T}\hat{X}), \nonumber \\
&\text{and } A_{\hat{\Gamma}}(\hat{X}) {\coloneqq} A_{11}{-}B_{1}(D{+}D^{T})^{-1}(C_{1}{-}B_{1}^{T}\hat{X}),\label{eq:ahgd}
\end{align}
it follows from case (i) that $\hat{X}_{-} \geq 0$, $\hat{\Gamma}(\hat{X}_{-}) = 0$, and $\text{spec}(A_{\hat{\Gamma}}(\hat{X}_{-})) \in \overbar{\mathbb{C}}_{-}$. Also, it can be verified that $\Gamma(X_{-}) = T^{T}\text{diag}\begin{pmatrix}\hat{\Gamma}(\hat{X}_{-})& 0\end{pmatrix}T$; and
\begin{equation}
TA_{\Gamma}(X_{-})T^{-1} = \begin{bmatrix}A_{\hat{\Gamma}}(\hat{X}_{-}) & 0\\ \hat{A}_{21} & A_{22}\end{bmatrix},\label{eq:agosf}
\end{equation}
where $\hat{A}_{21} \coloneqq A_{21} {-} B_{2}(D{+}D^{T})^{-1}(C_{1}^{T}{-}B_{1}^{T}\hat{X}_{-})$. Now, suppose $\lambda \in \mathbb{C}_{+}$ and $\mathbf{z} \in \mathbb{C}^{d}$ satisfy $A_{\Gamma}(X_{-})\mathbf{z} = \lambda \mathbf{z}$, and let $T_{1}$ be as in note \ref{lem:osf}. Since $\lambda I - A_{\hat{\Gamma}}(\hat{X}_{-})$ is nonsingular for all $\lambda \in \mathbb{C}_{+}$, then (\ref{eq:agosf}) implies that $T_{1}\mathbf{z} = 0$, and it is then easily shown that $V_{o}\mathbf{z} = 0$. 

It remains to show that if $X_{-}$ satisfies condition \ref{nl:ptnsc2}, then $S_{a}^{\sigma_{p}}(\mathbf{x}_{0}) = \tfrac{1}{2}\mathbf{x}_{0}^{T}X_{-}\mathbf{x}_{0}$. To prove this, we assume that $(C,A)$ is observable, and we show that if $X \in \mathbb{R}^{d \times d}_{s}$ satisfies $X \geq 0$, $X \neq X_{-}$, and $\Gamma(X) = 0$, then $\text{spec}(A_{\Gamma}(X)) \not\in \overbar{\mathbb{C}}_{-}$. The case with $(C,A)$ not observable can then be shown by considering the observer staircase form as in the proof of case (ii) above. 

If $\Gamma(X) = 0$, then $\Omega(X) \geq 0$, so $Y \coloneqq X - X_{-} \geq 0$ by Theorem \ref{thm:pbtsc}. Also, by direct calculation, $A_{\Gamma}(X_{-})^{T}Y {+} YA_{\Gamma}(X_{-}) {+} YB(D{+}D^{T})^{-1}B^{T}Y = 0$. From before, $(A_{\Gamma}(X_{-}),B)$ is stabilizable, so from \cite[Proof of Lemma A.1]{MOLLQOC} we find that $A_{\Gamma}(X_{-}) + B(D{+}D^{T})^{-1}B^{T}Y = A_{\Gamma}(X)$ satisfies $\text{spec}(A_{\Gamma}(X)) \not\in \overbar{\mathbb{C}}_{-}$.
\end{IEEEproof}

\begin{remark}
\label{rem:prptlmirc}
From the proof of Theorems \ref{thm:pbtsc} and \ref{thm:pbtsced}, in order to find the matrix $X_{-}$ in those theorems, it suffices to find an $X_{-} \in \mathbb{R}^{d \times d}_{s}$ satisfying $\Gamma(X_{-}) = 0$ and $\text{spec}(A_{\Gamma}(X_{-})) \in \overbar{\mathbb{C}}_{-}$ for the case with $(C,A)$ observable. This can be obtained from the controller staircase form
\cite[Theorem 3.3.4]{AndVong}:
\begin{equation}
\hspace*{-0.4cm}TAT^{-1} {=} \begin{bmatrix}A_{11}& A_{12}\\ 0& A_{22}\end{bmatrix}\!, \hspace{0.05cm} TB {=} \begin{bmatrix}B_{1}\\ 0\end{bmatrix}\!, \hspace{0.05cm} CT^{-1} {=} \begin{bmatrix}C_{1}& C_{2}\end{bmatrix}\!,\label{eq:css}
\end{equation}
with $(A_{11}, B_{1})$ controllable. Since $(C,A)$ is observable, then so too is $(C_{1}, A_{11})$. Furthermore, $D + C_{1}(\xi I {-} A_{11})^{-1}B_{1} = D + C(\xi I {-} A)^{-1}B$, which is positive-real (see Remark \ref{rem:pmbs}). Thus, with $\hat{\Gamma}$ and $A_{\hat{\Gamma}}$ as in (\ref{eq:ahgd}), there exists a unique $X_{11} > 0$ satisfying $\hat{\Gamma}(X_{11}) = 0$ and $\text{spec}(A_{\hat{\Gamma}}(X_{11})) \in \overbar{\mathbb{C}}_{-}$ \cite[Lemma 2]{JWDSP2}. This can be efficiently computed using the methods in \cite[Chapter 6]{AndVong}. Next, note that $(A,B)$ is stabilizable since $(C,A)$ is observable (see Remarks \ref{rem:bsieco} and \ref{rem:pmbs}), so $\text{spec}(A_{22}) \in \mathbb{C}_{-}$ \cite[Corollary 5.2.31]{JWIMTSC}. Thus, from \cite[Theorem 3.7.4]{AndVong}, there exists a unique real $X_{12}^{T}$ satisfying the Sylvester equation:
\begin{multline*}
A_{22}^{T}X_{12}^{T} {+}X_{12}^{T}A_{\hat{\Gamma}}(X_{11}) \\= - A_{12}^{T}X_{11} - C_{2}^{T}(D{+}D^{T})^{-1}(C_{1}{-}B_{1}^{T}X_{11}),
\end{multline*}
and a unique real $Z \geq 0$ satisfying the Lyapunov equation:
\begin{multline*}
{-}(A_{22}^{T}Z {+} Z A_{22}) \\= (C_{2}^{T}{-}X_{12}^{T}X_{11}^{-1}C_{1}^{T})(D{+}D^{T})^{-1}(C_{2}{-}C_{1}X_{11}^{-1}X_{12}).
\end{multline*}
Then, with the notation
\begin{align*}
\hat{A}_{12} &\coloneqq A_{12}{-}B_{1}(D{+}D^{T})^{-1}(C_{2}{-}B_{1}^{T}X_{12}), \\
\text{and } X_{-} &= T^{T}\begin{bmatrix}X_{11}& X_{12}\\ X_{12}^{T}& Z + X_{12}^{T}X_{11}^{-1}X_{12}\end{bmatrix}T,
\end{align*}
it can be verified that $X_{-} \geq 0$, $\Gamma(X_{-}) = 0$, and
\begin{equation*}
TA_{\Gamma}(X_{-})T^{-1} = \begin{bmatrix}A_{\hat{\Gamma}}(X_{11})& \hat{A}_{12}\\0& A_{22}\end{bmatrix}.
\end{equation*}
This implies that $\text{spec}(A_{\Gamma}(X_{-})) = \text{spec}(A_{\hat{\Gamma}}(X_{11})) \cup \text{spec}(A_{22}) \in \overbar{\mathbb{C}}_{-}$, so $X_{-}$ is the matrix in Theorem \ref{thm:pbtsced}.\hfill$\triangle$
\end{remark}

To finish this section we prove Theorem \ref{thm:pbtnsc}.

\begin{IEEEproof}[Proof of Theorem \ref{thm:pbtnsc}]
That \ref{nl:pts2c2} $\Rightarrow$ \ref{nl:pts2c1} is immediate from Theorem \ref{thm:pbtsc}, since $X_{-} \geq 0$ satisfies $\Omega(X_{-}) \geq 0$. It remains to show that \ref{nl:pts2c1} $\Rightarrow$ \ref{nl:pts2c2}, and if $X_{-}$ has the properties indicated in condition \ref{nl:pts2c2} then $S_{a}^{\sigma_{p}}(\mathbf{x}_{0}) = \tfrac{1}{2}\mathbf{x}_{0}^{T}X_{-}\mathbf{x}_{0}$. 
We will prove this for the cases: (i) $(C,A)$ observable and $D {+} D^{T} > 0$; (ii) $(C,A)$ observable; and finally (iii) $(C,A)$ not observable. 

{\bf Case (i) $\boldsymbol{(C,A)}$ observable and $\boldsymbol{D{+}D^{T} > 0}$.} \hspace{0.3cm} It suffices to show that $X_{-}$ satisfies condition \ref{nl:ptnsc2} in Theorem \ref{thm:pbtsced} if and only if $X_{-}$ satisfies condition \ref{nl:pts2c2} in the present theorem. 

First, let $X_{-}$ satisfy condition \ref{nl:ptnsc2} in Theorem \ref{thm:pbtsced}. Since $D+D^{T} > 0$, then there exists a nonsingular $W$ satisfying $W^{T}W = D + D^{T}$. We let $L \coloneqq (W^{T})^{-1}(C - B^{T}X_{-})$, and we obtain $-A^{T}X_{-} - X_{-}^{T}A - L^{T}L = \Gamma(X_{-}) = 0$. 

Now, let $Z(\xi) {\coloneqq} W {+} L(\xi I {-} A)^{-1}B$. From Theorems \ref{thm:pbtsc} and \ref{thm:pbtsced}, $\text{spec}(A) {\in} \overbar{\mathbb{C}}_{-}$ and $\text{spec}(A_{\Gamma}(X_{-})) {\in} \overbar{\mathbb{C}}_{-}$. Also, 
\begin{equation}
\hspace*{-0.3cm}\begin{bmatrix}\lambda I {-} A& -B\\ L& W\end{bmatrix} = \begin{bmatrix}\lambda I {-} A_{\Gamma}(X_{-})& {-}B\\ 0& W\end{bmatrix} \begin{bmatrix}I& 0\\ W^{-1}L& I\end{bmatrix}.\label{eq:sfvssc}
\end{equation}
The matrices in (\ref{eq:sfvssc}) have full row rank for all $\lambda \in \mathbb{C}_{+}$, so $Z$ is a spectral factor for $H + H^{\star}$ by Lemma \ref{lem:sfymd}. 

Next, let $X_{-}$ satisfy condition \ref{nl:pts2c2} in the present theorem. Since $W + L(\xi I - A)^{-1}B$ is a spectral factor of $H + H^{\star}$, then $W$ is nonsingular. Thus, $L = (W^{T})^{-1}(C-B^{T}X_{-})$, and $\Gamma(X_{-}) = -A^{T}X_{-} - X_{-}A - L^{T}L = 0$. As before, $\text{spec}(A) \in \overbar{\mathbb{C}}_{-}$, so the matrices in (\ref{eq:sfvssc}) have full row rank for all $\lambda \in \mathbb{C}_{+}$ by Lemma \ref{lem:sfymd}, and so $\text{spec}(A_{\Gamma}(X_{-})) \in \overbar{\mathbb{C}}_{-}$.

{\bf Case (ii) $\boldsymbol{(C,A)}$ observable.} \hspace{0.3cm} Let $P$ and $Q$ be as in Theorem \ref{thm:pbtsc}, and let $P_{1}\coloneqq P$ and $Q_{1}\coloneqq Q$. If $P_{1}$ and $Q_{1}$ do not satisfy the conditions of case (i), then we will construct $P_{m}, Q_{m} \in \mathbb{R}^{n_{m} \times n_{m}}[\xi]$ that do. Specifically, we consider the following four statements:
\begin{remunerate}
\labitem{(R\arabic{muni})}{nl:ip1} $P_{i}, Q_{i} \in \mathbb{R}^{n_{i} \times n_{i}}[\xi]$ where $(P_{i}, Q_{i})$ is a positive-real pair and $Q_{i}^{-1}P_{i}$ is proper.
\labitem{(R\arabic{muni})}{nl:ip2} $D_{i} \coloneqq \lim_{\xi \rightarrow \infty}(Q_{i}^{-1}P_{i}(\xi))$ is symmetric.
\labitem{(R\arabic{muni})}{nl:ip3} $P_{i}$ is nonsingular and $D_{i} = \text{diag}\begin{pmatrix}I_{r_{i}}& 0\end{pmatrix}$.
\labitem{(R\arabic{muni})}{nl:ip4} $D_{i} = I$ or $n_{i} = 0$.
\end{remunerate}
By Theorem \ref{thm:pbtsc}, $P_{1}$ and $Q_{1}$ satisfy condition \ref{nl:ip1}. Then, using Lemmas \ref{lem:gprl1}--\ref{lem:lrz}, we construct $P_{2}, \ldots, P_{m}$, $Q_{2}, \ldots, Q_{m}$ such that condition \ref{nl:ip1} is satisfied, $n_{i} {\leq} n_{i-1}$, and $\deg{(\det{(Q_{i})})} {\leq} \deg{(\det{(Q_{i-1})})}$, for $i = 2, \ldots , m$; and
\begin{enumerate}[leftmargin=0.5cm]
\item If, for $i = k-1$, \ref{nl:ip2} is not satisfied, then \ref{nl:ip2} is satisfied for $i = k$ (Lemma \ref{lem:gprl1}).
\item If, for $i = k-1$, \ref{nl:ip2} is satisfied but \ref{nl:ip3} is not, then \ref{nl:ip2} and \ref{nl:ip3} are satisfied for $i = k$; and if $P_{k{-}1}$ is singular then $n_{k} < n_{k{-}1}$ (Lemma \ref{lem:gprl2}).
\item If, for $i = k{-}1$, \ref{nl:ip2} and \ref{nl:ip3} are satisfied but \ref{nl:ip4} is not, then $\deg{(\det{(Q_{k})})} < \deg{(\det{(Q_{k{-}1})})}$ (Lemma \ref{lem:lrz}).
\end{enumerate}
This inductive procedure terminates in a finite number of steps with matrices $P_{m}$ and $Q_{m}$ that satisfy conditions \ref{nl:ip1}--\ref{nl:ip4}. The procedure is inspired by the sequence of transformations outlined in \cite[Section 8.4]{AndVong}. In contrast to \cite{AndVong}, we also consider the case of uncontrollable systems.

Next, we consider the following four statements:
\begin{remunerate}
\setlength{\itemsep}{3pt}
\labitem{(S\arabic{muni})}{nl:csn1} There exist polynomial matrices $M_{i}, N_{i}, U_{i}, V_{i}, E_{i}, F_{i}$\smallskip 

such that $\begin{bmatrix}M_{i}& N_{i} \\ U_{i}& V_{i}\end{bmatrix}\begin{bmatrix}-D_{i}& I& -C_{i}\\ -B_{i}& 0& \mathcal{A}_{i}\end{bmatrix} = \begin{bmatrix}-P_{i}& Q_{i}& 0\\ -E_{i}& -F_{i}& I\end{bmatrix}$,\smallskip 

where $\mathcal{A}_{i}(\xi) \coloneqq \xi I - A_{i}$, and the leftmost matrix is unimodular.
\labitem{(S\arabic{muni})}{nl:csn2-} With $\Omega_{i}(X) \coloneqq \!\begin{bmatrix}-A_{i}^{T}X {-} XA_{i}& C_{i}^{T}{-}XB_{i}\\ C_{i}{-}B_{i}^{T}X& D_{i} {+} D_{i}^{T}\end{bmatrix}\!$, then $X_{i}$\smallskip

is a real matrix that satisfies (i) $X_{i} \geq 0$; (ii) $\Omega_{i}(X_{i}) \geq 0$; and (iii) if $X$ is a real matrix that satisfies $X \geq 0$ and $\Omega_{i}(X) \geq 0$, then $X_{i} \leq X$.

\labitem{(S\arabic{muni})}{nl:csn2} $X_{i}, L_{i}$ and $W_{i}$ are real matrices such that $X_{i} \geq 0$ and\smallskip

$\Omega_{i}(X_{i}) {=} \!\begin{bmatrix}-A_{i}^{T}X_{i} {-} X_{i}A_{i}& C_{i}^{T}{-}X_{i}B_{i}\\ C_{i}{-}B_{i}^{T}X_{i}& D_{i} {+} D_{i}^{T}\end{bmatrix}\! {=} \!\begin{bmatrix}L_{i}^{T}\\ W_{i}^{T}\end{bmatrix}\begin{bmatrix}L_{i}& W_{i}\end{bmatrix}$. 
\labitem{(S\arabic{muni})}{nl:csn3} $\begin{bmatrix}\lambda I - A_{i}& {-}B_{i}\\ L_{i}& W_{i}\end{bmatrix}\!$ has full row rank for all $\lambda \in \mathbb{C}_{+}$.
\end{remunerate}
From notes \ref{lem:osf}--\ref{lem:orl}, there exist real matrices $A_{m}, B_{m}, C_{m}, D_{m}$ such that condition \ref{nl:csn1} holds. Then, from case (i), there is a unique $X_{m}$ for which there exist $L_{m}$ and $W_{m}$ that satisfy conditions \ref{nl:csn2} and \ref{nl:csn3}. Furthermore, by Theorem \ref{thm:pbtsc}, this $X_{m}$ also satisfies condition \ref{nl:csn2-}. Then, using Lemmas \ref{lem:gprl1}--\ref{lem:lrz}, we find that there are unique $X_{i}$ for which there exist $L_{i}$ and $W_{i}$ that satisfy conditions \ref{nl:csn2} and \ref{nl:csn3}, and these $X_{i}$ also satisfy condition \ref{nl:csn2-} ($i = m-1, \ldots , 1$). Now, let
\begin{align*}&\hat{\mathcal{B}}_{s} {\coloneqq} \lbrace  (\mathbf{u}, \mathbf{y}, \mathbf{x}_{1}) {\in} \mathcal{L}_{2}^{\text{loc}}\left(\mathbb{R}, \mathbb{R}^{n}\right) {\times} \mathcal{L}_{2}^{\text{loc}}\left(\mathbb{R}, \mathbb{R}^{n}\right) {\times} \mathcal{L}_{2}^{\text{loc}}\left(\mathbb{R}, \mathbb{R}^{d_{1}}\right) \mid \\ &\hspace{1.5cm} \tfrac{d\mathbf{x}_{1}}{dt} = A_{1}\mathbf{x}_{1} + B_{1}\mathbf{u} \text{ and } \mathbf{y} = C_{1}\mathbf{x}_{1} + D_{1}\mathbf{u}\rbrace.
\end{align*}
Since $P = P_{1}$ and $Q = Q_{1}$, then from note \ref{lem:orl} we conclude that $(C_{1},A_{1})$ is observable and $\hat{\mathcal{B}}_{s}^{(\mathbf{u}, \mathbf{y})} = \mathcal{B}_{s}^{(\mathbf{u}, \mathbf{y})}$. Thus, from note \ref{lem:orl2}, there exists a nonsingular $T \in \mathbb{R}^{d \times d}$ such that (\ref{eq:atd}) holds. It can then be verified that $X_{-} \coloneqq T^{T}X_{1}T$, $L \coloneqq L_{1}T$, and $W \coloneqq W_{1}$ satisfy condition \ref{nl:pts2c2} in the present theorem statement; and $X_{-}$ satisfies (a) $\Omega(X_{-}) \geq 0$; and (b) if $X \in \mathbb{R}^{d \times d}_{s}$ satisfies $X \geq 0$ and $\Omega(X) \geq 0$, then $X_{-} \leq X$. Since $X_{-}$ is uniquely determined by conditions (a)--(b), then $S_{a}^{\sigma_{p}}(\mathbf{x}_{0}) = \tfrac{1}{2}\mathbf{x}_{0}^{T}X_{-}\mathbf{x}_{0}$ by Theorem \ref{thm:pbtsc}.

{\bf Case (iii) $\boldsymbol{(C,A)}$ not observable.} \hspace{0.3cm} Consider the observer staircase form (see note \ref{lem:osf}), so $D + C_{1}(\xi I {-} A_{11})^{-1}B_{1} = D + C(\xi I {-} A)^{-1}B$, and let $T$ be as in note \ref{lem:osf} and $\hat{\mathcal{B}}_{s}$ and $\hat{S}_{a}^{\sigma}$ be as in Lemma \ref{lem:saop} (for the case $\sigma(\mathbf{u}, \mathbf{y}) = \mathbf{u}^{T}\mathbf{y}$). It follows from Lemma \ref{lem:saop} that $X_{-} = T^{T}\text{diag}\begin{pmatrix}\hat{X}_{-}& 0\end{pmatrix}T$ where $\hat{X}_{-} \in \mathbb{R}^{\hat{d} \times \hat{d}}_{s}$ with $\tfrac{1}{2}\hat{\mathbf{x}}_{0}^{T}\hat{X}_{-}\hat{\mathbf{x}}_{0} = \hat{S}_{a}^{\sigma_{p}}(\hat{\mathbf{x}}_{0})$ for all $\hat{\mathbf{x}}_{0} \in \mathbb{R}^{\hat{d}}$. From case (ii), $\hat{X}_{-}$ is the unique real matrix satisfying (a) $\hat{X}_{-} \geq 0$; and (b) there exist real matrices $\hat{L}$, $\hat{W}$ such that 
\begin{itemize}[leftmargin=0.9cm]
\item[(b1)] $-A_{11}^{T}\hat{X}_{-} {-} \hat{X}_{-}A_{11} {=} \hat{L}^{T}\hat{L}$, $C_{1}{-}B_{1}^{T}\hat{X}_{-} {=} \hat{W}^{T}\hat{L}$, and $D + D^{T} = \hat{W}^{T}\hat{W}$; and 
\item[(b2)] $\hat{W} {+} \hat{L}(\xi I {-} A_{11})^{-1}B_{1}$ is a spectral factor of $H {+} H^{\star}$.
\end{itemize}
Then, with $L \coloneqq [\hat{L} \hspace{0.15cm} 0]T$, and $W \coloneqq \hat{W}$, it can be verified that condition \ref{nl:pts2c2} of the present theorem statement holds. Also, if $X_{-}$, $L$ and $W$ are real matrices satisfying condition \ref{nl:pts2c2}, then  $X_{-} = T^{T}\text{diag}\begin{pmatrix}\hat{X}_{-}& 0\end{pmatrix}T$ for some $0 \leq \hat{X}_{-} \in \mathbb{R}^{\hat{d} \times \hat{d}}_{s}$ with
\begin{align*}
&\begin{bmatrix}(T^{-1})^{T}& 0\\ 0& I\end{bmatrix}\begin{bmatrix}{-}A^{T}X_{-} {-} X_{-}A& C^{T}{-}X_{-}B\\ C{-}B^{T}X_{-}& D {+} D^{T}\end{bmatrix}\begin{bmatrix}(T^{-1})& 0\\ 0& I\end{bmatrix} \\
&\hspace{1cm}=\left[\begin{array}{cc|c}-A_{11}^{T}\hat{X}_{-} {-} \hat{X}_{-}A_{11}& 0& C_{1}^{T}{-}\hat{X}_{-}B_{1}\\
0& 0& 0\\ \hline  \rule{0pt}{1.05\normalbaselineskip}
C_{1} {-} B_{1}^{T}\hat{X}_{-}& 0& D{+}D^{T}\end{array}\right].
\end{align*}
This implies that $L = [\hat{L} \hspace{0.15cm} 0]T$, and $W = \hat{W}$ where $\hat{L}$ and $\hat{W}$ satisfy the aforementioned conditions (b1) and (b2). Then, from case (ii) and Lemma \ref{lem:saop}, $S_{a}^{\sigma_{p}}(\mathbf{x}_{0}) = \hat{S}_{a}^{\sigma_{p}}(T_{1}\mathbf{x}_{0}) = \tfrac{1}{2}\mathbf{x}_{0}^{T}T_{1}^{T}\hat{X}_{-}T_{1}\mathbf{x}_{0} = \tfrac{1}{2}\mathbf{x}_{0}^{T}X_{-}\mathbf{x}_{0}$ for all $\mathbf{x}_{0} \in \mathbb{R}^{d}$.
\end{IEEEproof}

We conclude this section with a remark about computing the optimal control.

\begin{remark}
\label{rem:oc}
If $\mathcal{B}_{s}$ in (\ref{eq:sssg}) satisfies $m=n$ and $D{+}D^{T} > 0$, and $A_{\Gamma}(X_{-})$ in Theorem \ref{thm:pbtsced} satisfies $\text{spec}(A_{\Gamma}(X_{-})) \in \mathbb{C}_{-}$, then $\mathbf{u} \coloneqq -(D{+}D^{T})^{-1}(C{-}B^{T}X_{-})\mathbf{x}$ and $(\mathbf{u},\mathbf{y},\mathbf{x}) \in \mathcal{B}_{s}$ imply $\smallint_{t_{0}}^{t_{1}}{-}(\mathbf{u}^{T}\mathbf{y})(t)dt = {-}\tfrac{1}{2}\left[(\mathbf{x}^{T}X_{-}\mathbf{x})(t)\right]_{t_{0}}^{t_{1}}$ and $\tfrac{d\mathbf{x}}{dt} {=} A_{\Gamma}(X_{-})\mathbf{x}$. Thus, if $\mathbf{x}(t_{0}) = \mathbf{x}_{0}$, then $\smallint_{t_{0}}^{\infty}-(\mathbf{u}^{T}\mathbf{y})(t)dt = S_{a}^{\sigma_{p}}(\mathbf{x}_{0})$.

If, on the other hand, $D{+}D^{T}$ is singular or $\text{spec}(A_{\Gamma}(X_{-})) \not\in \mathbb{C}_{-}$, then there still exists a linear state feedback law such that, with $\mathbf{x}_{0} \coloneqq \mathbf{x}(t_{0})$, then $\smallint_{t_{0}}^{\infty}-(\mathbf{u}^{T}\mathbf{y})(t)dt$ comes arbitrarily close to the supremum $S_{a}^{\sigma_{p}}(\mathbf{x}_{0})$. This can be constructed as follows. First, it follows from note \ref{lem:osf} and Lemma \ref{lem:saop} that no generality is lost in assuming $(C,A)$ is observable. We then let $\epsilon > 0$, and we note that $(I+\epsilon D)$ is necessarily nonsingular. We define
\begin{align*}
A_{\epsilon} &{\coloneqq} A - B(I+\epsilon D)^{-1}\epsilon C, \hspace{0.1cm} B_{\epsilon} {\coloneqq} B(I+\epsilon D)^{-1}\sqrt{1+\epsilon^{2}},\\
C_{\epsilon} &{\coloneqq} \tfrac{(1-\epsilon^{2})}{\sqrt{1+\epsilon^{2}}}(I+ \epsilon D)^{-1}C, \hspace{0.1cm} D_{\epsilon} {\coloneqq} (D+\epsilon I)(I+\epsilon D)^{-1}, \text{ and} \\
\mathcal{B}_{s}^{\epsilon} &{\coloneqq} \lbrace (\mathbf{u}_{\epsilon}, \mathbf{y}_{\epsilon}, \mathbf{x}) {\in} \mathcal{L}_{2}^{\text{loc}}\left(\mathbb{R}, \mathbb{R}^{n}\right) {\times} \mathcal{L}_{2}^{\text{loc}}\left(\mathbb{R}, \mathbb{R}^{n}\right) {\times} \mathcal{L}_{2}^{\text{loc}}\left(\mathbb{R}, \mathbb{R}^{d}\right) \mid \\
&\hspace{1.5cm} \tfrac{d\mathbf{x}}{dt} = A_{\epsilon}\mathbf{x} + B_{\epsilon}\mathbf{u}_{\epsilon} \text{ and } \mathbf{y}_{\epsilon} = C_{\epsilon}\mathbf{x} + D_{\epsilon}\mathbf{u}_{\epsilon}\rbrace,
\end{align*}
so $\mathbf{u}_{\epsilon} = (\mathbf{u}+\epsilon \mathbf{y})/\sqrt{1+\epsilon^{2}}$ and $\mathbf{y}_{\epsilon} = (\mathbf{y}+\epsilon \mathbf{u})/\sqrt{1+\epsilon^{2}}$ satisfy
\begin{multline}
\label{eq:uytflb}
\smallint_{t_{0}}^{t_{1}}(\mathbf{u}_{\epsilon}^{T}\mathbf{y}_{\epsilon})(t)dt = \smallint_{t_{0}}^{t_{1}}(\mathbf{u}^{T}\mathbf{y})(t)dt \\ + \tfrac{\epsilon}{\sqrt{1 - \epsilon^{2}}}\smallint_{t_{0}}^{t_{1}}(\mathbf{u}_{\epsilon}^{T}\mathbf{u}_{\epsilon} + \mathbf{y}_{\epsilon}^{T}\mathbf{y}_{\epsilon})(t)dt,
\end{multline}
and $(\mathbf{u},\mathbf{y},\mathbf{x}) \in \mathcal{B}_{s}$ if and only if $(\mathbf{u}_{\epsilon},\mathbf{y}_{\epsilon},\mathbf{x}) \in \mathcal{B}_{s}^{\epsilon}$. Also, with $H(\xi) \coloneqq D {+} C(\xi I {-}A)^{-1}B$ and $H_{\epsilon}(\xi) \coloneqq D_{\epsilon} {+} C_{\epsilon}(\xi I {-}A_{\epsilon})^{-1}B_{\epsilon}$, then $H_{\epsilon} = (H{+} \epsilon I)(I {+} \epsilon H)^{-1}$. It can then be verified that $H_{\epsilon}(j\omega) + H_{\epsilon}(-j\omega)^{T} > 0$ for all $\omega \in \mathbb{R}$, $D_{\epsilon} + D_{\epsilon}^{T} > 0$, and $H_{\epsilon}$ has no poles in $\overbar{\mathbb{C}}_{+}$. Since, in addition, $(C,A)$ is observable and $(A,B)$ is stabilizable, then it can be shown that $\text{spec}(A_{\epsilon}) \in \mathbb{C}_{-}$. It then follows from \cite{SKSPRCPLTI} that there exists $X_{-}^{\epsilon} \in \mathbb{R}^{d \times d}_{s}$ such that 
\begin{align*}
&{-}A_{\epsilon}^{T}X_{-}^{\epsilon}{-}X_{-}^{\epsilon}A_{\epsilon} {-} (C_{\epsilon}^{T}{-}X_{-}^{\epsilon}B_{\epsilon})(D_{\epsilon}{+}D_{\epsilon}^{T})^{-1}(C_{\epsilon}{-}B_{\epsilon}^{T}X_{-}^{\epsilon}) {=} 0, \\
&\text{and } \text{spec}(A_{\epsilon}-B_{\epsilon}(D_{\epsilon}+D_{\epsilon}^{T})^{-1}(C_{\epsilon}-B_{\epsilon}^{T}X_{-}^{\epsilon})) \in \mathbb{C}_{-},
\end{align*}
and it follows that if $\mathbf{u}_{\epsilon} \coloneqq -(D_{\epsilon}{+}D_{\epsilon}^{T})^{-1}(C_{\epsilon}{-}B_{\epsilon}^{T}X_{-}^{\epsilon})\mathbf{x}$ and $(\mathbf{u}_{\epsilon}, \mathbf{y}_{\epsilon}, \mathbf{x}) \in \mathcal{B}_{s}^{\epsilon}$, then $\mathbf{x}(t) \rightarrow 0$ as $t \rightarrow \infty$, and $\smallint_{t_{0}}^{\infty}-(\mathbf{u}_{\epsilon}^{T}\mathbf{y}_{\epsilon})(t)dt = \tfrac{1}{2}\mathbf{x}(t_{0})^{T}X_{-}^{\epsilon}\mathbf{x}(t_{0})$. Thus, if $\mathbf{u} = (I+\epsilon D)^{-1}(\sqrt{1+ \epsilon^{2}} \mathbf{u}_{\epsilon} - \epsilon C\mathbf{x})$ and $(\mathbf{u}, \mathbf{y}, \mathbf{x}) \in \mathcal{B}_{s}$, then $\mathbf{u}_{\epsilon} = (\mathbf{u}+\epsilon \mathbf{y})/\sqrt{1+\epsilon^{2}}$ and $\mathbf{y}_{\epsilon} = (\mathbf{y}+\epsilon \mathbf{u})/\sqrt{1+\epsilon^{2}}$, so $\mathbf{x}(t) \rightarrow 0$ as $t \rightarrow \infty$ and $\smallint_{t_{0}}^{\infty}-(\mathbf{u}^{T}\mathbf{y})(t)dt \geq \tfrac{1}{2}\mathbf{x}(t_{0})^{T}X_{-}^{\epsilon}\mathbf{x}(t_{0})$ by (\ref{eq:uytflb}). Finally, it can be verified that $X_{-}^{\epsilon} \rightarrow X_{-}$ as $\epsilon \rightarrow 0$, so $\smallint_{t_{0}}^{\infty}-(\mathbf{u}^{T}\mathbf{y})(t)dt$ can be made arbitrarily close to the supremum $S_{a}^{\sigma_{p}}(\mathbf{x}_{0})$ by taking $\epsilon$ sufficiently small.

A similar argument holds for non-expansive behaviors (considered in the next three sections). In this case, we let $A_{\epsilon} {\coloneqq} A, B_{\epsilon} {\coloneqq} B, C_{\epsilon} {\coloneqq} (1{-}\epsilon)C, D_{\epsilon} {\coloneqq} (1{-}\epsilon)D$.\hfill$\triangle$
\end{remark}

\section{Non-expansive systems}
\label{sec:brl}
In addition to the results on passive systems, we also extend the famous bounded-real lemma to systems that are neither observable nor controllable. This lemma is concerned with \emph{non-expansive} systems, defined as follows.
\begin{definition}[Non-expansive system]
\label{def:sabrl}
Let $\mathcal{B}_{s}$ be as in (\ref{eq:sssg}). For any given $\mathbf{x}_{0} \in \mathbb{R}^{d}$, let
\begin{multline*}
\hspace*{-0.3cm}\mathcal{E}_{+}^{\sigma_{g}}(\mathbf{x}_{0}) = \lbrace \smallint_{t_{0}}^{t_{1}}{ (\mathbf{y}^{T}\mathbf{y}-\mathbf{u}^{T}\mathbf{u})(t) dt} \mid t_{1} \geq t_{0}, (\mathbf{u}, \mathbf{y}, \mathbf{x}) \in \mathcal{B}_{s}, \\
\text{and } \mathbf{x}(t_{0}) =  \mathbf{x}_{0}\rbrace.
\end{multline*}
Then the available storage $S_{a}^{\sigma_{g}}$ satisfies (i) $S_{a}^{\sigma_{g}}(\mathbf{x}_{0}) = \sup (\mathcal{E}_{+}^{\sigma_{g}}(\mathbf{x}_{0}))$ if $\mathcal{E}_{+}^{\sigma_{g}}(\mathbf{x}_{0})$ is bounded above; and (ii) $S_{a}^{\sigma_{g}}(\mathbf{x}_{0}) = \infty$ otherwise. If $S_{a}^{\sigma_{g}}(\mathbf{x}_{0}) < \infty$ for all $\mathbf{x}_{0} \in \mathbb{R}^{d}$, then $\mathcal{B}_{s}$ is called \emph{non-expansive}.
\end{definition}

In our results, the following new concept of a \emph{bounded-real pair} plays a central role.
\begin{definition}[Bounded-real pair]
\label{def:brp}
Let $P \in \mathbb{R}^{m \times n}[\xi]$ and $Q \in \mathbb{R}^{m \times m}[\xi]$. 
We call $(P, Q)$ a \emph{bounded-real pair} if the following hold:
\begin{enumerate}[label=(\alph*)]
\item $Q(\lambda)Q(\bar{\lambda})^{T} - P(\lambda)P(\bar{\lambda})^{T} \geq 0$ for all $\lambda \in \overbar{\mathbb{C}}_{+}$. \label{thm:nebc1}
\item $\text{rank}(\begin{bmatrix}P& {-}Q\end{bmatrix}(\lambda)) = m$ for all $\lambda \in \overbar{\mathbb{C}}_{+}$.\label{thm:nebc2}
\item If $\mathbf{p} \in \mathbb{R}^{m}[\xi]$ and $\lambda \in \mathbb{C}$ satisfy $\mathbf{p}^{T}(QQ^{\star} - PP^{\star}) = 0$ and $\mathbf{p}(\lambda)^{T}\begin{bmatrix}P& {-}Q\end{bmatrix}(\lambda) = 0$, then $\mathbf{p}(\lambda) = 0$. \label{thm:nebc3}
\end{enumerate}
\end{definition}

\begin{remark}
It can be shown that, if $(P,Q)$ is a bounded-real pair, then $Q$ is nonsingular and $\|Q^{-1}P\|_{\infty} {\leq} 1$. But the converse is not true. For example, if $P(\xi) = Q(\xi) = \xi {+} 1$, then $\|Q^{-1}P\|_{\infty} {=} 1$, and condition \ref{thm:nebc2} in Definition \ref{def:brp} holds, but not condition \ref{thm:nebc3}, so $(P,Q)$ is not a bounded-real pair.\hfill$\triangle$
\end{remark}

In this section, we provide necessary and sufficient conditions for a system to be non-expansive (in the absence of any controllability and observability assumptions). These relate 
(a) the existence of matrices $X \in \mathbb{R}_{s}^{d \times d}$ such that $X \geq 0$ and
\begin{equation}
\label{eq:lmineb}
\hspace*{-0.3cm} \Lambda(X) \coloneqq 
\begin{bmatrix}-A^{T}X - XA-C^{T}C& -C^{T}D-XB\\ -D^{T}C-B^{T}X& I - D^{T}D\end{bmatrix}
\end{equation}
satisfies $\Lambda(X) \geq 0$; and (b) the bounded-real pair concept. Also, if $I-D^{T}D > 0$, then, with the notation
\begin{align}
\hspace*{-0.3cm}\Pi(X) \coloneqq -A^{T}X-XA-C^{T}C \hspace*{3.5cm} \nonumber \\  - (C^{T}D+XB)(I-D^{T}D)^{-1}(D^{T}C+B^{T}X), \label{eq:areneb} \\
\text{and } A_{\Pi}(X) \coloneqq A + B(I-D^{T}D)^{-1}(D^{T}C+B^{T}X),\label{eq:amneb}
\end{align}
conditions (a)--(b) also relate to the spectral properties of $A_{\Pi}(X)$ for solutions $X$ to the ARE $\Pi(X) = 0$. The results in this section are presented in the next three theorems, which we prove in Sections \ref{sec:nebas}--\ref{sec:compas}. 

\begin{theorem}
\label{thm:nebtsc}
Let $\mathcal{B}_{s}, \mathcal{B}_{s}^{(\mathbf{u}, \mathbf{y})}, V_{o}$ and $\Lambda$ be as in (\ref{eq:sssg}), (\ref{eq:ebbs}), (\ref{eq:vo}) and (\ref{eq:lmineb}), respectively; and let $S_{a}^{\sigma_{g}}$ be as in Definition \ref{def:sabrl}. The following are equivalent:
\begin{enumerate}[label=\arabic*., ref=\arabic*, leftmargin=0.5cm]
\item $S_{a}^{\sigma_{g}}(\mathbf{x}_{0}) < \infty$ for all $\mathbf{x}_{0} \in \mathbb{R}^{d}$ (i.e., $\mathcal{B}_{s}$ is non-expansive). \label{nl:netc1}
\item The external behavior $\mathcal{B}_{s}^{(\mathbf{u}, \mathbf{y})}$ takes the form of (\ref{eq:bsio}), where $(P, Q)$ is a bounded-real pair.\label{nl:netc3}
\item There exists $X \in \mathbb{R}^{d \times d}_{s}$ such that $X \geq 0$ and $\Lambda(X) \geq 0$.
\label{nl:netc4}
\item $S_{a}^{\sigma_{g}}(\mathbf{x}_{0}) = \mathbf{x}_{0}^{T}X_{-}\mathbf{x}_{0}$, where $X_{-} \in \mathbb{R}^{d \times d}_{s}$ satisfies (i) $X_{-} \geq 0$; (ii) $\Lambda(X_{-}) \geq 0$; (iii) if $\mathbf{z} \in \mathbb{R}^{d}$, then $V_{o}\mathbf{z} = 0 \iff X_{-}\mathbf{z} = 0$; and (iv) if $X \in \mathbb{R}^{d \times d}_{s}$ satisfies $X \geq 0$ and $\Lambda(X) \geq 0$, then $X_{-} \leq X$.\label{nl:netc5}
\end{enumerate}
Moreover, if $(C,A)$ is observable and the above conditions hold, then (i) $\text{spec}(A) \in \mathbb{C}_{-}$; and (ii) if $X \in \mathbb{R}_{s}^{d \times d}$ satisfies $\Lambda(X) \geq 0$, then $X_{-} \leq X$.
\end{theorem}

\begin{remark}
From \cite[Theorems 3--6]{MOLLQOC}, if $(A,B)$ is controllable, then (i) for a system to be non-expansive it is necessary and sufficient for the $\mathcal{H}_{\infty}$ norm of the system's transfer function to be bounded above by one; and (ii) the set of solutions to the LMI in the bounded-real lemma (condition \ref{nl:netc4} in Theorem \ref{thm:nebtsc}) is bounded. However, both of these conditions can fail to hold when $(A,B)$ is not controllable.\hfill$\triangle$
\end{remark} 

Theorem \ref{thm:nebtnsc} provides an explicit solution to the optimal control problem in Definition \ref{def:sabrl} in the case $I - D^{T}D > 0$.
\begin{theorem}
\label{thm:nebtnsc}
Let $\mathcal{B}_{s}, V_{o}, H, \Pi$ and $A_{\Pi}$ be as in (\ref{eq:sssg}), (\ref{eq:vo}), (\ref{eq:gd}), (\ref{eq:areneb}) and (\ref{eq:amneb}), respectively; let $S_{a}^{\sigma_{g}}$ be as in Definition \ref{def:sabrl}; and let $I - D^{T}D > 0$. The following are equivalent
\begin{enumerate}[label=\arabic*., ref=\arabic*, leftmargin=0.5cm]
\item $S_{a}^{\sigma_{g}}(\mathbf{x}_{0}) < \infty$ for all $\mathbf{x}_{0} \in \mathbb{R}^{d}$ (i.e., $\mathcal{B}_{s}$ is non-expansive). \label{nl:netnsc1}
\item There exists $X_{-} \in \mathbb{R}^{d \times d}_{s}$ satisfying (i) $X_{-} \geq 0$; (ii) $\Pi(X_{-}) = 0$; (iii) if $\mathbf{z} \in \mathbb{R}^{d}$ satisfies $V_{o}\mathbf{z} = 0$, then $X_{-}\mathbf{z} = 0$; and (iv) if $\lambda \in \mathbb{C}_{+}$ and $\mathbf{z} \in \mathbb{C}^{d}$ satisfy $A_{\Pi}(X_{-})\mathbf{z} = \lambda \mathbf{z}$, then $V_{o}\mathbf{z} = 0$.\label{nl:netnsc2} 
\end{enumerate}
Moreover, if these conditions hold, then $S_{a}^{\sigma_{g}}(\mathbf{x}_{0}) = \mathbf{x}_{0}^{T}X_{-}\mathbf{x}_{0}$.
\end{theorem}

Theorem \ref{thm:nebtsc2} solves the optimal control problem in Definition \ref{def:sabrl} in the general case.
\begin{theorem}
\label{thm:nebtsc2}
Let $\mathcal{B}_{s}, V_{o}$ and $H$ be as in (\ref{eq:sssg}), (\ref{eq:vo}) and (\ref{eq:gd}), respectively; and let $S_{a}^{\sigma_{g}}$ be as in Definition \ref{def:sabrl}. The following are equivalent:
\begin{enumerate}[label=\arabic*., ref=\arabic*, leftmargin=0.5cm]
\item $S_{a}^{\sigma_{g}}(\mathbf{x}_{0}) < \infty$ for all $\mathbf{x}_{0} \in \mathbb{R}^{d}$ (i.e., $\mathcal{B}_{s}$ is non-expansive). \label{nl:nets2c1}
\item There exists $X_{-} \in \mathbb{R}^{d \times d}_{s}$ satisfying (i) $X_{-} \geq 0$; (ii) if $\mathbf{z} \in \mathbb{R}^{d}$ satisfies $V_{o}\mathbf{z} = 0$, then $X_{-}\mathbf{z} = 0$; and (iii) there exist real matrices $L$ and $W$ such that 
\begin{itemize}[leftmargin=0.8cm]
\item[(iiia)] $-A^{T}X_{-} - X_{-}A-C^{T}C = L^{T}L$, $-D^{T}C-B^{T}X_{-} = W^{T}L$, and $I - D^{T}D = W^{T}W$; and
\item[(iiib)] $Z(\xi) \coloneqq W + L(\xi I {-} A)^{-1}B$ is a spectral factor of $I - H^{\star}H$.
\end{itemize}
\end{enumerate}
Moreover, if these conditions hold, then $S_{a}^{\sigma_{g}}(\mathbf{x}_{0}) = \mathbf{x}_{0}^{T}X_{-}\mathbf{x}_{0}$.
\end{theorem}

\begin{remark}
As is the case with the positive-real lemma, there have been many notable attempts to relax the controllability and observability assumptions in the bounded-real lemma. A particularly well known result is the so-called strictly bounded-real lemma \cite[Lemma 5.6.5]{DSAC}. This lemma proves that, if $\mathcal{B}_{s}$ is as in (\ref{eq:sssg}) and $\text{spec}(A) \in \mathbb{C}_{-}$, and $H, \Pi$ and $A_{\Pi}$ are as in (\ref{eq:gd}), (\ref{eq:areneb}) and (\ref{eq:amneb}) then $\|H\|_{\infty} < 1$ if and only if $I - D^{T}D > 0$ and there exists $X \geq 0$ such that $\Pi(X) = 0$ and $\text{spec}(A_{\Pi}(X)) \in \mathbb{C}_{-}$.\hfill$\triangle$
\end{remark}

\section{Non-expansive systems and the available storage}
\label{sec:nebas}
To prove Theorem \ref{thm:nebtsc}, we will employ transformations that relate non-expansive and passive systems, and similar transformations that relate positive-real and bounded-real pairs.

\begin{IEEEproof}[Proof of Theorem \ref{thm:nebtsc}]
We will first show the two chains of implications \ref{nl:netc1} $\Rightarrow$ \ref{nl:netc5} $\Rightarrow$ \ref{nl:netc4} $\Rightarrow$ \ref{nl:netc1}, and \ref{nl:netc5} $\Rightarrow$ \ref{nl:netc3} $\Rightarrow$ \ref{nl:netc4}.

{\bf \ref{nl:netc1} $\boldsymbol{\Rightarrow}$ \ref{nl:netc5} $\boldsymbol{\Rightarrow}$ \ref{nl:netc4} $\boldsymbol{\Rightarrow}$ \ref{nl:netc1}.} \hspace{0.3cm} First, let $\mathbf{z} \in \mathbb{R}^{d}$, and let $\tilde{\mathbf{x}}(t) = e^{A(t-t_{0})}\mathbf{z}$ for all $t \in \mathbb{R}$, $\tilde{\mathbf{u}} = 0$, and $\tilde{\mathbf{y}} = C\tilde{\mathbf{x}}$. Then $(\tilde{\mathbf{u}}, \tilde{\mathbf{y}}, \tilde{\mathbf{x}}) \in \mathcal{B}_{s}$, $\tilde{\mathbf{x}}(t_{0}) = \mathbf{z}$, and $\smallint_{t_{0}}^{t_{1}}{(\tilde{\mathbf{y}}^{T}\tilde{\mathbf{y}} - \tilde{\mathbf{u}}^{T}\tilde{\mathbf{u}})(t)dt} \geq 0$. Second, note that $\smallint_{t_{0}}^{t_{1}}{(\mathbf{u}^{T}\mathbf{u} - \mathbf{y}^{T}\mathbf{y})(t)dt} - \left[\mathbf{x}^{T}X\mathbf{x}\right]_{t_{0}}^{t_{1}} = \smallint_{t_{0}}^{t_{1}}{(\begin{bmatrix}\mathbf{x}^{T}& \mathbf{u}^{T}\end{bmatrix}\Lambda(X)\text{col}\begin{pmatrix}\mathbf{x}& \mathbf{u}\end{pmatrix})(t)dt}$. With these two observations, the present implications can be shown in a similar manner to the corresponding implications in Theorem \ref{thm:pbtsc}. 

{\bf \ref{nl:netc3} $\boldsymbol{\Rightarrow}$ \ref{nl:netc4}.} \hspace{0.3cm} Consider the observer staircase form (see note \ref{lem:osf}), and let
\begin{equation*}
\hat{\Lambda}(\hat{X}) \coloneqq \begin{bmatrix}-\hat{A}_{11}^{T}\hat{X} {-} \hat{X}\hat{A}_{11}{-}\hat{C}_{1}^{T}\hat{C}_{1}& -\hat{C}_{1}^{T}\hat{D}{-}\hat{X}\hat{B}_{1}\\ -\hat{D}^{T}\hat{C}_{1}{-}\hat{B}_{1}^{T}\hat{X}& I {-} \hat{D}^{T}\hat{D}\end{bmatrix}.
\end{equation*}
If there exists $\hat{X} \in \mathbb{R}^{\hat{d} \times \hat{d}}_{s}$ satisfying $\hat{X} \geq 0$ and $\hat{\Lambda}(\hat{X}) \geq 0$, then $X \coloneqq T^{T}\text{diag}\begin{pmatrix}\hat{X}& 0\end{pmatrix}T$ satisfies $X \geq 0$ and $\Lambda(X) \geq 0$. Thus, it suffices to prove this implication for the case with $(C,A)$ observable. We will prove this for the cases: (i) $m = n$, (ii) $n < m$, and (iii) $m < n$.

{\bf Case (i): m = n.} \hspace{0.3cm} Let $\mathcal{A}(\xi) \coloneqq \xi I - A$, and let $M, N, U, V, E, F$ and $G$ be polynomial matrices satisfying conditions (a) and (b) in note \ref{lem:orl}. From note \ref{nl:pprbrp5}, there exists a signature matrix $\Sigma$ and matrices 
\begin{equation}
\hat{Q} \coloneqq \tfrac{1}{2}(Q - P\Sigma) \text{ and } \hat{P} \coloneqq \tfrac{1}{2}(P\Sigma + Q) \label{eq:hqhpd}
\end{equation}
such that $\hat{Q}$ is nonsingular and $\hat{Q}^{-1}\hat{P}$ is proper. Now, let $D \coloneqq \lim_{\xi \rightarrow \infty}(Q^{-1}P(\xi))$ and $\hat{D} \coloneqq \lim_{\xi \rightarrow \infty}(\hat{Q}^{-1}\hat{P}(\xi))$. Note that $(I - Q^{-1}P\Sigma)\hat{Q}^{-1}\hat{P} = I+Q^{-1}P\Sigma$, so by taking the limit as $\xi \rightarrow \infty$ we obtain $(I-D\Sigma)\hat{D} = I+D\Sigma$. Thus, if $\mathbf{z} \in \mathbb{R}^{m}$ and $\mathbf{z}^{T}(I - D\Sigma) = 0$, then $\mathbf{z}^{T}(I+D\Sigma) = 0$, so $\mathbf{z} = 0$. Hence, $(I-D\Sigma)$ is nonsingular, and
\begin{equation}
\hat{D} = (I-D\Sigma)^{-1}(I+D\Sigma) = 2(I-D\Sigma)^{-1} - I.\label{eq:hdd}
\end{equation}
Now, let
\begin{equation*}
\begin{bmatrix}\hat{M}& \hat{N}\\ \hat{U} & \hat{V}\end{bmatrix} \coloneqq \begin{bmatrix}\tfrac{1}{2}I& 0\\ 0& \tfrac{1}{\sqrt{2}}I\end{bmatrix}\begin{bmatrix}M& N \\ U& V\end{bmatrix}\begin{bmatrix}I {-} D\Sigma& 0\\ -B\Sigma& \sqrt{2}I\end{bmatrix},
\end{equation*}
so all of the above matrices are unimodular. Then, with
\begin{align}
&\hat{A} \coloneqq A + B\Sigma(I-D\Sigma)^{-1}C, \hspace{0.2cm} \hat{B} \coloneqq \sqrt{2}B\Sigma(I-D\Sigma)^{-1}, \nonumber \\
&\hat{C} \coloneqq \sqrt{2}(I-D\Sigma)^{-1}C, \hspace{0.2cm} \hat{\mathcal{A}}(\xi) \coloneqq \xi I - \hat{A}, \nonumber \\&\hat{E} \coloneqq \tfrac{1}{\sqrt{2}}(E\Sigma-F), \text{ and } \hat{F} \coloneqq \tfrac{1}{\sqrt{2}}(E\Sigma+F),
 \label{eq:ahad}
\end{align} 
it can be verified that $(\hat{C},\hat{A})$ is observable, and
\begin{equation}
\begin{bmatrix}\hat{M}& \hat{N} \\ \hat{U}& \hat{V}\end{bmatrix}\begin{bmatrix}-\hat{D}& I& -\hat{C}\\ -\hat{B}& 0& \hat{\mathcal{A}}\end{bmatrix} = \begin{bmatrix}-\hat{P}& \hat{Q}& 0\\ -\hat{E}& -\hat{F}& I\end{bmatrix}.\label{eq:ore2}
\end{equation}
Hence, $(\hat{A},\hat{B},\hat{C},\hat{D})$ is an observable realization for $(\hat{P},\hat{Q})$ (see note \ref{lem:orl}). Since $\hat{Q} \coloneqq \tfrac{1}{2}(Q - P\Sigma)$ and $\hat{P} \coloneqq \tfrac{1}{2}(P\Sigma + Q)$, then it follows from notes \ref{nl:pprbrp1}--\ref{nl:pprbrp5} that $(\hat{P}, \hat{Q})$ is a positive-real pair. Thus, from Lemma \ref{lem:ssr} and Theorem \ref{thm:pbtsc}, there exists $X \in \mathbb{R}^{d \times d}_{s}$ such that $X > 0$ and 
\begin{equation}
\hat{\Omega}(X) \coloneqq \begin{bmatrix}-\hat{A}^{T}X {-} X\hat{A}& \hat{C}^{T}{-}X\hat{B}\\ \hat{C}{-}\hat{B}^{T}X& \hat{D} {+} \hat{D}^{T}\end{bmatrix} \label{eq:homx} 
\end{equation}
satisfies $\hat{\Omega}(X) \geq 0$. Furthermore, with
\begin{equation}
 S = \begin{bmatrix}I& 0\\ -\tfrac{1}{\sqrt{2}}C& \tfrac{1}{\sqrt{2}}(I-D\Sigma)\Sigma\end{bmatrix}, \label{eq:std}
 \end{equation}
then it can be verified that $S^{T} \hat{\Omega}(X) S = \Lambda(X)$, which is non-negative definite since $\hat{\Omega}(X)$ is. This proves case (i).

{\bf Case (ii): m > n.} \hspace{0.3cm} Let $\hat{P}\coloneqq\begin{bmatrix}P& 0_{m \times (m-n)}\end{bmatrix}$ and $\hat{Q} \coloneqq Q$. It is easily shown from note \ref{nl:pprbrp0} that $(\hat{P}, \hat{Q})$ is a bounded-real pair. Also, with $\hat{A} {=} A$, $\hat{B} {=} \begin{bmatrix}B& 0_{d \times (m-n)}\end{bmatrix}$, $\hat{C} {=} C$, and $\hat{D} {=} \begin{bmatrix}D& 0_{m \times (m-n)}\end{bmatrix}$, it can be verified that $(A, B, C, D)$ is an observable realization for $(P, Q)$ if and only if $(\hat{A}, \hat{B}, \hat{C}, \hat{D})$ is an observable realization for $(\hat{P}, \hat{Q})$. With
\begin{equation}
\hat{\Lambda}(X) \coloneqq \!\begin{bmatrix}\! {-}\hat{A}^{T}X {-} X\hat{A}{-}\hat{C}^{T}\hat{C}& {-}\hat{C}^{T}\hat{D}{-}X\hat{B}\\ {-}\hat{D}^{T}\hat{C}{-}\hat{B}^{T}X& I {-} \hat{D}^{T}\hat{D}\!\end{bmatrix}\! ,\label{eq:hlxd}
\end{equation}
then $\hat{\Lambda}(X) = \text{diag}\begin{pmatrix}\Lambda(X)& I\end{pmatrix}$. From case (i), there exists $X > 0$ such that $\hat{\Lambda}(X) \geq 0$. This $X$ also satisfies $\Lambda(X) \geq 0$.

{\bf Case (iii): m < n.} \hspace{0.3cm}. In this case, let $\hat{P} \coloneqq \text{col}\begin{pmatrix}P& 0_{(n-m) \times n}\end{pmatrix}$, $\hat{Q} \coloneqq \text{diag}\begin{pmatrix}Q& I_{(n-m) \times (n-m)}\end{pmatrix}$, $\hat{A} {=} A$, $\hat{B} {=} B$, $\hat{C} {=} \text{col}\begin{pmatrix}C& 0_{(n-m) \times d}\end{pmatrix}$, and $\hat{D} {=} \text{col}\begin{pmatrix}D& 0_{(n-m) \times n}\end{pmatrix}$; and let $\Lambda(X)$ and $\hat{\Lambda}(X)$ be as in (\ref{eq:lmineb}) and (\ref{eq:hlxd}), respectively. Then $(\hat{P}, \hat{Q})$ is a bounded-real pair (this is easily shown from note \ref{nl:pprbrp0}), $\Lambda(X) = \hat{\Lambda}(X)$, and the proof is similar to case (ii). 

{\bf \ref{nl:netc5} $\boldsymbol{\Rightarrow}$ \ref{nl:netc3}.} \hspace{0.3cm} We will prove this for the two cases (i) $(C,A)$ observable; and (ii) $(C,A)$ not observable.

{\bf Case (i): $\boldsymbol{(C,A)}$ observable.} \hspace{0.3cm} We consider the case $m = n$. The proofs for the cases $m > n$ and $m < n$ are then similar to the corresponding cases in the proof of \ref{nl:netc3} $\Rightarrow$ \ref{nl:netc4}. Let $\Sigma, \hat{P}, \hat{Q}, \hat{A}, \hat{B}, \hat{C}, \hat{D}$, and $\hat{\Omega}(X)$ be as in case (i) of the proof of \ref{nl:netc3} $\boldsymbol{\Rightarrow}$ \ref{nl:netc4}. Then, from that proof, $(\hat{A}, \hat{B}, \hat{C}, \hat{D})$ is an observable realization of $(\hat{P}, \hat{Q})$, and $\hat{\Omega}(X_{-}) \geq 0$. Thus, $(\hat{P}, \hat{Q})$ is a positive-real pair by Theorem \ref{thm:pbtsc}, so $(P,Q)$ is a bounded-real pair by notes  \ref{nl:pprbrp1}--\ref{nl:pprbrp5}.

{\bf Case (ii): $\boldsymbol{(C,A)}$ not observable.} \hspace{0.3cm} Consider the observer staircase form (see note \ref{lem:osf}), and let $\hat{\Lambda}$ be as in (\ref{eq:hlxd}). Then $X_{-} = T^{T}\text{diag}\begin{pmatrix}\hat{X}_{-}& 0\end{pmatrix}T$ where $\hat{X}_{-} \in \mathbb{R}_{s}^{\hat{d} {\times} \hat{d}}$, $\hat{\Lambda}(\hat{X}_{-}) \geq 0$ and $\hat{X}_{-} \geq 0$. Also, with $\hat{\mathcal{B}}_{s}$ as in note \ref{lem:osf}, then $\mathcal{B}_{s}^{(\mathbf{u},\mathbf{y})} = \hat{\mathcal{B}}_{s}^{(\mathbf{u},\mathbf{y})}$ as shown in that note. Condition \ref{nl:netc3} then follows from case (i).

It remains to prove conditions (i)--(ii) in the final paragraph of the present theorem statement. To see (i), let $\lambda \in \overbar{\mathbb{C}}_{+}$ and $\mathbf{z} \in \mathbb{C}^{d}$ satisfy $(\lambda I - A)\mathbf{z} = 0$, and note that $\bar{\mathbf{z}}^{T}(A^{T}X+XA)\mathbf{z} = (\lambda{+}\bar{\lambda})\bar{\mathbf{z}}^{T}X\mathbf{z}$. Since $-A^{T}X - XA - C^{T}C \geq 0$, then $\mathbf{\bar{z}}^{T}C^{T}C\mathbf{z} \leq -2\Re{(\lambda)}\mathbf{\bar{z}}^{T}X\mathbf{z} \leq 0$, so $C\mathbf{z} = 0$. If $(C,A)$ is observable, then $\mathbf{z} = 0$, so $\text{spec}(A) \in \mathbb{C}_{-}$. The proof of condition (ii) is similar to the corresponding condition in Theorem \ref{thm:pbtsc}, using the observations in the second paragraph of this proof.
\end{IEEEproof}

\section{Explicit characterisation of the available storage for a non-expansive system}
\label{sec:compas}
This section contains the proofs of Theorems \ref{thm:nebtnsc} and \ref{thm:nebtsc2}. The proofs provide methods for calculating the available storage for a non-expansive system by using the results in Section \ref{sec:cae}.
\begin{IEEEproof}[Proof of Theorem \ref{thm:nebtnsc}]
{\bf \ref{nl:netnsc2} $\boldsymbol{\Rightarrow}$ \ref{nl:netnsc1}.} \hspace{0.3cm} This follows from Theorem \ref{thm:nebtsc}, since $X_{-} \geq 0$ and $\Lambda(X_{-}) \geq 0$.

{\bf \ref{nl:netnsc1} $\boldsymbol{\Rightarrow}$ \ref{nl:netnsc2}.} \hspace{0.3cm} First, we note from Theorem \ref{thm:nebtsc} that $(P,Q)$ is a bounded-real pair since $S_{a}^{\sigma_{g}}(\mathbf{x}_{0}) < \infty$. We will show that this implies condition \ref{nl:netnsc2} for the cases: (i) $(C,A)$ observable and $m=n$; (ii) $(C,A)$ observable and $m > n$; (iii) $(C,A)$ observable and $m < n$; then finally (iv) $(C,A)$ not observable. 

{\bf Case (i) $\boldsymbol{(C,A)}$ observable, m = n.} \hspace{0.3cm}  Let $\Sigma, \hat{P}, \hat{Q}, \hat{A}, \hat{B}, \hat{C}$, and $\hat{D}$ be as in case (i) in the proof of \ref{nl:netc3} $\Rightarrow$ \ref{nl:netc4} in Theorem \ref{thm:nebtsc}. From that proof, $(\hat{P}, \hat{Q})$ is a positive-real pair, and $(\hat{A}, \hat{B}, \hat{C}, \hat{D})$ is an observable realization of $(\hat{P}, \hat{Q})$. From Theorem \ref{thm:pbtsced}, with the notation
\begin{align*}
&\hat{\Gamma}(X) \coloneqq  -\hat{A}^{T}X{-}X\hat{A} {-} (\hat{C}^{T}{-}X\hat{B})(\hat{D}{+}\hat{D}^{T})^{-1}(\hat{C}{-}\hat{B}^{T}X),\\
&\text{and } A_{\hat{\Gamma}}(X) \coloneqq \hat{A}-\hat{B}(\hat{D}+\hat{D}^{T})^{-1}(\hat{C}-\hat{B}^{T}X),
\end{align*}
there exists $X \in \mathbb{R}^{d \times d}_{s}$ such that $X \geq 0$, $\hat{\Gamma}(X) = 0$, and $\text{spec}(A_{\hat{\Gamma}}(X)) \in \overbar{\mathbb{C}}_{-}$. It can then be verified that $\Pi(X) = \hat{\Gamma}(X)$ and $A_{\Pi}(X) =A_{\hat{\Gamma}}(X)$, so condition \ref{nl:netnsc2} holds.

{\bf Case (ii) $\boldsymbol{(C,A)}$ observable, m > n.} \hspace{0.3cm} Let $\hat{P}$, $\hat{Q}$, $\hat{A}, \hat{B}, \hat{C}$, and $\hat{D}$ be as in case (ii) in the proof of \ref{nl:netc3} $\Rightarrow$ \ref{nl:netc4} in Theorem \ref{thm:nebtsc}; so $(\hat{P}, \hat{Q})$ is a bounded-real pair, and $(A, B, C, D)$ is an observable realization for $(P, Q)$ if and only if $(\hat{A}, \hat{B}, \hat{C}, \hat{D})$ is an observable realization for $(\hat{P}, \hat{Q})$. Also, let
\begin{align}
&\hspace*{-0.3cm}\hat{\Pi}(X) \coloneqq -\hat{A}^{T}X-X\hat{A}-\hat{C}^{T}\hat{C} \nonumber \\  &\hspace{0.5cm} - (\hat{C}^{T}\hat{D}+X\hat{B})(I-\hat{D}^{T}\hat{D})^{-1}(\hat{D}^{T}\hat{C}+\hat{B}^{T}X), \label{eq:arenheb} \\
&\hspace*{-0.3cm}\text{and } \hat{A}_{\hat{\Pi}}(X) \coloneqq \hat{A} + \hat{B}(I-\hat{D}^{T}\hat{D})^{-1}(\hat{D}^{T}\hat{C}+\hat{B}^{T}X).\label{eq:amnheb}
\end{align}
It can be verified that $\hat{\Pi}(X) = \Pi(X)$ and $\hat{A}_{\hat{\Pi}}(X) = A_{\Pi}(X)$, so this case follows from case (i).

{\bf Case (iii) $\boldsymbol{(C,A)}$ observable, m < n.} \hspace{0.3cm} In this case, we let  $\hat{P}, \hat{Q}, \hat{A}, \hat{B}, \hat{C}$, and $\hat{D}$ be as in case (iii) in the proof of \ref{nl:netc3} $\Rightarrow$ \ref{nl:netc4} in Theorem \ref{thm:nebtsc}. Then, with $\hat{\Pi}(X)$ and $\hat{A}_{\hat{\Pi}}(X)$ as in (\ref{eq:arenheb})--(\ref{eq:amnheb}), we obtain $\hat{\Pi}(X) = \Pi(X)$ and $\hat{A}_{\hat{\Pi}}(X) = A_{\Pi}(X)$. The proof then follows the argument in case (ii).

{\bf Case (iv) $\boldsymbol{(C,A)}$ not observable.} \hspace{0.25cm} This can be proved in the manner of case (ii) in the proof of \ref{nl:netnsc1} $\Rightarrow$ \ref{nl:netnsc2} in Theorem \ref{thm:pbtsced}.

Finally, with a similar proof to the corresponding implication in Theorem \ref{thm:pbtsced}, we find that if $X_{-}$ satisfies condition \ref{nl:netnsc2} of the present theorem, then $S_{a}^{\sigma_{g}}(\mathbf{x}_{0}) = \mathbf{x}_{0}^{T}X_{-}\mathbf{x}_{0}$.
\end{IEEEproof}

\begin{IEEEproof}[Proof of Theorem \ref{thm:nebtsc2}]
{\bf \ref{nl:netnsc2} $\boldsymbol{\Rightarrow}$ \ref{nl:netnsc1}.} \hspace{0.3cm} This follows from Theorem \ref{thm:nebtsc}, since $X_{-} \geq 0$ and $\Lambda(X_{-}) \geq 0$.

For the remainder of the proof, we let $(C,A)$ be observable and $m = n$.  The cases $m > n$ and $m < n$ can be shown by augmenting to the case $m = n$ as in the proof of Theorem \ref{thm:nebtnsc}. The case $(C,A)$ not observable can be shown with a similar argument to the corresponding implication in Theorem \ref{thm:pbtnsc}.

{\bf \ref{nl:netnsc1} $\boldsymbol{\Rightarrow}$ \ref{nl:netnsc2}.} \hspace{0.3cm} Since $S_{a}^{\sigma_{g}}(\mathbf{x}_{0}) < \infty$ for all $\mathbf{x}_{0} \in \mathbb{R}^{d}$, then $(P,Q)$ is a bounded-real pair by Theorem \ref{thm:nebtsc}. Next, let $\Sigma, \hat{P}, \hat{Q}, \hat{A}, \hat{B}, \hat{C}$, and $\hat{D}$ be as in case (i) in the proof of \ref{nl:netc3} $\Rightarrow$ \ref{nl:netc4} in Theorem \ref{thm:nebtsc} (so $I-D\Sigma$ is nonsingular and $(\hat{C},\hat{A})$ is observable), and let $\hat{H}(\xi) \coloneqq \hat{D} + \hat{C}(\xi I - \hat{A})^{-1}\hat{B}$. Then $(\hat{P}, \hat{Q})$ is a positive-real pair, so from Theorems \ref{thm:pbtsc}, \ref{thm:pbtnsc} there exist real matrices $X_{-}, \hat{L}$, and $\hat{W}$ with $X_{-} \geq 0$ such that 
\begin{enumerate}[label=(\alph*), leftmargin=0.5cm]
\item $-\hat{A}^{T}X_{-} - X_{-}\hat{A} = \hat{L}^{T}\hat{L}$, $\hat{C}-\hat{B}^{T}X_{-} = \hat{W}^{T}\hat{L}$, $\hat{D} + \hat{D}^{T} = \hat{W}^{T}\hat{W}$; and\label{nl:brprc1}
\item $\hat{Z}(\xi) {\coloneqq} \hat{W} {+} \hat{L}(\xi I {-} \hat{A})^{-1}\hat{B}$ is a spectral factor of $\hat{H} {+} \hat{H}^{\star}$.\label{nl:brprc2}
\end{enumerate}
Then, let $L \coloneqq \hat{L} {-} \tfrac{1}{\sqrt{2}}\hat{W}C$ and $W \coloneqq \tfrac{1}{\sqrt{2}}\hat{W}(I{-}D\Sigma)\Sigma$, and it can be verified that condition (iiia) holds. Also, 
\begin{equation*}
\begin{bmatrix}\! \lambda I {-} \hat{A}& -\hat{B}\\ \hat{L}& \hat{W} \!\end{bmatrix}\! \!\begin{bmatrix}\! I& 0\\ {-}\tfrac{1}{\sqrt{2}}C& \tfrac{1}{\sqrt{2}}(I{-}D\Sigma)\Sigma \end{bmatrix}\!  {=} \!\begin{bmatrix}\! \lambda I {-} A& -B\\ L& W \!\end{bmatrix}\! .\label{eq:omtlams}
 \end{equation*}
From Theorem \ref{thm:pbtsc}, $\text{spec}(\hat{A}) \in \overbar{\mathbb{C}}_{-}$. Also, from Theorem \ref{thm:nebtsc}, $\text{spec}(A) \in \mathbb{C}_{-}$. Since, in addition, $(I{-}D\Sigma)$ is nonsingular, then a similar argument to the proof of Lemma \ref{lem:sfymd} shows that $Z$ is a spectral factor of $I - H^{\star}H$.
 
Finally, we prove that if $X_{-}$ satisfies condition \ref{nl:netnsc2}, then $S_{a}^{\sigma_{g}}(\mathbf{x}_{0}) = \mathbf{x}_{0}^{T}X_{-}\mathbf{x}_{0}$. It suffices to show that $X_{-}$ is uniquely determined by condition \ref{nl:netnsc2}. To show this, we let $\Sigma, \hat{P}, \hat{Q}, \hat{A}, \hat{B}, \hat{C}, \hat{D}$ and $\hat{H}$ be as in the previous paragraph. Following that paragraph, if $X_{-}$ satisfies condition \ref{nl:netnsc2}, then $\hat{L} \coloneqq L + W\Sigma (I-D\Sigma)^{-1}C$ and $\hat{W} \coloneqq \sqrt{2}W\Sigma(I-D\Sigma)^{-1}$ satisfy the aforementioned conditions \ref{nl:brprc1} and \ref{nl:brprc2}. From Theorem \ref{thm:pbtnsc}, these conditions uniquely determine $X_{-}$.
\end{IEEEproof}
 
\bibliographystyle{IEEEtran}
\bibliography{IEEEabrv,passive_behav_va_refs}

\appendices
\section{Observable realizations of behaviors}
\label{sec:abas}
In this appendix, we present several results on observable realizations which are used in the proofs of the main theorems. These results build on Lemmas \ref{lem:ssr} and \ref{lem:ssr2}.
\begin{remunerate}
\labitem{\ref{sec:abas}{.}\arabic{muni}}{lem:osf}
Let $\mathcal{B}_{s}$ and $V_{o}$ be as in (\ref{eq:sssg}) and (\ref{eq:vo}); let the columns of $S_{2} \in \mathbb{R}^{d \times (d - \hat{d})}$ be a basis for the nullspace of $V_{o}$; let $S = \begin{bmatrix}S_{1}& S_{2}\end{bmatrix}$ be nonsingular; and let $S^{-1} \eqqcolon T = \text{col}\begin{pmatrix}T_{1}& T_{2}\end{pmatrix}$ (partitioned compatibly with $S$). Then,
\begin{equation*}
\begin{bmatrix}T_{1}\\ T_{2}\end{bmatrix}A\begin{bmatrix}S_{1}& S_{2}\end{bmatrix}=\begin{bmatrix}A_{11}& 0\\A_{21}& A_{22}\end{bmatrix}, \hspace{0.1cm} C\begin{bmatrix}S_{1}& S_{2}\end{bmatrix}=\begin{bmatrix}C_{1}& 0\end{bmatrix},
\end{equation*}
and $(C_{1}, A_{11})$ is observable \cite[Corollary 5.3.14]{JWIMTSC}. Furthermore, with the notation $B_{1} \coloneqq T_{1}B$, $B_{2} \coloneqq T_{2}B$, and
\begin{align*}
&\hspace*{-0.1cm} \hat{\mathcal{B}}_{s} {=} \lbrace (\mathbf{u}, \mathbf{y}, \hat{\mathbf{x}}) \in \mathcal{L}_{2}^{\text{loc}}\left(\mathbb{R}, \mathbb{R}^{n}\right) {\times} \mathcal{L}_{2}^{\text{loc}}\left(\mathbb{R}, \mathbb{R}^{n}\right) {\times} \mathcal{L}_{2}^{\text{loc}}(\mathbb{R}, \mathbb{R}^{\hat{d}}) \nonumber\\
& \hspace{0.5cm} \text{such that } \tfrac{d\hat{\mathbf{x}}}{dt} = A_{11}\hat{\mathbf{x}} + B_{1}\mathbf{u} \text{ and } \mathbf{y} = C_{1}\hat{\mathbf{x}} + D\mathbf{u}\rbrace,
\end{align*}
then it is easily shown from the variation of the constants formula (\ref{eq:xvocf})--(\ref{eq:yvocf}) that $\mathcal{B}_{s}^{(\mathbf{u}, \mathbf{y})} = \hat{\mathcal{B}}_{s}^{(\mathbf{u}, \mathbf{y})}$. Thus, if $P$ and $Q$ are as in Lemma \ref{lem:ssr2}, then it follows from Lemmas \ref{lem:ssr}--\ref{lem:ssr2} that there exists an observable realization for $(P,Q)$.
\labitem{\ref{sec:abas}{.}\arabic{muni}}{lem:orl}
Let $\mathcal{B}_{s}$ be as in (\ref{eq:sssg}) and $\mathcal{A}(\xi) = \xi I - A$. Then $(A,B,C,D)$ is an observable realization for $(P,Q)$ if and only if $P \in \mathbb{R}^{m \times n}[\xi], Q \in \mathbb{R}^{m \times m}[\xi]$, and there exist polynomial matrices $M,N,U,V,E,F$ and $G$ such that (a) conditions \ref{nl:utsse1} and \ref{nl:utsse2} of Lemma \ref{lem:ssr} hold; and (b) $G = I_{d}$. To see this, note from the final block column in condition \ref{nl:utsse1} of Lemma \ref{lem:ssr} that, for any given $\lambda \in \mathbb{C}$ and $\mathbf{z} \in \mathbb{C}^{d}$, then $C\mathbf{z} = 0$ and $(\lambda I - A)\mathbf{z} = 0$ if and only if $G(\lambda)\mathbf{z} = 0$. It then follows from \cite[Theorem 5.3.7]{JWIMTSC} that $G$ in Lemma \ref{lem:ssr} is unimodular if and only if $(C,A)$ is observable. Furthermore, if $G$ is unimodular, then by pre-multiplying both sides in condition \ref{nl:utsse1} of Lemma \ref{lem:ssr} by $\text{diag}\begin{pmatrix}I& G^{-1}\end{pmatrix}$ we obtain polynomial matrices satisfying conditions (a) and (b).
\labitem{\ref{sec:abas}{.}\arabic{muni}}{lem:orl2}
Let $P$ and $Q$ be as in Lemma \ref{lem:ssr2}. If $(A,B,C,D)$ and $(\hat{A},\hat{B},\hat{C},\hat{D})$ are two observable realizations of $(P,Q)$, then there exists  a nonsingular $T {\in} \mathbb{R}^{d \times d}$ such that
\begin{equation}
\hspace*{-0.3cm}\hat{A} = TAT^{-1}, \hspace{0.1cm} \hat{B} = TB, \hspace{0.1cm} \hat{C} = CT^{-1},\text{ and }\hat{D} = D.\label{eq:atd}
\end{equation}
To see this, let $A \in \mathbb{R}^{d \times d}$ and $\hat{A} \in \mathbb{R}^{\hat{d} \times \hat{d}}$; let $V_{o}$ be as in (\ref{eq:vo}); and let $\hat{V}_{o} \coloneqq \text{col}(\hat{C} \hspace{0.15cm} \hat{C}\hat{A} \hspace{0.15cm} \cdots  \hspace{0.15cm} \hat{C}\hat{A}^{\hat{d}-1})$. It follows from the variation of the constants formula (\ref{eq:xvocf})--(\ref{eq:yvocf}) that, for any given $\mathbf{z} \in \mathbb{R}^{d}$, there exists $\hat{\mathbf{z}} \in \mathbb{R}^{\hat{d}}$ such that $\hat{C}e^{\hat{A}t}\hat{\mathbf{z}} = Ce^{At}\mathbf{z}$ for all $t \in \mathbb{R}$. Suppose initially that $\hat{d} \leq d$. Since $\mathbf{z}$ is arbitrary, there must exist $T \in \mathbb{R}^{\hat{d} \times d}$ such that $\hat{C}\hat{A}^{k}T = CA^{k}$ ($k = 0, 1, \ldots$). In particular, $V_{o} = \hat{V}_{o}T$. As $(C, A)$ and $(\hat{C}, \hat{A})$ are observable, then $V_{o}$ and $\hat{V}_{o}$ have full column rank, so $\hat{d} = d$ and $T = (\hat{V}_{o}^{T}\hat{V}_{o})^{-1}\hat{V}_{o}^{T}V_{o}$, which is nonsingular (with $T^{-1} =  (V_{o}^{T}V_{o})^{-1}V_{o}^{T}\hat{V}_{o}$). In particular, $\hat{C} = CT^{-1}$. Also, since $V_{o}A = \hat{V}_{o}\hat{A}T$, then $\hat{A} = (\hat{V}_{o}^{T}\hat{V}_{o})^{-1}\hat{V}_{o}^{T}V_{o}AT^{-1} = TAT^{-1}$. Finally, from the variation of the constants formula (\ref{eq:xvocf})--(\ref{eq:yvocf}), we require  $V_{o}B = \hat{V}_{o}\hat{B}$, so $\hat{B} = TB$. A similar argument applies when $\hat{d} \geq d$, and completes the proof.
\end{remunerate}

\section{Storage functions}
\label{sec:sf}

The storage function concept features in many classical proofs of the positive-real lemma, e.g., \cite{JWDSP1,JWDSP2}. Here, in contrast to \cite{JWDSP1,JWDSP2}, we present results on storage functions without any controllability assumptions. 

We consider the following optimal control problem.
\begin{defapp}
\label{def:sagd}
Let $\mathcal{B}_{s}$ be as in (\ref{eq:sssg}); let $\sigma(\mathbf{u},\mathbf{y}) \coloneqq \mathbf{u}^{T}\Sigma_{11}\mathbf{u} + 2\mathbf{u}^{T}\Sigma_{12}\mathbf{y} + \mathbf{y}^{T}\Sigma_{22}\mathbf{y}$ for some $\Sigma_{11} \in \mathbb{R}^{n \times n}_{s}, \Sigma_{12} \in \mathbb{R}^{n \times m}$ and $\Sigma_{22} \in \mathbb{R}^{m \times m}_{s}$; and, for any given $\mathbf{x}_{0} \in \mathbb{R}^{d}$, let
\begin{multline*}
\mathcal{E}_{+}^{\sigma}(\mathbf{x}_{0}) = \lbrace \smallint_{t_{0}}^{t_{1}}{-\sigma(\mathbf{u},\mathbf{y})(t) dt} \mid t_{1} \geq t_{0}, (\mathbf{u}, \mathbf{y}, \mathbf{x}) \in \mathcal{B}_{s}, \\
\text{and } \mathbf{x}(t_{0}) =  \mathbf{x}_{0}\rbrace.
\end{multline*}
Then the \emph{available storage} $S_{a}^{\sigma}$ with respect to the supply rate $\sigma$ satisfies (i) $S_{a}^{\sigma}(\mathbf{x}_{0}) = \sup (\mathcal{E}_{+}^{\sigma}(\mathbf{x}_{0}))$ if $\mathcal{E}_{+}^{\sigma}(\mathbf{x}_{0})$ is bounded above; and (ii) $S_{a}^{\sigma}(\mathbf{x}_{0}) = \infty$ otherwise.
\end{defapp}

Note, with $\Sigma_{11} = I$, $\Sigma_{12} = 0$ and $\Sigma_{22} = -I$ (resp., $\Sigma_{11} = \Sigma_{22} = 0$, $\Sigma_{12} = \tfrac{1}{2}I$), then $S_{a}^{\sigma} = S_{a}^{\sigma_{g}}$ (resp., $S_{a}^{\sigma} = S_{a}^{\sigma_{p}}$). As in \cite{JWDSP1}, we define a storage function with respect to $\sigma$ as follows.

\begin{defapp}
\label{def:sfd}
Let $\mathcal{B}_{s}$ be as in (\ref{eq:sssg}), and let $\sigma$ be as in Definition \ref{def:sagd}. We say $S$ is a \emph{storage function} with respect to the \emph{supply rate} $\sigma$ if (i) $S(\mathbf{x}_{0}) \in \mathbb{R}$ and $S(\mathbf{x}_{0}) \geq 0$ for all $\mathbf{x}_{0} \in \mathbb{R}^{d}$; (ii) $S(0) = 0$; and (iii) if $(\mathbf{u}, \mathbf{y}, \mathbf{x}) \in \mathcal{B}_{s}$ and $t_{1} \geq t_{0} \in \mathbb{R}$, then $S(\mathbf{x}(t_{1})) \leq \smallint_{t_{0}}^{t_{1}}{(\sigma(\mathbf{u},\mathbf{y}))(t) dt}+ S(\mathbf{x}(t_{0}))$.
\end{defapp}

The next lemma proves that the boundedness of the available storage is equivalent to the existence of a storage function.
\begin{lemapp}
\label{lem:avstor}
Let $\mathcal{B}_{s}$ be as in (\ref{eq:sssg}); and let  $\sigma$ and $S_{a}^{\sigma}$ be as in Definition \ref{def:sagd}. The following hold:
\begin{enumerate}[label=\arabic*., ref=\arabic*, leftmargin=0.5cm]
\item If $S_{a}^{\sigma}(\mathbf{x}_{0}) < \infty$ for all $\mathbf{x}_{0} \in \mathbb{R}^{d}$, then $S_{a}^{\sigma}$ is a storage function with respect to $\sigma$.
\item If there exists a storage function with respect to $\sigma$ (denoted $S$), then $S_{a}^{\sigma}(\mathbf{x}_{0}) \leq S(\mathbf{x}_{0}) < \infty$ for all $\mathbf{x}_{0} \in \mathbb{R}^{d}$.
\end{enumerate}
\end{lemapp}

\begin{IEEEproof}
See \cite[Theorem 1]{JWDSP1}.
\end{IEEEproof}

In the next lemma, we prove that the available storage $S_{a}^{\sigma}(\mathbf{x}_{0})$ is a quadratic form in $\mathbf{x}_{0}$, under an assumption which is satisfied by both passive and non-expansive systems.
\begin{lemapp}
\label{lem:sfqao}
Let $\mathcal{B}_{s}$ be as in (\ref{eq:sssg}); and let $\sigma$ and $S_{a}^{\sigma}$ be as in Definition \ref{def:sagd}. Also, for any given $\mathbf{z} \in \mathbb{R}^{d}$ and $t_{0} \in \mathbb{R}$, let there exist $t_{1} \geq t_{0}$, and $(\mathbf{u}, \mathbf{y}, \mathbf{x}) \in \mathcal{B}_{s}$ with $\mathbf{x}(t_{0}) = \mathbf{z}$, such that $-\smallint_{t_{0}}^{t_{1}}{(\sigma(\mathbf{u},\mathbf{y}))(t) dt} \geq 0$. If $S_{a}^{\sigma}(\mathbf{x}_{0}) < \infty$ for all $\mathbf{x}_{0} \in \mathbb{R}^{d}$, then there exists $X \in \mathbb{R}^{d \times d}_{s}$ with $X \geq 0$ such that $S_{a}^{\sigma}(\mathbf{x}_{0}) = \mathbf{x}_{0}^{T}X\mathbf{x}_{0}$ for all $\mathbf{x}_{0} \in \mathbb{R}^{d}$. 
\end{lemapp}
\begin{IEEEproof}
For any given $\mathbf{x}_{1}, \mathbf{x}_{2} \in \mathbb{R}^{d}$, we let $W(\mathbf{x}_{1}, \mathbf{x}_{2}) \coloneqq \tfrac{1}{4}(S_{a}^{\sigma}(\mathbf{x}_{1}+\mathbf{x}_{2}) - S_{a}^{\sigma}(\mathbf{x}_{1}-\mathbf{x}_{2}))$. We then let $\mathbf{e}_{j}$ denote the $j$th column of the identity matrix $I_{d}$, we let the $ij$th entry of $X$ be defined as $X_{ij} \coloneqq W(\mathbf{e}_{i},\mathbf{e}_{j})$ ($i,j = 1, \ldots , d$), and we will show that $X$ is symmetric and $S_{a}^{\sigma}(\mathbf{z}) = \mathbf{z}^{T}X\mathbf{z}$ for all $\mathbf{z} \in \mathbb{R}^{d}$. To prove this, we will show that, for any given $\mathbf{x}_{1}, \mathbf{x}_{2} \in \mathbb{R}^{d}$ and $\lambda \in \mathbb{R}$, then 
\begin{enumerate}[label=(\roman*)]
\item $W(\lambda \mathbf{x}_{1}, \mathbf{x}_{2}) = \lambda W(\mathbf{x}_{1}, \mathbf{x}_{2})$; and
\item $S_{a}^{\sigma}(\mathbf{x}_{1} + \mathbf{x}_{2}) + S_{a}^{\sigma}(\mathbf{x}_{1}-\mathbf{x}_{2}) = 2(S_{a}^{\sigma}(\mathbf{x}_{1}) + S_{a}^{\sigma}(\mathbf{x}_{2}))$.
\end{enumerate}
From \cite[Lemma 3]{MOLNNQF}, condition (ii) implies that, for any given $\mathbf{x}_{1}, \mathbf{x}_{2}, \mathbf{z} \in \mathbb{R}^{d}$, then (iia) $W(\mathbf{x}_{1}, \mathbf{x}_{2}) = W(\mathbf{x}_{2}, \mathbf{x}_{1})$; (iib) $W(\mathbf{x}_{1}+\mathbf{x}_{2}, \mathbf{z}) = W(\mathbf{x}_{1}, \mathbf{z}) + W(\mathbf{x}_{2}, \mathbf{z})$; and (iic) $S_{a}^{\sigma}(\mathbf{z}) = W(\mathbf{z}, \mathbf{z})$. Together with condition (i), we conclude that $W$ is a symmetric bilinear form, and $X$ is symmetric. We then let $\mathbf{z} \in \mathbb{R}^{d}$ and we denote the $i$th entry of $\mathbf{z}$ by $z_{i}$, and it follows that $S_{a}^{\sigma}(\mathbf{z}) = S_{a}^{\sigma}(\sum_{i = 1}^{d}z_{i}\mathbf{e}_{i}) = W(\sum_{i = 1}^{d}z_{i}\mathbf{e}_{i}, \sum_{j = 1}^{d}z_{j}\mathbf{e}_{j}) = \sum_{i = 1}^{d}\sum_{j = 1}^{d}z_{i}W(\mathbf{e}_{i},\mathbf{e}_{j})z_{j} = \sum_{i = 1}^{d}\sum_{j = 1}^{d}z_{i}X_{ij}z_{j} = \mathbf{z}^{T}X\mathbf{z}$.

It remains to show conditions (i) and (ii). We first show that, for any given $t_{1} \geq t_{0}$,
\begin{align}
&S_{a}^{\sigma}(\mathbf{x}_{0}) {=} \sup_{\mathbf{u} \in \mathcal{L}_{2}^{\text{loc}}\left(\mathbb{R}, \mathbb{R}^{n}\right), t_{2} \geq t_{1}} \smallint_{t_{0}}^{t_{2}}{-(\sigma(\mathbf{u},\mathbf{y}))(t) dt}, \nonumber \\ 
& \hspace{2.5cm} \text{such that } (\mathbf{u}, \mathbf{y}, \mathbf{x}) \in \mathcal{B}_{s}, \mathbf{x}(t_{0}) =  \mathbf{x}_{0}. \label{eq:saad1}
\end{align}
To see this, let $t_{1} \geq t_{0}$ and $(\mathbf{u}, \mathbf{y}, \mathbf{x}) \in \mathcal{B}_{s}$ with $\mathbf{x}(t_{0}) = \mathbf{x}_{0}$ satisfy $\smallint_{t_{0}}^{t_{1}}{-(\sigma(\mathbf{u},\mathbf{y}))(t) dt} = S_{a}^{\sigma}(\mathbf{x}_{0}) - \epsilon$ for some $\epsilon > 0$. Then, from the conditions in the lemma statement, there exist $t_{2} \geq t_{1}$ and $(\tilde{\mathbf{u}}, \tilde{\mathbf{y}}, \tilde{\mathbf{x}}) \in \mathcal{B}_{s}$ such that $\tilde{\mathbf{u}}(t) = \mathbf{u}(t)$, $\tilde{\mathbf{y}}(t) = \mathbf{y}(t)$, and $\tilde{\mathbf{x}}(t) = \mathbf{x}(t)$ for all $t_{0} \leq t \leq t_{1}$; and $\smallint_{t_{1}}^{t_{2}}{-(\sigma(\tilde{\mathbf{u}},\tilde{\mathbf{y}}))(t) dt} \geq 0$. It follows that $\smallint_{t_{0}}^{t}{-(\sigma(\tilde{\mathbf{u}},\tilde{\mathbf{y}}))(t) dt} \geq S_{a}^{\sigma}(\mathbf{x}_{0}) - \epsilon$ for all $t \geq t_{1}$. But $S_{a}^{\sigma}(\mathbf{x}_{0}) \geq \smallint_{t_{0}}^{t}{-(\sigma(\tilde{\mathbf{u}},\tilde{\mathbf{y}}))(t) dt}$, and $\epsilon > 0$ can be made arbitrarily small by choosing $t_{1}$ and $\mathbf{u}$. This proves (\ref{eq:saad1}).

To prove (i), we let $\mathbf{x}_{1}, \mathbf{x}_{2} \in \mathbb{R}^{d}$ and $\lambda \in \mathbb{R}$ be fixed but arbitrary, and we show that $S_{a}^{\sigma}(\lambda \mathbf{x}_{1} + \mathbf{x}_{2}) + \lambda S_{a}^{\sigma}(\mathbf{x}_{1} - \mathbf{x}_{2}) \leq S_{a}^{\sigma}(\lambda \mathbf{x}_{1} - \mathbf{x}_{2}) + \lambda S_{a}^{\sigma}(\mathbf{x}_{1} + \mathbf{x}_{2})$. To see this, suppose instead that there exists $\epsilon > 0$ such that
\begin{multline}
S_{a}^{\sigma}(\lambda \mathbf{x}_{1} {+} \mathbf{x}_{2}) {+} \lambda S_{a}^{\sigma}(\mathbf{x}_{1} {-} \mathbf{x}_{2}) \\ = S_{a}^{\sigma}(\lambda \mathbf{x}_{1} {-} \mathbf{x}_{2}) {+} \lambda S_{a}^{\sigma}(\mathbf{x}_{1} {+} \mathbf{x}_{2}) + \epsilon.\label{eq:sals0}
\end{multline}
There exist $(\mathbf{u}_{a}, \mathbf{y}_{a}, \mathbf{x}_{a}) \in \mathcal{B}_{s}$ with $\mathbf{x}_{a}(t_{0}) = \lambda \mathbf{x}_{1} + \mathbf{x}_{2}$, $(\mathbf{u}_{b}, \mathbf{y}_{b}, \mathbf{x}_{b}) \in \mathcal{B}_{s}$ with $\mathbf{x}_{b}(t_{0}) = \mathbf{x}_{1} - \mathbf{x}_{2}$, and $t_{1} \geq t_{0}$, with
\begin{multline}
S_{a}^{\sigma}(\lambda \mathbf{x}_{1} + \mathbf{x}_{2}) + \lambda S_{a}^{\sigma}(\mathbf{x}_{1} - \mathbf{x}_{2}) \\ \leq \smallint_{t_{0}}^{t_{1}}(-\sigma(\mathbf{u}_{a},\mathbf{y}_{a}) - \lambda(\sigma(\mathbf{u}_{b},\mathbf{y}_{b})))(t) + \epsilon/2.\label{eq:sals1}
\end{multline} 
Now, let $(\mathbf{u}_{c}, \mathbf{y}_{c}, \mathbf{x}_{c}) \coloneqq (\lambda-1)/(1+\lambda)(\mathbf{u}_{a}, \mathbf{y}_{a}, \mathbf{x}_{a}) + 2\lambda/(1+\lambda)(\mathbf{u}_{b}, \mathbf{y}_{b}, \mathbf{x}_{b})$ and $(\mathbf{u}_{d}, \mathbf{y}_{d}, \mathbf{x}_{d}) \coloneqq 2/(1+\lambda)(\mathbf{u}_{a}, \mathbf{y}_{a}, \mathbf{x}_{a}) + (1-\lambda)/(1+\lambda)(\mathbf{u}_{b}, \mathbf{y}_{b}, \mathbf{x}_{b})$. It can be verified that $\sigma(\mathbf{u}_{a},\mathbf{y}_{a}) + \lambda \sigma(\mathbf{u}_{b},\mathbf{y}_{b}) = \sigma(\mathbf{u}_{c},\mathbf{y}_{c}) + \lambda \sigma(\mathbf{u}_{d},\mathbf{y}_{d})$, $\mathbf{x}_{c}(t_{0}) = \lambda \mathbf{x}_{1} - \mathbf{x}_{2}$ and $\mathbf{x}_{d}(t_{0}) = \mathbf{x}_{1} + \mathbf{x}_{2}$. It follows from (\ref{eq:sals0})--(\ref{eq:sals1}) that
\begin{align*}
&S_{a}^{\sigma}(\lambda \mathbf{x}_{1} + \mathbf{x}_{2}) + \lambda S_{a}^{\sigma}(\mathbf{x}_{1} - \mathbf{x}_{2}) \\
&\hspace{1.5cm} \leq \smallint_{t_{0}}^{t_{1}}(-\sigma(\mathbf{u}_{c},\mathbf{y}_{c}) - \lambda(\sigma(\mathbf{u}_{d},\mathbf{y}_{d})))(t) + \epsilon/2 \\
&\hspace{1.5cm} \leq S_{a}^{\sigma}(\lambda \mathbf{x}_{1} {-} \mathbf{x}_{2}) {+} \lambda S_{a}^{\sigma}(\mathbf{x}_{1} {+} \mathbf{x}_{2}) + \epsilon/2 \\
&\hspace{1.5cm} = S_{a}^{\sigma}(\lambda \mathbf{x}_{1} + \mathbf{x}_{2}) + \lambda S_{a}^{\sigma}(\mathbf{x}_{1} - \mathbf{x}_{2}) - \epsilon/2,
\end{align*}
a contradiction. Substituting $-\mathbf{x}_{2}$ for $\mathbf{x}_{2}$ in the above argument gives $S_{a}^{\sigma}(\lambda \mathbf{x}_{1} {-} \mathbf{x}_{2}) {+} \lambda S_{a}^{\sigma}(\mathbf{x}_{1} {+} \mathbf{x}_{2}) \leq S_{a}^{\sigma}(\lambda \mathbf{x}_{1} {+} \mathbf{x}_{2}) {+} \lambda S_{a}^{\sigma}(\mathbf{x}_{1} {-} \mathbf{x}_{2})$, and completes the proof of (i).

To see (ii), suppose instead that there exists $\epsilon > 0$ such that
\begin{equation*}
S_{a}^{\sigma}(\mathbf{x}_{1} {+} \mathbf{x}_{2}) {+} S_{a}^{\sigma}(\mathbf{x}_{1}{-}\mathbf{x}_{2}) {+} \epsilon = 2(S_{a}^{\sigma}(\mathbf{x}_{1}) {+} S_{a}^{\sigma}(\mathbf{x}_{2})).
\end{equation*}
Let $t_{1} \geq t_{0}$ and $(\mathbf{u}_{a}, \mathbf{y}_{a}, \mathbf{x}_{a}), (\mathbf{u}_{b}, \mathbf{y}_{b}, \mathbf{x}_{b}) \in \mathcal{B}_{s}$ with $\mathbf{x}_{a}(t_{0}) {=}  \mathbf{x}_{1}$ and $\mathbf{x}_{b}(t_{0}) {=}  \mathbf{x}_{2}$ be such that
\begin{align*}
& \displaystyle \smallint_{t_{0}}^{t_{1}}{-(\sigma(\mathbf{u}_{a},\mathbf{y}_{a}))(t) dt} + \tfrac{\epsilon}{8} > S_{a}^{\sigma}(\mathbf{x}_{1}), \text{ and} \\
& \smallint_{t_{0}}^{t_{1}}{-(\sigma(\mathbf{u}_{b},\mathbf{y}_{b}))(t) dt} + \tfrac{\epsilon}{8} > S_{a}^{\sigma}(\mathbf{x}_{2}).
\end{align*}
Similar to \cite[p.\ 796]{MOLNNQF}, we let $(\mathbf{\tilde{u}}_{a}, \mathbf{\tilde{y}}_{a}, \mathbf{\tilde{x}}_{a}) = (\mathbf{u}_{a},\mathbf{y}_{a},\mathbf{x}_{a})+(\mathbf{u}_{b},\mathbf{y}_{b},\mathbf{x}_{b})$ and  $(\mathbf{\tilde{u}}_{b}, \mathbf{\tilde{y}}_{b}, \mathbf{\tilde{x}}_{b}) = (\mathbf{u}_{a},\mathbf{y}_{a},\mathbf{x}_{a})-(\mathbf{u}_{b},\mathbf{y}_{b},\mathbf{x}_{b})$. Then $(\mathbf{\tilde{u}}_{a}, \mathbf{\tilde{y}}_{a}, \mathbf{\tilde{x}}_{a}), (\mathbf{\tilde{u}}_{b}, \mathbf{\tilde{y}}_{b}, \mathbf{\tilde{x}}_{b}) \in \mathcal{B}_{s}$, $\mathbf{\tilde{x}}_{a}(t_{0}) {=}  \mathbf{x}_{1} {+} \mathbf{x}_{2}$, and $\mathbf{\tilde{x}}_{b}(t_{0}) {=}  \mathbf{x}_{1} {-} \mathbf{x}_{2}$, whence 
\begin{align*}
&\hspace{-0.3cm} S_{a}^{\sigma}(\mathbf{x}_{1} {+} \mathbf{x}_{2}) {+} S_{a}^{\sigma}(\mathbf{x}_{1}{-}\mathbf{x}_{2}) \\
&\geq \smallint_{t_{0}}^{t_{1}}{{-}(\sigma(\mathbf{\tilde{u}}_{a},\mathbf{\tilde{y}}_{a}))(t) dt} + \smallint_{t_{0}}^{t_{1}}{{-}(\sigma(\mathbf{\tilde{u}}_{b},\mathbf{\tilde{y}}_{b}))(t) dt} \\
&= 2(\smallint_{t_{0}}^{t_{1}}{{-}(\sigma(\mathbf{u}_{a},\mathbf{y}_{a}))(t) dt} + \smallint_{t_{0}}^{t_{1}}{{-}(\sigma(\mathbf{u}_{b},\mathbf{y}_{b}))(t) dt}) \\
& > 2(S_{a}^{\sigma}(\mathbf{x}_{1}) {+} S_{a}^{\sigma}(\mathbf{x}_{2}) {-} \tfrac{\epsilon}{4}) = S_{a}^{\sigma}(\mathbf{x}_{1} {+} \mathbf{x}_{2}) {+} S_{a}^{\sigma}(\mathbf{x}_{1}{-}\mathbf{x}_{2}) {+} \tfrac{\epsilon}{2},
\end{align*}
a contradiction. Thus, $S_{a}^{\sigma}(\mathbf{x}_{1} {+} \mathbf{x}_{2}) {+} S_{a}^{\sigma}(\mathbf{x}_{1} {-} \mathbf{x}_{2}) \leq 2(S_{a}^{\sigma}(\mathbf{x}_{1}) {+} S_{a}^{\sigma}(\mathbf{x}_{2}))$. A similar argument shows that $S_{a}^{\sigma}(\mathbf{x}_{1} {+} \mathbf{x}_{2}) {+} S_{a}^{\sigma}(\mathbf{x}_{1} {-} \mathbf{x}_{2}) \geq 2(S_{a}^{\sigma}(\mathbf{x}_{1}) {+} S_{a}^{\sigma}(\mathbf{x}_{2}))$, and completes the proof of (ii).
\end{IEEEproof}

We next consider a related optimal control problem concerning the \emph{observer staircase form} in note \ref{lem:osf}: 

\begin{lemapp}
\label{lem:saop}
Let $\mathcal{B}_{s}$ be as in (\ref{eq:sssg}); let $\sigma$ and $S_{a}^{\sigma}$ be as in Definition \ref{def:sagd}; let $S_{a}^{\sigma}(\mathbf{x}_{0}) < \infty$ for all $\mathbf{x}_{0} \in \mathbb{R}^{d}$; let $T_{1}$ and $\hat{\mathcal{B}}_{s}$ be as in note \ref{lem:osf}; and let
\begin{align}
\label{eq:hsad}
&\hat{S}_{a}^{\sigma}(\hat{\mathbf{x}}_{0}) \coloneqq \sup_{t_{1} \geq t_{0}, \mathbf{u} \in \mathcal{L}_{2}^{\text{loc}}\left(\mathbb{R}, \mathbb{R}^{n}\right)} \smallint_{t_{0}}^{t_{1}}{-(\sigma(\mathbf{u},\mathbf{y}))(t) dt}, \nonumber \\
&\hspace{2.4cm} \text{such that } (\mathbf{u}, \mathbf{y}, \mathbf{\hat{x}}) \in \hat{\mathcal{B}}_{s}, \mathbf{\hat{x}}(t_{0}) =  \hat{\mathbf{x}}_{0}.
\end{align}
Then $S_{a}^{\sigma}(\mathbf{x}_{0}) = \hat{S}_{a}^{\sigma}(T_{1}\mathbf{x}_{0})$ for all $\mathbf{x}_{0} \in \mathbb{R}^{d}$. In particular, with $V_{o}$ as in (\ref{eq:vo}), then  $\mathbf{z} \in \mathbb{R}^{d}$ and $V_{o}\mathbf{z} = 0$ imply $S_{a}^{\sigma}(\mathbf{z}) = 0$. 
\end{lemapp}

\begin{IEEEproof}
Let $T = \text{col}\begin{pmatrix}T_{1}& T_{2}\end{pmatrix}$ be as in note \ref{lem:osf}. It can be shown from the variation of the constants formula (\ref{eq:xvocf})--(\ref{eq:yvocf}) that (i) if $(\mathbf{u}, \mathbf{y}, \mathbf{x}) \in \mathcal{B}_{s}$ satisfies $\mathbf{x}(t_{0}) = \mathbf{x}_{0}$, then there exists $(\mathbf{u}, \mathbf{y}, \hat{\mathbf{x}}) \in \hat{\mathcal{B}}_{s}$ with $\hat{\mathbf{x}}(t_{0}) = T_{1}\mathbf{x}_{0}$; and (ii) if $(\mathbf{u}, \mathbf{y}, \hat{\mathbf{x}}) \in \hat{\mathcal{B}}_{s}$ satisfies  $\hat{\mathbf{x}}(t_{0}) = \hat{\mathbf{x}}_{0}$, and $\hat{\mathbf{x}}_{1} \in \mathbb{R}^{d - \hat{d}}$, then there exists $(\mathbf{u}, \mathbf{y}, \mathbf{x}) \in \mathcal{B}_{s}$ with $\mathbf{x}(t_{0}) = T^{-1}\text{col}\begin{pmatrix}\hat{\mathbf{x}}_{0}& \hat{\mathbf{x}}_{1}\end{pmatrix}$. Now, consider a fixed but arbitrary $\mathbf{x}_{0} \in \mathbb{R}^{d}$. It follows from (i) that
\begin{align*}
&S_{a}^{\sigma}(\mathbf{x}_{0}) \leq \sup_{t_{1} \geq t_{0}, \mathbf{u} \in \mathcal{L}_{2}^{\text{loc}}\left(\mathbb{R}, \mathbb{R}^{n}\right)}\smallint_{t_{0}}^{t_{1}}{-(\sigma(\mathbf{u},\mathbf{y}))(t) dt}, \\
&\hspace{2.5cm} \text{such that } (\mathbf{u}, \mathbf{y}, \mathbf{\hat{x}}) \in \hat{\mathcal{B}}_{s}, \mathbf{\hat{x}}(t_{0}) =  T_{1}\mathbf{x}_{0},
\end{align*}
i.e., $S_{a}^{\sigma}(\mathbf{x}_{0}) \leq \hat{S}_{a}^{\sigma}(T_{1}\mathbf{x}_{0})$. Similarly, from (ii), it can be shown that $S_{a}^{\sigma}(\mathbf{x}_{0}) \geq \hat{S}_{a}^{\sigma}(T_{1}\mathbf{x}_{0})$, so $S_{a}^{\sigma}(\mathbf{x}_{0}) = \hat{S}_{a}^{\sigma}(T_{1}\mathbf{x}_{0})$. Finally, if $V_{o}\mathbf{z} = 0$, then it can be shown that $T_{1}\mathbf{z} = 0$.  As $\hat{S}_{a}^{\sigma}$ is a storage function by Lemma \ref{lem:avstor}, then $S_{a}^{\sigma}(\mathbf{z}) = \hat{S}_{a}^{\sigma}(0) = 0$.
\end{IEEEproof}

\section{Positive-real and bounded-real pairs}
\label{app:prbrp}
Here, we provide several results relating to the new concepts of positive-real and bounded-real pairs. 
\begin{remunerate}
\labitem{\ref{app:prbrp}{.}\arabic{muni}}{nl:pprbrp0} 
Let $P \in \mathbb{R}^{m \times n}[\xi]$ and $Q \in \mathbb{R}^{m \times m}[\xi]$; and let
\begin{align}
J_{n} &\coloneqq \frac{1}{2}\begin{bmatrix}0& I_{n}\\ I_{n} & 0\end{bmatrix}, \hspace{0.15cm} \Sigma_{m,n} \coloneqq \begin{bmatrix}I_{n} & 0\\ 0&  -I_{m}\end{bmatrix}, \label{eq:jnsn} \\
\Psi(\eta, \xi) &\coloneqq \begin{bmatrix}P& -Q\end{bmatrix}(\eta)\Sigma_{m,n}\begin{bmatrix}P& -Q\end{bmatrix}(\xi)^{T}, \text{ and} \\
\Phi(\eta, \xi) &\coloneqq \begin{bmatrix}P& -Q\end{bmatrix}(\eta)J_{n}\begin{bmatrix}P& -Q\end{bmatrix}(\xi)^{T} \text{ if } m=n.
\end{align}
Then $(P,Q)$ is a positive-real pair (resp., bounded-real pair) if and only if (i) $\Phi(\lambda, \bar{\lambda}) \leq 0$ (resp., $\Psi(\lambda, \bar{\lambda}) \leq 0$) for all $\lambda \in \mathbb{C}_{+}$; (ii) $\text{rank}(\begin{bmatrix}P& -Q\end{bmatrix}(\lambda)) = n$ for all $\lambda \in \overbar{\mathbb{C}}_{+}$; and (iii) if $\mathbf{p} \in \mathbb{R}^{n}[\xi]$ and $\lambda \in \mathbb{C}$ satisfy $\mathbf{p}(\xi)^{T}\Phi(\xi, -\xi) = 0$ (resp., $\mathbf{p}(\xi)^{T}\Psi(\xi, -\xi) = 0$) and $\mathbf{p}(\lambda)^{T}\begin{bmatrix}P& -Q\end{bmatrix}(\lambda) = 0$, then $\mathbf{p}(\lambda) = 0$.
\labitem{\ref{app:prbrp}{.}\arabic{muni}}{nl:pprbrp1} Let $P, Q \in \mathbb{R}^{n \times n}[\xi]$; let $J_{n}$ be as in (\ref{eq:jnsn}); let $Y \in \mathbb{R}^{n \times n}[\xi]$ and $S \in \mathbb{R}^{2n \times 2n}$ be nonsingular with $S J_{n} S^{T} = J_{n}$; and let $\hat{P}, \hat{Q} \in \mathbb{R}^{n \times n}[\xi]$ satisfy $\begin{bmatrix}\hat{P}& -\hat{Q}\end{bmatrix} \coloneqq Y\begin{bmatrix}P& -Q\end{bmatrix}S$. Then $(P,Q)$ is a positive-real pair if and only if $(\hat{P}, \hat{Q})$ is a positive-real pair (this follows from note \ref{nl:pprbrp0}).
\labitem{\ref{app:prbrp}{.}\arabic{muni}}{nl:pprbrp4} Let $P, Q \in \mathbb{R}^{n \times n}[\xi]$; let $J_{n}$ and $\Sigma_{n,n}$ be as in (\ref{eq:jnsn}); let $Y \in \mathbb{R}^{n \times n}[\xi]$ and $S \in \mathbb{R}^{2n \times 2n}$ be nonsingular with $S J_{n} S^{T} = \Sigma_{n,n}$; and let $\hat{P}, \hat{Q} \in \mathbb{R}^{n \times n}[\xi]$ satisfy $\begin{bmatrix}\hat{P}& -\hat{Q}\end{bmatrix} \coloneqq Y\begin{bmatrix}P& -Q\end{bmatrix}S$. Then $(P,Q)$ is a bounded-real pair if and only if $(\hat{P}, \hat{Q})$ is a positive-real pair  (this follows from note \ref{nl:pprbrp0}).
\labitem{\ref{app:prbrp}{.}\arabic{muni}}{nl:pprbrp5} Let $\Sigma \in \mathbb{R}^{n \times n}$ be a signature matrix (i.e., $\Sigma$ is diagonal with diagonal entries $\pm 1$), let $P, Q \in \mathbb{R}^{n \times n}[\xi]$, and let $\hat{Q} \coloneqq \tfrac{1}{2}(Q - P\Sigma)$ and $\hat{P} \coloneqq \tfrac{1}{2}(P\Sigma + Q)$. Then $(P,Q)$ is a bounded-real pair if and only if $(\hat{P}, \hat{Q})$ is a positive-real pair (this follows from note \ref{nl:pprbrp4}). 
Also, if $P, Q \in \mathbb{R}^{n \times n}[\xi]$ and $Q^{-1}P$ is proper, then there necessarily exists a signature matrix $\Sigma$ and matrices $\hat{Q} \coloneqq \tfrac{1}{2}(Q {-} P\Sigma)$ and $\hat{P} \coloneqq \tfrac{1}{2}(P\Sigma {+} Q)$ such that $\hat{Q}$ is nonsingular and $\hat{Q}^{-1}\hat{P}$ is proper. To obtain such matrices $\Sigma, \hat{P}$ and $\hat{Q}$, we let $\tilde{P} \coloneqq \tfrac{1}{2}(P + Q)$ and $\tilde{Q} \coloneqq \tfrac{1}{2}(Q - P)$, so $P = \tilde{P}-\tilde{Q}$ and $Q = \tilde{P} + \tilde{Q}$. We then let $S_{1}$ and $S_{2} \in \mathbb{R}^{n \times n}$ be matrices that select columns from $\tilde{P}$ and $\tilde{Q}$ to achieve the maximal determinantal degree. I.e., (i) $S_{1}$ and $S_{2}$ are diagonal matrices with all entries either $0$ or $1$; (ii) $S_{1} + S_{2} = I$; and (iii) $\text{deg}(\det{(\tilde{P}S_{1} + \tilde{Q}S_{2})})$ takes its maximum value among all matrices $S_{1}$ and $S_{2}$ that satisfy (i) and (ii). We then let $\hat{P} \coloneqq \tilde{P}S_{2} + \tilde{Q}S_{1}$, $\hat{Q} \coloneqq \tilde{P}S_{1} + \tilde{Q}S_{2}$, and $\Sigma \coloneqq S_{2}-S_{1}$, so $\Sigma$ is a signature matrix. The method in \cite[Proof of Theorem 9]{THTPLSNA} then proves that $\hat{Q}^{-1}\hat{P}$ is proper.
\end{remunerate}

\section{Explicit characterisation of the available energy: supplementary lemmas}
In this final appendix, we present four supplementary lemmas used in the proof of Theorem \ref{thm:pbtnsc}.
\label{sec:ecaesl}
\begin{lemapp}
\label{lem:sfymd}
Let $\mathcal{B}_{s}$ and $H$ be as in (\ref{eq:sssg}) and (\ref{eq:gd}) with $m = n$; let $\text{spec}(A) \in \overbar{\mathbb{C}}_{-}$; let $X_{-}, L$ and $W$ be real matrices that satisfy condition \ref{nl:pts2c2}(iiia) of Theorem \ref{thm:pbtnsc}; and let
\begin{equation*}
Z(\xi)\coloneqq W+L(\xi I - A)^{-1}B \text{ and }Y(\xi) \coloneqq \begin{bmatrix}\xi I - A& -B\\ L& W\end{bmatrix}.
\end{equation*}
Then $Z^{\star}Z {=} H {+} H^{\star}$, and $Z$ is a spectral factor for $H + H^{\star}$ if and only if $Y(\lambda)$ has full row rank for all $\lambda \in \mathbb{C}_{+}$.
\end{lemapp}

\begin{IEEEproof}
That $Z^{\star}Z = H+H^{\star}$ follows by pre-multiplying $\Omega(X)$ in (\ref{eq:lmipb}) by $\begin{bmatrix}B^{T}(-\xi I - A^{T})^{-1}& I\end{bmatrix}$ and post-multiplying by $\text{col}\begin{pmatrix}(\xi I - A)^{-1}B& I\end{pmatrix}$. Since $\text{spec}(A) \in \overbar{\mathbb{C}}_{-}$, then $Z$ is analytic in $\mathbb{C}_{+}$. Finally, consider a fixed but arbitrary $\lambda \in \mathbb{C}_{+}$, so $\lambda I - A$ is nonsingular. It remains to show that $Z(\lambda)$ has full row rank if and only if $Y(\lambda)$ does. This follows from
\begin{equation*}
\begin{bmatrix}0& \lambda I - A\\ Z(\lambda)& L\end{bmatrix} = \begin{bmatrix}\lambda I - A& -B\\ L& W\end{bmatrix}\begin{bmatrix}(\lambda I - A)^{-1}B& I\\ I & 0\end{bmatrix},
\end{equation*}
since the rightmost matrix in this equation is nonsingular.
\end{IEEEproof}

The final three lemmas relate to the decomposition in case (ii) in the proof of Theorem \ref{thm:pbtnsc}. We refer back to that proof for definitions of conditions \ref{nl:ip1}--\ref{nl:ip4} and \ref{nl:csn1}--\ref{nl:csn3}.
\begin{lemapp}
\label{lem:gprl1}
Let $P_{k{-}1}, Q_{k{-}1}$ satisfy \ref{nl:ip1} for $i = k{-}1$, and let $D_{k{-}1} \coloneqq \lim_{\xi \rightarrow \infty}(Q_{k{-}1}^{-1}P_{k{-}1}(\xi))$. The following hold.
\begin{enumerate}[label=\arabic*., ref=\arabic*, leftmargin=0.5cm]
\item Let $P_{k} \coloneqq P_{k{-}1} - \tfrac{1}{2}Q_{k{-}1}(D_{k{-}1}-D_{k{-}1}^{T})$ and $Q_{k} \coloneqq Q_{k{-}1}$. Then \ref{nl:ip1} and \ref{nl:ip2} hold for $i = k$.\label{nl:ms1}
\item Let $A_{k}, B_{k}, C_{k}, D_{k}$ satisfy \ref{nl:csn1} for $i {=} k$; and let $A_{k{-}1}{\coloneqq} A_{k}$, $B_{k{-}1}{\coloneqq}B_{k}$, and $C_{k{-}1}{\coloneqq} C_{k}$.
Then:\label{nl:msa2}
\begin{enumerate}
\item \ref{nl:csn1} holds for $i {=} k{-}1$. \label{nl:ms2}
\item Let $X_{k{-}1}, L_{k{-}1}, W_{k{-}1}, X_{k}, L_{k}$, and $W_{k}$ be real matrices with $X_{k} = X_{k{-}1} \geq 0$, $L_{k}=L_{k{-}1}$, and $W_{k}=W_{k{-}1}$. Then (i) \ref{nl:csn2} holds for $i = k{-}1$ if and only if \ref{nl:csn2} holds for $i = k$; and (ii) \ref{nl:csn3} holds for $i = k{-}1$ if and only if \ref{nl:csn3} holds for $i = k$.\label{nl:ms3}
\end{enumerate}
\end{enumerate}
\end{lemapp}

\begin{IEEEproof}
{\bf Condition \ref{nl:ms1}.} \hspace{0.3cm} Clearly, $Q_{k}^{-1}P_{k} = Q_{k{-}1}^{-1}P_{k{-}1} - \tfrac{1}{2}(D_{k{-}1}{-}D_{k{-}1}^{T})$, so $Q_{k}^{-1}P_{k}$ is proper and $\lim_{\xi \rightarrow \infty}(Q_{k}^{-1}P_{k}(\xi)) = \lim_{\xi \rightarrow \infty}(Q_{k{-}1}^{-1}P_{k{-}1}(\xi)) - \tfrac{1}{2}(D_{k{-}1}{-}D_{k{-}1}^{T}) = \tfrac{1}{2}(D_{k{-}1}{+}D_{k{-}1}^{T})$. Next, let
\begin{equation*}
S = \begin{bmatrix}I & 0\\ \tfrac{1}{2}(D_{k{-}1}{-}D_{k{-}1}^{T})& I\end{bmatrix}. 
\end{equation*}
Then $\begin{bmatrix}P_{k}& {-}Q_{k}\end{bmatrix} = \begin{bmatrix}P_{k{-}1}& {-}Q_{k{-}1}\end{bmatrix}S$, and it follows from note \ref{nl:pprbrp1} that $(P_{k}, Q_{k})$ is a positive-real pair. 

{\bf Condition \ref{nl:msa2}.} \hspace{0.3cm} Let $\mathcal{A}_{k}, M_{k}, N_{k}, U_{k}, V_{k}, E_{k}, F_{k}$ be as in \ref{nl:csn1} for the case $i=k$; and let $E_{k{-}1} \coloneqq E_{k} - \tfrac{1}{2}F_{k}(D_{k{-}1} - D_{k{-}1}^{T})$, $F_{k{-}1} \coloneqq F_{k}$, $M_{k{-}1} \coloneqq M_{k}$, $N_{k{-}1}\coloneqq N_{k}$, $U_{k{-}1} \coloneqq U_{k}$, and $V_{k{-}1} \coloneqq V_{k}$. By post-multiplying both sides of the relationship in \ref{nl:csn1} for the case $i{=}k$ by $\text{diag}\begin{pmatrix}S^{-1}& I\end{pmatrix}$, we find that \ref{nl:csn1} holds for $i{=}k{-}1$. Finally, condition \ref{nl:ms3} follows since $D_{k}+D_{k}^{T} = D_{k{-}1}+D_{k{-}1}^{T}$. Thus, with $\Omega_{i}(X_{i})$ as in \ref{nl:csn2}, then $\Omega_{k}(X_{k}) = \Omega_{k{-}1}(X_{k{-}1})$.
\end{IEEEproof}

\begin{lemapp}
\label{lem:gprl2}
Let $P_{k{-}1}, Q_{k{-}1}$ satisfy \ref{nl:ip1}--\ref{nl:ip2} for $i {=} k{-}1$, and let $n_{k} \coloneqq \text{normalrank}(P_{k{-}1})$, $m_{k} \coloneqq  n_{k{-}1} - n_{k}$, and $r_{k} {\coloneqq} \text{rank}(D_{k{-}1})$. The following hold.
\begin{enumerate}[label=\arabic*., ref=\arabic*, leftmargin=0.5cm]
\item There exists a nonsingular $T \in \mathbb{R}^{n_{k{-}1} \times n_{k{-}1}}$; unimodular $Y \in \mathbb{R}^{n_{k{-}1} \times n_{k{-}1}}[\xi]$ and $\tilde{Q}_{22} \in \mathbb{R}^{m_{k} {\times} m_{k}}[\xi]$; $\tilde{Q}_{12} \in \mathbb{R}^{n_{k} {\times} m_{k}}[\xi]$; and $P_{k}, Q_{k}$ satisfying \ref{nl:ip1}--\ref{nl:ip3} for $i {=} k$, with\label{nl:fns1}
\begin{equation}
\hspace*{-0.3cm} YP_{k{-}1}T {=} \begin{bmatrix}P_{k}& 0 \\0 & 0\end{bmatrix}, YQ_{k{-}1}(T^{-1})^{T} {=} \begin{bmatrix}Q_{k}& \tilde{Q}_{12}\\0& \tilde{Q}_{22}\end{bmatrix}.\label{eq:tmfoz}
\end{equation}
\item Let $A_{k}, B_{k}, C_{k}, D_{k}$ satisfy \ref{nl:csn1} for $i {=} k$; and $A_{k{-}1} {\coloneqq} A_{k}$, $B_{k{-}1} {\coloneqq} \begin{bmatrix}B_{k}& 0\end{bmatrix}T^{-1}$, $C_{k{-}1} {\coloneqq} (T^{-1})^{T}\text{col}\begin{pmatrix}C_{k} & 0\end{pmatrix}$, and $D_{k{-}1} {\coloneqq} (T^{-1})^{T}\text{diag}\begin{pmatrix}D_{k}& 0\end{pmatrix}T^{-1}$. Then:\label{nl:fnsa2}
\begin{enumerate}
\item  \ref{nl:csn1} holds for $i=k{-}1$.\label{nl:fns2}
\item Let $X_{k{-}1}, L_{k{-}1}$ and $W_{k{-}1}$ satisfy \ref{nl:csn2} for $i = k{-}1$; partition $T$ as $T \eqqcolon \begin{bmatrix}T_{1}& T_{2}\end{bmatrix}$ with $T_{1} \in \mathbb{R}^{n_{k{-}1} \times n_{k}}$ and $T_{2} \in \mathbb{R}^{n_{k{-}1} \times m_{k}}$; and let $X_{k} \coloneqq X_{k{-}1}, L_{k} \coloneqq L_{k{-}1}$, and $W_{k} \coloneqq W_{k{-}1}T_{1}$. Then (i) $W_{k{-}1}T_{2} = 0$; (ii) \ref{nl:csn2} holds for $i = k$; and (iii) if \ref{nl:csn3} holds for $i = k{-}1$, then \ref{nl:csn3} holds for $i = k$.\label{nl:fns3}
\item Let $X_{k}, L_{k}$ and $W_{k}$ satisfy \ref{nl:csn2} for $i = k$; and let $X_{k{-}1} \coloneqq X_{k}, L_{k{-}1} \coloneqq L_{k}$, and $W_{k{-}1} \coloneqq \begin{bmatrix}W_{k}& 0\end{bmatrix}T^{-1}$. Then (i) \ref{nl:csn2} holds for $i = k{-}1$; and (ii) if \ref{nl:csn3} holds for $i = k$, then \ref{nl:csn3} holds for $i = k{-}1$.\label{nl:fns4}
\end{enumerate}
\end{enumerate}
\end{lemapp}

\begin{IEEEproof}
{\bf Condition \ref{nl:fns1}.} \hspace{0.3cm} Since $(P_{k{-}1}, Q_{k{-}1})$ is a positive-real pair and $Q_{k{-}1}$ is nonsingular, then $H \coloneqq Q_{k{-}1}^{-1}P_{k{-}1}$ is positive-real (see Remark \ref{rem:pmbs}). Since, in addition, $D_{k{-}1}$ is symmetric, then $D_{k{-}1} \geq 0$ by Theorem \ref{thm:pbtsc}, so there exists $T_{1a} \in \mathbb{R}^{n_{k{-}1} \times r_{k}}$ such that $T_{1a}^{T}D_{k{-}1}T_{1a} = I_{r_{k}}$ by Sylvester's law of inertia (as $\text{rank}(D_{k{-}1}) = r_{k}$). Now, let the columns of $T_{2} \in \mathbb{R}^{n_{k{-}1} \times m_{k}}$ be a basis for the nullspace of $H$ (i.e., $T_{2}$ has full column rank and $HT_{2} = 0$). Then, since the nullspace of $H$ is contained in the nullspace of $D_{k{-}1}$, there exists $T_{1b} \in \mathbb{R}^{n_{k{-}1} {\times} (n_{k}{-}r_{k})}$ such that the columns of $\begin{bmatrix}T_{1b}& T_{2}\end{bmatrix}$ are a basis for the nullspace of $D_{k{-}1}$. With $T_{1} = \begin{bmatrix}T_{1a}& T_{1b}\end{bmatrix}$, then $T = \begin{bmatrix}T_{1}& T_{2}\end{bmatrix}$ is nonsingular and $T^{T}D_{k{-}1}T = \text{diag}\begin{pmatrix}I_{r_{k}} & 0\end{pmatrix}$. Also, from \cite[Theorem 8.4.1]{AndVong}, $T^{T}HT = \text{diag}\begin{pmatrix}\hat{H} & 0\end{pmatrix}$, where $\hat{H} \in \mathbb{R}^{n_{k} \times n_{k}}(\xi)$ is positive-real and nonsingular.

From \cite[Theorem B.1.1]{JWIMTSC}, there exists a unimodular $Y \in \mathbb{R}^{n_{k{-}1} \times n_{k{-}1}}[\xi]$ such that $\tilde{Q} \coloneqq YQ_{k{-}1}(T^{-1})^{T}$ is upper triangular. Let $\tilde{P} \coloneqq YP_{k{-}1}T$, and note that $\tilde{Q}$ is nonsingular with $\tilde{Q}^{-1}\tilde{P} = T^{T}HT = \text{diag}\begin{pmatrix}\hat{H} & 0\end{pmatrix}$. Since $\tilde{P} = \tilde{Q}\text{diag}\begin{pmatrix}\hat{H} & 0\end{pmatrix}$, then $\tilde{P}$ and $\tilde{Q}$ (partitioned compatibly with $\text{diag}\begin{pmatrix}\hat{H} & 0\end{pmatrix}$) take the form indicated in (\ref{eq:tmfoz}). To show that $\tilde{Q}_{22}$ in (\ref{eq:tmfoz}) is unimodular, we let $\lambda \in \mathbb{C}$ and $\hat{\mathbf{p}} \in \mathbb{R}^{m_{k}}[\xi]$ satisfy $\hat{\mathbf{p}}(\lambda)^{T}\tilde{Q}_{22}(\lambda) = 0$, and $\mathbf{p}^{T} \coloneqq \begin{bmatrix}0& \mathbf{\hat{p}}^{T}\end{bmatrix}Y$. It can be verified that $\mathbf{p}^{T}(P_{k{-}1}Q_{k{-}1}^{\star} + Q_{k{-}1}P_{k{-}1}^{\star}) = 0$ and $\mathbf{p}(\lambda)^{T}\begin{bmatrix}P_{k{-}1}& -Q_{k{-}1}\end{bmatrix}(\lambda) = 0$. Since $(P_{k{-}1},Q_{k{-}1})$ is a positive-real pair, this implies $\mathbf{p}(\lambda) = 0$. Since, in addition, $Y$ is unimodular, then $\hat{\mathbf{p}}(\lambda) = 0$, and it follows that $\tilde{Q}_{22}$ is unimodular. It is then easily shown from notes \ref{nl:pprbrp0} and \ref{nl:pprbrp1} that $(P_{k}, Q_{k})$ is a positive-real pair. Moreover, $T^{T}HT = T^{T}Q_{k{-}1}^{-1}P_{k{-}1}T = \text{diag}\begin{pmatrix}Q_{k}^{-1}P_{k}& 0\end{pmatrix} = \text{diag}\begin{pmatrix}\hat{H}& 0\end{pmatrix}$ where $\hat{H}$ is nonsingular and $\lim_{\xi \rightarrow \infty}(\hat{H}(\xi)) = T^{T}D_{k{-}1}T = \text{diag}\begin{pmatrix}I_{r_{k}}& 0\end{pmatrix}$. Thus, $P_{k}, Q_{k}$ satisfy \ref{nl:ip1}--\ref{nl:ip3} for $i = k$. 

{\bf Condition \ref{nl:fnsa2}.} \hspace{0.15cm} Let $\mathcal{A}_{k}, M_{k}, N_{k}, U_{k}, V_{k}, E_{k}, F_{k}$ be as in \ref{nl:csn1} for the case $i=k$; and let
\begin{equation*}
\left[\!\begin{smallmatrix}M_{k{-}1} & N_{k{-}1}\\ U_{k{-}1} & V_{k{-}1}\end{smallmatrix}\!\right] {\coloneqq} \left[\!\begin{smallmatrix}Y^{-1}& 0\\ 0& I\end{smallmatrix}\!\right]\! \left[\!\begin{smallarray}{cc|c}M_{k}& \tilde{Q}_{12} & N_{k} \\ 0 & \tilde{Q}_{22} & 0 \\ \hline \rule{0pt}{1.1\normalbaselineskip}
 U_{k} & 0 & V_{k}\end{smallarray}\!\right]\! \left[\!\begin{smallmatrix}T^{T}& 0\\ 0& I\end{smallmatrix}\!\right].
\end{equation*}
It can be verified that each of these four matrices is unimodular. Also, with $\mathcal{A}_{i}$ as in \ref{nl:csn1} for $i = k-1$ and $i=k$, then
\begin{equation*}
\!\left[\!\begin{smallmatrix}T^{T}& 0\\0& I\end{smallmatrix}\!\right]\! \!\left[\!\begin{smallmatrix}-D_{k{-}1}& I& -C_{k{-}1}\\ -B_{k{-}1}& 0& \mathcal{A}_{k{-}1}\end{smallmatrix}\!\right]\! = \!\left[\!\begin{smallarray}{cc|c|c}\begin{smallarray}{c}-D_{k}\\0\end{smallarray}& \begin{smallarray}{c}0\\0\end{smallarray}& I & \begin{smallarray}{c}-C_{k}\\ 0\end{smallarray}\\ \hline \rule{0pt}{1.05\normalbaselineskip} -B_{k}& 0& 0& \mathcal{A}_{k}\end{smallarray}\!\right]\! \!\left[\!\begin{smallmatrix}T^{-1}& 0& 0\\ 0& T^{T}& 0\\ 0& 0& I\end{smallmatrix}\!\right]\!.
\end{equation*}
Thus, with $E_{k{-}1} \coloneqq \begin{bmatrix}E_{k}& 0\end{bmatrix}T^{-1}$ and $F_{k{-}1} \coloneqq \begin{bmatrix}F_{k}& 0\end{bmatrix}T^{T}$, it can be verified that \ref{nl:csn1} holds for $i{=}k{-}1$. To see \ref{nl:fns3}, note initially that $T_{2}^{T}W_{k{-}1}^{T}W_{k{-}1}T_{2} = T_{2}^{T}(D_{k{-}1} + D_{k{-}1}^{T})T_{2} = 0$, so $W_{k{-}1}T_{2} = 0$. Next, note that
\begin{equation}
\label{eq:sfl2}
\hspace*{-0.5cm}\left[\begin{smallarray}{c|cc}\! \xi I {-} A_{k}& -B_{k}& 0\\ L_{k}& W_{k}& 0\!\end{smallarray}\right]\! {=} \!\left[\!\begin{smallmatrix}\xi I {-} A_{k{-}1}& -B_{k{-}1}\\ L_{k{-}1}& W_{k{-}1}\end{smallmatrix}\!\right]\!\left[\!\begin{smallmatrix}I& 0\\ 0& T\end{smallmatrix}\!\right]\!.
\end{equation}
We denote the rightmost matrix in (\ref{eq:sfl2}) by $S$; we let $\Omega_{k{-}1}(X_{k{-}1})$ and $\Omega_{k}(X_{k})$ be as in \ref{nl:csn2}, and we note that $S^{T}\Omega_{k{-}1}(X_{k{-}1})S = \text{diag}\begin{pmatrix}\Omega_{k}(X_{k}) & 0\end{pmatrix}$. This shows \ref{nl:fns3}(ii). Also, since $S$ is nonsingular, then \ref{nl:fns3}(iii) holds. The proof of \ref{nl:fns4} is similar, noting that (\ref{eq:sfl2}) also holds in this case.
\end{IEEEproof}

\begin{lemapp}
\label{lem:lrz}
Let   $P_{k{-}1}, Q_{k{-}1}$ satisfy \ref{nl:ip1}--\ref{nl:ip3} for $i {=} k{-}1$, with $m_{k} \coloneqq  n_{k{-}1} - r_{k{-}1} > 0$. The following hold.
\begin{enumerate}[label=\arabic*., ref=\arabic*, leftmargin=0.5cm]
\item There exists $0 < K \in \mathbb{R}^{m_{k} \times m_{k}}$ such that $\lim_{\xi \rightarrow \infty}(\tfrac{1}{\xi}P_{k{-}1}^{-1}Q_{k{-}1}(\xi)) = \text{diag}\begin{pmatrix}0& K\end{pmatrix}$.
\label{nl:lrz1} 
\item Let $P_{k}(\xi) \coloneqq Q_{k{-}1}(\xi) - P_{k{-}1}(\xi)\text{diag}\begin{pmatrix}0 & K\xi\end{pmatrix}$, and $Q_{k} \coloneqq P_{k{-}1}$. Then \ref{nl:ip1} holds for $i = k$; $\deg{(\det{(Q_{k})})} < \deg{(\det{(Q_{k{-}1})})}$; and there exist $\hat{D}_{12} \in \mathbb{R}^{r_{k{-}1}{\times}m_{k}}, \hat{D}_{21} \in \mathbb{R}^{m_{k}{\times}r_{k{-}1}}, \hat{D}_{22} \in  \mathbb{R}^{m_{k}{\times}m_{k}}$ such that\label{nl:lrz2}
\begin{equation}
\lim_{\xi \rightarrow \infty}(Q_{k}^{-1}P_{k}(\xi)) \eqqcolon D_{k} = \begin{bmatrix}I_{r_{k{-}1}} & \hat{D}_{12}\\ \hat{D}_{21}& \hat{D}_{22}\end{bmatrix}.\label{eq:digf} 
\end{equation}
\item Let $A_{k}, B_{k}, C_{k}, D_{k}$ satisfy \ref{nl:csn1} for $i = k$; partition $B_{k}, C_{k}$ compatibly with $D_{k}$ as $B_{k} = \begin{bmatrix}\hat{B}_{1}& \hat{B}_{2}\end{bmatrix}$, $C_{k} = \text{col}\begin{pmatrix}\hat{C}_{1}& \hat{C}_{2}\end{pmatrix}$; and let 
\begin{align*}
&A_{k{-}1} \coloneqq \begin{bmatrix} A_{k} - \hat{B}_{1} \hat{C}_{1} & \hat{B}_{2} K^{-1} - \hat{B}_{1} \hat{D}_{12} K^{-1}\\  \hat{D}_{21} \hat{C}_{1} - \hat{C}_{2} & \hat{D}_{21} \hat{D}_{12} K^{-1} - \hat{D}_{22} K^{-1}\end{bmatrix}, \\ 
&B_{k{-}1} \coloneqq \begin{bmatrix} \hat{B}_{1} & 0 \\  - \hat{D}_{21} & I\end{bmatrix}, \text{ and } C_{k{-}1} \coloneqq \begin{bmatrix} - \hat{C}_{1} &  - \hat{D}_{12} K^{-1} \\ 0 & K^{-1}\end{bmatrix}. 
\end{align*}
Then:\label{nl:lrza3}
\begin{enumerate}
\item \ref{nl:csn1} holds for $i = k{-}1$.\label{nl:lrz4}
\item Let $X_{k{-}1}, L_{k{-}1}$ and $W_{k{-}1}$ satisfy \ref{nl:csn2} for $i = k{-}1$; partition $L_{k{-}1}$ and $W_{k{-}1}$ compatibly with $A_{k{-}1}, B_{k{-}1}, C_{k{-}1}$ and $D_{k{-}1}$ as $L_{k{-}1} \coloneqq \begin{bmatrix}\tilde{L}_{1}& \tilde{L}_{2}\end{bmatrix}$ and $W_{k{-}1} \coloneqq \begin{bmatrix}\tilde{W}_{1}& \tilde{W}_{2}\end{bmatrix}$; and let $L_{k} \coloneqq \tilde{L}_{1} + \tilde{W}_{1}\hat{C}_{1}$, and $W_{k} \coloneqq \begin{bmatrix}\tilde{W}_{1} & \tilde{L}_{2}K+\tilde{W}_{1}\hat{D}_{12}\end{bmatrix}$. Then (i) $\tilde{W}_{2} = 0$; (ii) $X_{k{-}1}$ has the form $X_{k{-}1} = \text{diag}\begin{pmatrix}X_{k}& K^{-1}\end{pmatrix}$; (iii) with $X_{k}$ as in condition (ii), then \ref{nl:csn2} holds for $i = k$; and (iv) if \ref{nl:csn3} holds for $i = k{-}1$, then \ref{nl:csn3} holds for $i = k$.\label{nl:lrz5}
\item Let $X_{k}, L_{k}$ and $W_{k}$ satisfy \ref{nl:csn2} for $i = k$; partition $W_{k}$ compatibly with $D_{k}$ as $W_{k} = \begin{bmatrix}\hat{W}_{1}& \hat{W}_{2}\end{bmatrix}$; and let $L_{k{-}1} \coloneqq \begin{bmatrix}L_{k} - \hat{W}_{1}\hat{C}_{1}& (\hat{W}_{2}-\hat{W}_{1}\hat{D}_{12})K^{-1}\end{bmatrix}$, $W_{k{-}1} \coloneqq \begin{bmatrix}\hat{W}_{1}& 0\end{bmatrix}$ and $X_{k{-}1} \coloneqq \text{diag}\begin{pmatrix}X_{k}& K^{-1}\end{pmatrix}$. Then (i) \ref{nl:csn2} holds for $i = k{-}1$; and (ii) if \ref{nl:csn3} holds for $i = k$, then \ref{nl:csn3} holds for $i = k{-}1$.\label{nl:lrz6}
\end{enumerate}
\end{enumerate}
\end{lemapp}

\begin{IEEEproof}
{\bf Condition \ref{nl:lrz1}.} \hspace{0.3cm} Since $(P_{k{-}1}, Q_{k{-}1})$ is a positive-real pair and $P_{k{-}1}$ is nonsingular, then $P_{k{-}1}^{-1}Q_{k{-}1}$ is positive-real. Hence, if $P_{k{-}1}^{-1}Q_{k{-}1}$ has a pole at infinity, then it is simple and the residue matrix $J \coloneqq \lim_{\xi \rightarrow \infty}(\tfrac{1}{\xi}(P_{k{-}1}^{-1}Q_{k{-}1})(\xi))$ is real and non-negative definite \cite[Theorem 2.7.2]{AndVong}. Thus, there exist real matrices $J, \hat{D}$ and strictly proper real-rational matrices $G, H$ (partitioned compatibly with $D_{k{-}1}$) such that 
\begin{align}
P_{k{-}1}^{-1}Q_{k{-}1}(\xi) &= \left[\begin{smallmatrix}J_{11}& J_{12}\\ J_{12}^{T}& J_{22}\end{smallmatrix}\right] \xi + \!\left[\begin{smallmatrix}\hat{D}_{11}& \hat{D}_{12}\\ \hat{D}_{21}& \hat{D}_{22}\end{smallmatrix}\right]\! + \!\left[\begin{smallmatrix}G_{11}& G_{12}\\ G_{21}& G_{22}\end{smallmatrix}\right]\!(\xi)\nonumber\\
Q_{k{-}1}^{-1}P_{k{-}1}(\xi) &= \left[\begin{smallmatrix}I_{r_{k}{-}1}& 0\\0& 0\end{smallmatrix}\right] + \left[\begin{smallmatrix}H_{11}& H_{12}\\ H_{21}& H_{22}\end{smallmatrix}\right](\xi).\label{eq:lip}
\end{align}

By considering the first block row in the equation $0 = \lim_{\xi \rightarrow \infty}(\tfrac{1}{\xi}(Q_{k{-}1}^{-1}P_{k{-}1}P_{k{-}1}^{-1}Q_{k{-}1})(\xi))$, we obtain $J_{11} = 0$ and $J_{12} = 0$. Then, by considering the bottom right block in the equation $\lim_{\xi \rightarrow \infty}((P_{k{-}1}^{-1}Q_{k{-}1}Q_{k{-}1}^{-1}P_{k{-}1})(\xi)) = I$, we find that $J_{22}\lim_{\xi \rightarrow \infty}(\xi H_{22}(\xi)) = I$, which implies that $J_{22}$ is nonsingular. By letting $K \coloneqq J_{22}$, we obtain condition \ref{nl:lrz1}.

{\bf Condition \ref{nl:lrz2}.} \hspace{0.3cm}  Since $Q_{k}^{-1}P_{k}(\xi) = P_{k{-}1}^{-1}Q_{k{-}1}(\xi) - \text{diag}\begin{pmatrix}0& K\xi\end{pmatrix}$, then $Q_{k}^{-1}P_{k}$ is proper and positive-real by \cite[Theorem 8.4.3]{AndVong}, and $\lim_{\xi \rightarrow \infty}((Q_{k}^{-1}P_{k})(\xi))$ is equal to the matrix $\hat{D}$ in (\ref{eq:lip}). Then, the top left block in the equation $I = \lim_{\xi \rightarrow \infty}((P_{k{-}1}^{-1}Q_{k{-}1}Q_{k{-}1}^{-1}P_{k{-}1})(\xi))$ gives $\hat{D}_{11} = I_{r_{k{-}1}}$. Next, note that $\begin{bmatrix}P_{k}& -Q_{k}\end{bmatrix} = \begin{bmatrix}P_{k{-}1}& -Q_{k{-}1}\end{bmatrix}S$ where
\begin{equation*}
S = \begin{bmatrix}S_{11}& -I\\ -I& 0\end{bmatrix}, \text{ with } S_{11}(\xi) = \begin{bmatrix}0& 0\\0& -K\xi\end{bmatrix}.
\end{equation*}
With $J_{n}$ as defined in Appendix \ref{app:prbrp}, it can be verified that $S$ is unimodular and $SJ_{n}S^{\star} = J_{n}$, and it is then easily shown that $(P_{k}, Q_{k})$ satisfy conditions \ref{nl:prpc2} and \ref{nl:prpc3} in Definition \ref{def:prp}. Since, in addition, $Q_{k}^{-1}P_{k}$ is positive-real, then $(P_{k}, Q_{k})$ also satisfies condition \ref{nl:prpc1} in Definition \ref{def:prp}, so $(P_{k}, Q_{k})$ is a positive-real pair. 

Finally, that $\deg{(\det{(Q_{k})})} < \deg{(\det{(Q_{k{-}1})})}$ will follow from condition \ref{nl:lrz4}, noting from the final two block columns in \ref{nl:csn1} that $\deg{(\det{(Q_{i})})} = \deg{(\det{(\mathcal{A}_{i})})}$.

{\bf Condition \ref{nl:lrza3}.} \hspace{0.15cm}  Let $\mathcal{A}_{k}, M_{k}, N_{k}, U_{k}, V_{k}, E_{k}, F_{k}$ be as in \ref{nl:csn1} for the case $i=k$, partition these matrices compatibly as 
\begin{equation*}
\left[\!\begin{smallarray}{cc|c}\!\hat{M}_{11}& \hat{M}_{12}& \hat{N}_{1}\\ \hat{M}_{21}& \hat{M}_{22}& \hat{N}_{2}\\ \hline  \rule{0pt}{1.05\normalbaselineskip} \hat{U}_{1}& \hat{U}_{2}& \hat{V}\end{smallarray}\!\right]\!, \text{ and } \!\left[\!\begin{smallarray}{cc|c|c}\!\begin{smallarray}{c}-I\\ -\hat{D}_{21}\end{smallarray}\!& \!\begin{smallarray}{c}-\hat{D}_{12}\\ -\hat{D}_{22}\end{smallarray}\!& I& \!\begin{smallarray}{c}-\hat{C}_{1}\\ -\hat{C}_{2}\end{smallarray}\! \\ \hline  \rule{0pt}{1.05\normalbaselineskip} -\hat{B}_{1}& -\hat{B}_{2}& 0& \mathcal{A}_{k} \!\end{smallarray}\!\right]\!,
\end{equation*}
and let 
\begin{equation*}
\left[\!\begin{smallmatrix}M_{k{-}1}& N_{k{-}1}\\ U_{k{-}1}& V_{k{-}1}\end{smallmatrix}\!\right]\! {=} \!\left[\! \begin{smallarray}{cc|cc}\hat{M}_{11} & \hat{M}_{12}& \hat{N}_{1}& 0 \\ \hat{M}_{21}& \hat{M}_{22} & \hat{N}_{2} & 0 \\ \noalign{\hrule} \rule{0pt}{1.05\normalbaselineskip} -\hat{U}_{1}& -\hat{U}_{2}& -\hat{V}& 0\\ 0& I& 0& -I\end{smallarray} \!\right]\! \!\left[\!\begin{smallarray}{cc|cc}I& \hat{D}_{12}& 0& 0\\ \hat{D}_{21} &  \hat{D}_{22} + K\xi & 0 & I\\ \noalign{\hrule} \rule{0pt}{1.05\normalbaselineskip} \hat{B}_{1} & \hat{B}_{2} & -I & 0\\ \hat{D}_{21}& \hat{D}_{22}+K(1+\xi) & 0 & I\end{smallarray}\!\right]\!.
\end{equation*}
It can be verified that each of the above matrices is unimodular (the modulus of the determinant of the rightmost matrix is equal to $\det{(K)}$). Also, with $E_{k{-}1} \coloneqq \text{col}\begin{pmatrix}F_{k}& 0\end{pmatrix}$ and $F_{k{-}1}(\xi) \coloneqq \text{col}\begin{pmatrix}E_{k}(\xi)& 0\end{pmatrix} + \text{col}\begin{pmatrix}\xi \hat{U}_{2}(\xi)& I\end{pmatrix}\begin{bmatrix}0& K\end{bmatrix}$, it can be verified that
\ref{nl:csn1} holds for $i=k{-}1$. Now, let $L_{k{-}1}, W_{k{-}1}$ be as in condition \ref{nl:lrz5}. Since $W_{k{-}1}^{T}W_{k{-}1} = D + D^{T} = \text{diag}\begin{pmatrix}2I_{r_{k{-}1}}& 0\end{pmatrix}$, then $\tilde{W}_{2}^{T}\tilde{W}_{2} = 0$, which proves \ref{nl:lrz5}(i). To show \ref{nl:lrz5}(ii), we partition $X_{k{-}1}$ compatibly with $A_{k{-}1}$ as 
\begin{equation*}
X_{k{-}1} = \begin{bmatrix}X_{11}& X_{12}\\ X_{12}^{T}& X_{22}\end{bmatrix}.
\end{equation*}
Since $C_{k{-}1}^{T} - X_{k{-}1}B_{k{-}1} = L_{k{-}1}^{T}W_{k{-}1} = \begin{bmatrix}L_{k{-}1}^{T}\tilde{W}_{1}& 0\end{bmatrix}$, then $X_{12} = 0$ and $X_{22} = K^{-1}$. Now, note that
\begin{equation*}
\left[\!\begin{smallarray}{cc|cc}\xi I {-} A_{k}& {-}\hat{B}_{1}& {-}\hat{B}_{2}& 0\\0 & 0& 0& {-}I\\ \hline  \rule{0pt}{1.05\normalbaselineskip} L_{k}& \hat{W}_{1}& \hat{W}_{2}& 0\end{smallarray}\!\right]\! {=} \!\left[\! \begin{smallmatrix}\xi I {-} A_{k{-}1}& {-}B_{k{-}1}\\ L_{k{-}1}& W_{k{-}1}\end{smallmatrix}\!\right]\! \!\left[\!\begin{smallarray}{cc|cc}I& 0& 0& 0\\ 0& 0& K& 0\\ \hline  \rule{0pt}{1.05\normalbaselineskip} \hat{C}_{1}& I& \hat{D}_{12}& 0\\ \hat{C}_{2} & \hat{D}_{21}& \xi K {+} \hat{D}_{22}& I\end{smallarray} \!\right]\!.\label{eq:sflrz}
\end{equation*}
We denote the rightmost matrix in this equation by $S$, we let $\Omega_{k{-}1}(X_{k{-}1})$ and $\Omega_{k}(X_{k})$ be as in \ref{nl:csn2}, and by direct calculation we obtain $S^{T}\Omega_{k{-}1}(X_{k{-}1})S = \text{diag}\begin{pmatrix}\Omega_{k}(X_{k})& 0\end{pmatrix}$. This proves \ref{nl:lrz5}(iii). Condition \ref{nl:lrz5}(iv) then follows since the rightmost matrix in the above displayed equation is nonsingular. 

Next, let $X_{k}, L_{k}, W_{k}, X_{k{-}1}, L_{k{-}1}$ and $W_{k{-}1}$ be as in condition \ref{nl:lrz6}. We recall that the rightmost matrix in the above displayed equation (denoted $S$) is nonsingular. We then find that $\Omega_{k{-}1}(X_{k{-}1}) {=} (S^{-1})^{T}\text{diag}\begin{pmatrix}\Omega_{k}(X_{k}) & 0\end{pmatrix}S^{-1}$, so \ref{nl:lrz6}(i) holds, and \ref{nl:lrz6}(ii) follows since $S^{-1}$ is nonsingular.
\end{IEEEproof}

\begin{IEEEbiography}[{\includegraphics[width=1in,height=1.25in,clip,keepaspectratio]{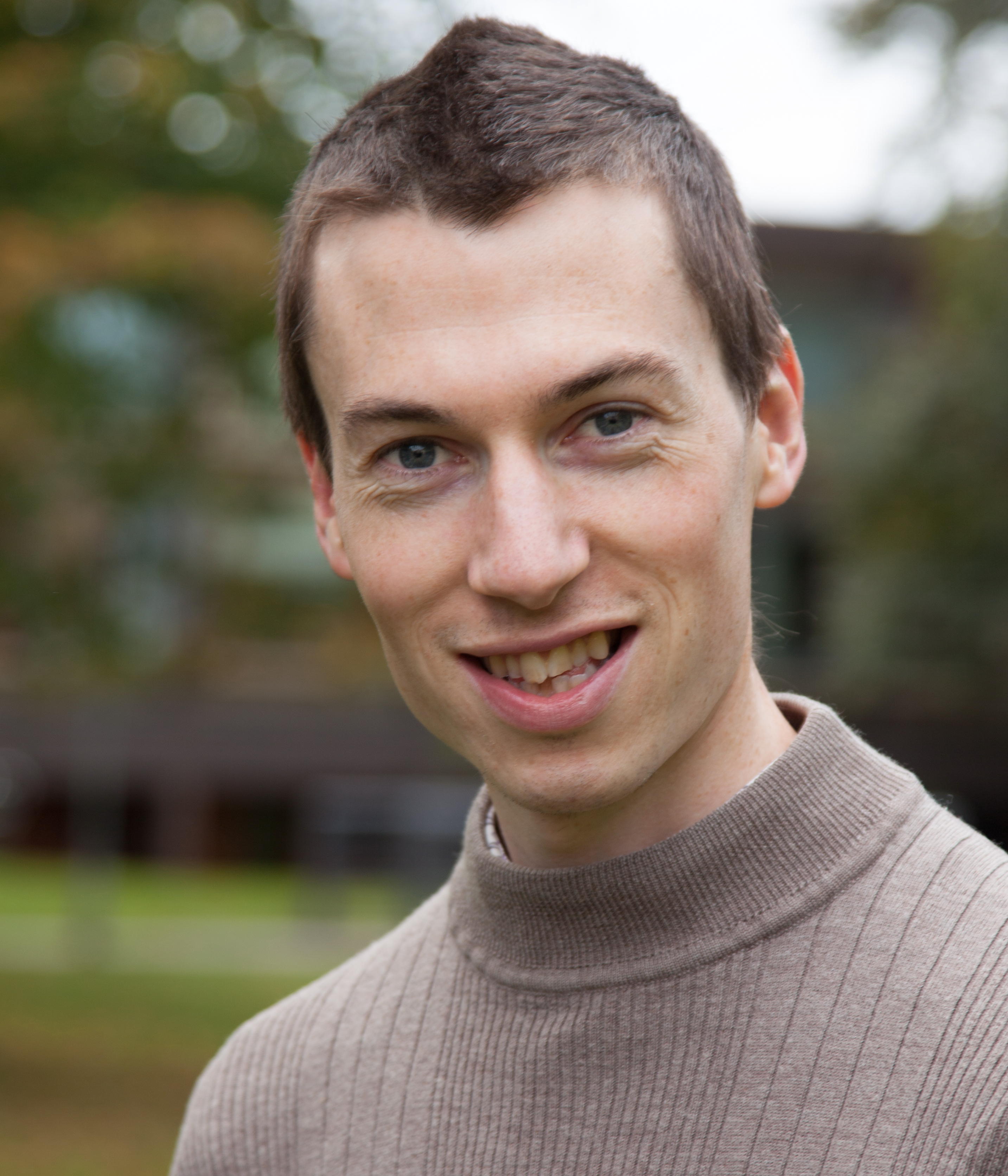}}]{Timothy
 H. Hughes}

 received the M.Eng.\ degree in mechanical engineering, and the Ph.D.\ degree in control engineering, from the University of Cambridge, U.K., in 2007 and 2014, respectively. From 2007 to 2010 he was employed as a mechanical engineer at The Technology Partnership, Hertfordshire, U.K; and from 2013 to 2017 he was a Henslow Research Fellow at the University of Cambridge, supported by the Cambridge Philosophical Society. He is currently a lecturer in the Department of Mathematics at the University of Exeter, U.K.
 \end{IEEEbiography}
 \vfill
\end{document}